%% file: flags_umap.tex
\documentclass[twocolumn]{openjournal}
\usepackage{natbib}
\usepackage{graphicx,amsmath,amssymb,amstext}
\usepackage{amsbsy,amsfonts,amsthm,color}
\usepackage[colorlinks,linkcolor=blue,citecolor=blue,urlcolor=blue ]{hyperref}
\usepackage[utf8]{inputenc}
\usepackage{float}
\usepackage[caption=false]{subfig}
\usepackage{upgreek}
\usepackage{lipsum}

\usepackage[usenames,dvipsnames]{xcolor}
\usepackage[normalem]{ulem} 
\usepackage{array}
\usepackage{booktabs} 
\usepackage{multirow}
\usepackage{orcidlink}

\newcommand{\beagle}{\mbox{\tt{BEAGLE}}}
\newcommand{\jwst}{\mbox{\emph{JWST}}}
\newcommand{\hst}{\mbox{\emph{HST}}}
\newcommand{\scsam}{\mbox{\sc sc-sam}}
\newcommand{\jaguar}{\mbox{\sc jaguar}}
\newcommand{\sage}{\mbox{\sc sage}}
\newcommand{\spritz}{\mbox{\sc spritz}}
\newcommand{\flags}{\mbox{\sc flags}}

\newcolumntype{C}[1]{>{\centering\arraybackslash}p{#1}}

\newcommand\blfootnote[1]{%
  \begingroup
  \renewcommand\thefootnote{}\footnote{#1}%
  \addtocounter{footnote}{-1}%
  \endgroup
}

\defcitealias{Turner_2026}{\mbox{\textsc{flags-i}}}
\newcommand{\flagsi}{\citetalias{Turner_2026}}

\renewcommand{\farcs}{.\!\!^{\prime\prime}}

\begin{document}

\title{FLAGS II: Constraining galaxy formation models with dimensionality reduction of direct observables\vspace{-1em}}

\author{Jack C. Turner$^{1\star \, \orcidlink{0000-0001-6247-041X}}$}
\author{Stephen M. Wilkins$^{1,2 \, \orcidlink{0000-0003-3903-6935}}$}
\author{William J. Roper$^{1 \, \orcidlink{0000-0002-3257-8806}}$}
\author{Aswin P. Vijayan$^{1 \, \orcidlink{0000-0002-1905-4194}}$}

\blfootnote{$^{\star}$\mbox{Corresponding author, email: \href{mailto:jt458@sussex.ac.uk}{jt458@sussex.ac.uk}}}

% List of institutions
\affiliation{$^1$Astronomy Centre, University of Sussex, Falmer, Brighton BN1 9QH, UK}
\affiliation{$^{2}$Institute of Space Sciences and Astronomy, University of Malta, Msida MSD 2080, Malta}

\begin{abstract}

Comparisons between observations of galaxies and theoretical predictions are regularly performed using physical properties, which are inferred by the often slow and biased process of SED fitting. Forward modelling facilitates a reliable alternative, whereby models are evaluated using direct observables alone. However, these datasets become high-dimensional when collating observations from multiple telescopes, leading to sparse sampling, memory intensity and visualisation difficulties. We show that 2D embeddings of \emph{JWST} and \emph{HST} photometric fluxes, constructed using the non-linear dimensionality reduction algorithm UMAP, preserve sufficient information to differentiate between five models. Using a simple $\chi^{2}$-like metric, we show that \jaguar\ reproduces the population of bright galaxies ($m_{\mathrm{AB}}<26$) in GOODS-S six times as well as \scsam\ and twelve times as well as \sage. By adjusting the hyperparameters, we quantify how well each model replicates the distribution of SED shapes. The template SED approach of \spritz\ and the lack of photoionisation in \sage\ cause significant discrepancies, highlighting the importance of comprehensive forward modelling. The embedded position of each galaxy can be identified $>100$ times faster than inferring its properties with Bayesian SED fitting, making this approach an ideal alternative for deriving statistical model constraints from large surveys such as LSST and \emph{Euclid}, and performing simulation-based inference with \textsc{camels}.

\end{abstract}

\maketitle

%%%%%%%%%%%%%%%%%%%%%%%%%%%%%%%%%%%%%%%%%%%%%%%%%%

%%%%%%%%%%%%%%%%% BODY OF PAPER %%%%%%%%%%%%%%%%%%

\input sections/1_intro

\input sections/2_data
\input sections/3_analysis
\input sections/4_results
\input sections/5_conclusion

\section*{Acknowledgements}

We wish to acknowledge contributors to the \jwst\ programmes with IDs 1180, 1210, 1283, 1895, 2079, 2198, 2514 and 6541, as well as \hst\ programmes 9425, 9500, 9793, 9803, 9978, 10086, 10189, 10258, 10530, 10632, 11144, 11359, 11563, 12007, 12060, 12061, 12062, 12099, 12177, 12461, 12498, 12534, 12866, 13779, 13868, 13872 and 15647, from which the imaging products utilised by this work are derived. 

Additional analysis is based on spectra collected by \jwst\ programmes 1180, 1210, 1212, 1286, 1287, 2198, 3215 and 6541.

These products were retrieved from the Dawn JWST Archive (DJA). DJA is an initiative of the Cosmic Dawn Center (DAWN), which is funded by the Danish National Research Foundation under grant DNRF140.

We thank the authors of the models used in this work for making their lightcone catalogues publicly available.

We also wish to acknowledge the following open source software packages used in the analysis: \texttt{}{Astropy} \citep{Astropy_2022}, \texttt{CMasher} \citep{van-der-Velden_2020}, \texttt{h5py}, \texttt{Matplotlib} \citep{Hunter_2007}, \texttt{Numpy} \citep{Harris_2020}, \texttt{scikit-learn} \citep{Pedregosa_2011} and \texttt{Scipy} \citep{Virtanen_2020}.

Jack C. Turner would like to acknowledge Jonathan Loveday and Vivienne Wild, whose comments greatly improved the manuscript.

JCT is supported by an STFC PhD studentship ST/X508822/1.
APV and SMW acknowledge support from the Sussex Astronomy Centre STFC Consolidated Grant (ST/X001040/1).

We list here the roles and contributions of the authors according to the Contributor Roles Taxonomy (CRediT)\footnote{\url{https://credit.niso.org/}}.
\textbf{JCT}: Conceptualisation, Data curation, Methodology, Investigation, Formal Analysis, Visualisation, Writing - original draft. \textbf{SMW}: Conceptualisation, Methodology, Supervision, Writing - review \& editing. \textbf{APV}, \textbf{WJR}: Writing - review \& editing.\\

\section*{Data Availability}

Links to the fiducial catalogues, embedded positions and associated analysis code will be added here upon acceptance of the manuscript. Processed imaging and additional catalogue variations will be provided upon reasonable request to the corresponding author.

%%%%%%%%%%%%%%%%%%%% REFERENCES %%%%%%%%%%%%%%%%%%

% The best way to enter references is to use BibTeX:
\bibliographystyle{mnras}
\bibliography{flags_umap} 
%%%%%%%%%%%%%%%%%%%%%%%%%%%%%%%%%%%%%%%%%%%%%%%%%%

%%%%%%%%%%%%%%%%% APPENDICES %%%%%%%%%%%%%%%%%%%%%

%\appendix
%\input sections/A1

\end{document}

%% file: sections/1_intro.tex
\section{Introduction}\label{sec:intro}

Recent years have seen an unprecedented increase in the number of both ground and space-based instruments capable of extragalactic imaging. Wide-area surveys performed by \emph{eROSITA}, \emph{Euclid} and LSST provide an unbiased view of the sky and contain the rarest bright and massive sources \citep{Bulbul_2024, Allen_2025}. These allow detailed studies of the mechanisms driving late-time galaxy evolution, such as stellar \citep{Dong_2018} and active galactic nuclei \citep[AGN; ][]{Wylezalek_2020} feedback, gas depletion \citep{Saintonge_2017} and mergers \citep{van-der-Wel_2009, Wilkinson_2022}. In contrast, deep pointed observations from \jwst\ and ALMA leverage greater sensitivity and spatial resolution to detect the faintest objects. Some of these galaxies exist beyond the new redshift frontier at $z>14$ \citep{Bakx_2023, Carniani_2024, Naidu_2025}, allowing us to constrain the physics of early cosmological accretion \citep{Heintz_2023, Tacchella_2023} and the formation of the first stars \citep{Maiolino_2024} and black holes \citep{Juodzbalis_2023}.

The range of observable rest-frame wavelengths is what ultimately determines the physics probed by a telescope. The star formation and metal enrichment history of a galaxy are encoded within its intrinsic spectral energy distribution (SED), alongside the contribution of AGN. Intervening gas and dust leave their own imprints, resulting in nebular emission lines and a reddening of the SED respectively \citep{Conroy_2013, Iyer_2025}. Therefore, if we wish to fully understand galaxies, we must constrain their SEDs with observations spanning a range of wavelengths. 

This multi-wavelength data is readily available from the combination of the aforementioned active observatories and a wealth of archival catalogues \citep{Galametz_2013, Kovlakas_2021, Weaver_2022}, but the question remains how best to analyse it. Under the assumption of a representative sample, these SEDs can be used to constrain theoretical models of galaxy formation. However, most models are unable to predict the observable properties of galaxies ab initio, and those that can offset increased computational cost with a smaller effective volume, lower resolution or restricted redshift limit \citep{Crain_2023}. Observers are therefore often tasked with inferring the physical properties of galaxies using SED fitting. Distribution functions of stellar masses, star formation rates and metallicities can then be used to constrain simulations \citep{Adams_2024, Baker_2025, Shuntov_2025}, from which they are a natural prediction. This can be considered a form of dimensionality reduction (DR), condensing multi-band information into a handful of key parameters. 

While both useful and physically meaningful, SED fitting is plagued by systematic uncertainties and biases that restrict the constraining power of inferred properties \citep{Pacifici_2023, Narayanan_2024, Turner_2026_flares}. This is exacerbated by the need for complex correction factors, which are applied to observed distribution functions to account for incompleteness \citep{Carrasco_2018, Leethochawalit_2022}. In addition to these statistical limitations, SED fitting can be a slow process. While simple template fitting approaches may process each galaxy in only a few seconds, more complex models employing MCMC or nested sampling may take minutes, hours or even days to fully explore the parameter space \citep{Harvey_2025_syn}. The dataset collected by modern surveys, or generated by simulation suites exploring cosmological and astrophysical parameter variation \citep{Villaescusa-Navarro_2021, Katz_2026}, therefore demand unfeasible computational resources.

Reassuringly, it is now routinely possible to forward model simulation data to predict observer-frame properties, such as spectra and photometry \citep{Camps_2015, Wilkins_2016, Fortuni_2023, Lovell_2025b, Roper_2025}. The standard approach is to create synthetic observations from individual simulation snapshots, but subsequent comparisons require the redshift of real sources to be known. While spectroscopy enables this reasonably robustly, unbiased photometric samples are highly model dependent \citep{Dahlen_2013, Steinhardt_2023}. Alternatively, given sufficient volume and snapshot cadence, lightcones with realistic redshift distributions can be used to create synthetic source catalogues or survey maps \citep{Yung_2022, Snyder_2023}. This data can, in principle, be compared directly to real SED samples, without the need for additional inference. Many biases introduced by SED fitting can be bypassed entirely, and completeness corrections can be avoided by applying consistent selection criteria. As the properties of simulated galaxies are known exactly, parameter space exploration is not inherently required, making forward modelling codes both fast and reliable \citep{Lovell_2025b}. This speed makes it feasible to quantify the impact of remaining uncertainties, introduced by SED modelling assumptions. 

With this data in hand, we can explore methods that evaluate each filter flux or spectral bin equally and maximise the constraining power of their distributions. However, we now encounter the curse of dimensionality, with distributions becoming sparsely sampled and computationally demanding as more features are considered. Most data-driven fields are afflicted by this curse, so several dimensionality reduction routines have been developed to overcome it \citep{Wang_2020, Jia_2022}. Largely unsupervised, these approaches identify the lower-dimensional latent space that best preserves the original input information. Astronomy has already adopted a variety of these tools, with Principal Component Analysis (PCA) often used as a means of spectral classification \citep{Deeming_1964, Galaz_1998, Wild_2014, Sharbaf_2023}. While scalable and fast, PCA cannot preserve complex nonlinear relationships between features. Slower non-linear approaches such as t-distributed stochastic neighbour embedding have been applied as unsupervised classifiers \citep{Steinhardt_2020, Dai_2023}, and Self-organising maps have seen a recent surge in popularity, particularly for estimating photometric redshifts \citep{Wright_2020, Kovacevic_2022, Holwerda_2022}. 

Uniform Manifold Approximation \citep[UMAP; ][]{McInnes_2018} is not only non-linear, but scales with dataset size almost as well as basic PCA. This scaling is also largely independent of the number of input dimensions, making it the perfect tool to process aggregated surveys covering a broad range of wavelengths. It has already seen use in astronomy, primarily as a classifier \citep{Rosito_2023, Cook_2024, Asadi_2025, Kamai_2025}, but also for photometric redshift estimation \citep{Zhou_2025}. The computational efficiency of UMAP makes it a promising candidate for analysing large datasets, such as those generated by wide-area surveys and physics varying simulation suites.

This work continues the mission of the \flags\ series, which aims to derive robust constraints on galaxy formation models using direct observables. We investigate whether embeddings generated from broadband photometry alone preserve sufficient information with which to statistically differentiate between models. By processing \jwst\ and \hst\ imaging, we create a multi-band dataset that can be compared to publicly available lightcone data from five models (\S\ref{sec:data}). We outline how UMAP can make such an analysis feasible and detail our implementation (\S\ref{sec:analysis}). We then quantify the accuracy of model predictions using a simple metric and suggest approaches for interpreting the underlying physics (\S\ref{sec:results}). In keeping with the \flags\ philosophy, we evaluate the impact of systematic uncertainties on these results (\S\ref{sec:systematics}).

%% file: sections/2_data.tex
\section{Data}\label{sec:data}

Data selection is influenced by the need for each sample (galaxy) to have data associated with each feature (photometric filter), if DR is to be performed. Non-detections caused by limited imaging depth can be handled naturally, but entirely missing coverage cannot. This cannot be substituted for placeholder information, which would introduce a meaningless connection between the affected inputs and bias the algorithm. Sample selection therefore requires a balance between expanding filter coverage to maximise the encoded information and the sky area to boost statistical significance. We are also limited by the effective volumes probed by simulations, which should be comparable to the observations.

\subsection{Observations}\label{sec:obs}

After considering the synergy between public datasets, we focus on \jwst\ and \hst\ imaging of the Southern Great Observatories Origins Deep Survey field (GOODS-S). Mosaics are sourced from the DAWN JWST Archive (DJA), where calibrated science and weight maps are publicly available. These have the desirable quality of being consistently reduced across survey instruments \citep{brammer_2023_msaexp, Valentino_2023}, removing a systematic which may otherwise need consideration when combining data from multiple sources \citep{Williams_2024}. The \jwst\ data was collected primarily as part of the JADES survey \citep{Eisenstein_2026}, and spans $\sim 100 \ \rm{arcmin}^{2}$. JADES utilised several wide and medium-band near-infrared (NIR) filters, from which we selected eight based on their availability in simulated datasets. This includes F090W, F115W, F150W and F200W in the short-wavelength and F277W, F356W, F410M and F444W in the long-wavelength \jwst/NIRCam channel. These are supplemented by F125W and F160W from \hst/WFC3 and optical data from \hst/ACS F435W, F606W and F814W. These filters probe the rest-frame optical at all redshifts, and the rest-frame UV at $z\gtrsim2.5$, providing strong constraints on star formation across cosmic time. All contributing programme IDs are listed in the acknowledgements.

\renewcommand{\arraystretch}{1.2}
\begin{table}
\centering
\begin{tabular}{cc}
\toprule
\textbf{Parameter} & \textbf{Value}\\

\midrule

BACK\_VALUE & 0.0\\
BACKPHOTO\_TYPE & LOCAL\\
BACKPHOTO\_THICK & 18\\

\midrule

WEIGHT\_TYPE & MAP\_RMS, MAP\_RMS\\
DETECT\_THRESH & 1.5\\
DETECT\_MINAREA & 5\\
FILTER\_NAME & gauss\_2.0\_3x3\\
DEBLEND\_MINCONT & 0.005\\
DEBLEND\_NTHRESH & 32\\
CLEAN & Y\\
CLEAN\_PARAM & 5.0\\

\midrule

PHOT\_AUTOPARAMS & 2.5, 1.1\\
MASK\_TYPE & BLANK\\

\bottomrule
\end{tabular}
\caption{The key \texttt{Source Extractor} parameters used to generate the photometry catalogue.}
\label{tab:fiducial}

\end{table}

Short wavelength DJA imaging is provided at $0.02 \ \mathrm{arcsec/pixel}$ resolution, which we rebin to match the long wavelengths at $0.04 \ \mathrm{arcsec/pixel}$. These images are then processed using the \texttt{CREST}\footnote{\url{https://github.com/jackcturner/crest}} library, following the approach of \citet[\textsc{flags-i}]{Turner_2026}. This begins with background subtraction, informed by iterative source masking based broadly on the \cite{Bagley_2023} routine. Point-spread functions (PSFs) are then measured empirically by selecting star candidates in the size-magnitude plane \citep{Skelton_2014, Whitaker_2019}, before re-centring, stacking and sigma-clipping their cutouts \citep{Weaver_2024}. The observing angle varies across the field, most notably in \hst\ imaging built up over decades, so these should be considered field-averaged PSFs. We then use \texttt{PyPHER} \citep{Boucaud_2016} to generate kernels with which to match the resolution of each image to that of F444W. A regularisation parameter of $r=10^{-4}$ is assumed, which minimises the residuals between true and convolved PSFs. We apply this PSF homogenisation to all NIRCam and ACS filters, but not to F125W or F160W, because their native PSFs are broader than F444W. We instead account for this with an aperture correction, described later in this section.

Galaxies are identified from an inverse-variance weighted stack of the F277W, F356W and F444W images to minimise spurious detection of noise fluctuations. We then use \mbox{\texttt{Source~Extractor~v2.28.0}} \citep[SE; ][]{Bertin_1996} to detect sources and measure photometry, adopting key parameters which are consistent with \flagsi\ and listed in Table \ref{tab:fiducial}. The sensitivity of our results to these parameter choices, as well as the choice of code, is quantified in Section \ref{sec:systematics}. Photometry is measured in Kron apertures \citep{Kron_1980} positioned and sized based on the stacked source shapes. When scaled by a factor of $2.5$, these apertures are expected to capture $\sim90\%$ of the flux \citep{Graham_2005}, so we apply an aperture correction to recover closer to the total. We construct the circular aperture that encloses an area equivalent to the Kron, and then scale the flux by the fraction of the PSF enclosed within. The F444W PSF is used for the homogenised mosaics, whereas the native PSFs are used for the WFC3 filters. The total fluxes are corrected for galactic extinction using the \cite{Schlafly_2011} dust map in the \hst\ or UKIRT filter that is closest in wavelength. Hot pixels are removed by comparing the FWHM measured from the detection image to that of the F444W PSF. Sources narrower than the PSF, within a $10\%$ tolerance to guard against the misclassification of quasars, are removed. Stars and their large diffraction spikes are masked semi-automatically based on GAIA classifications \citep{GAIA_2016, GAIA_2023}, and a second mask removes all sources within 40 pixels of an image edge. These masks are combined with a third that removes regions without coverage in all thirteen filters, resulting in a final unmasked sky area of $\sim 43 \ \rm{arcmin}^{2}$. 

\subsection{Models}\label{sec:models}

Our analysis is limited by the availability of public forward modelling data, derived from boxes with sufficient volume to generate representative lightcones. As a result, we use public lightcones from five models, which span a sufficient redshift range and provide photometry in the required bands. These models are quite dissimilar, which complicates interpretations of the underlying physics from a statistical observer-frame framework. Regardless, these models are sufficient to demonstrate the constraining power of DR, and we outline applications better suited to determining the physical drivers in Section \ref{sec:applications}.

\subsubsection{Semi-Analytic Models}

The Santa Cruz semi-analytical model \citep[\scsam; ][]{Somerville_2015, Yung_2019} and Semi-Analytic Galaxy Evolution \citep[\sage; ][]{Croton_2006, Croton_2016} are, as their names suggest, semi-analytic models (SAMs). Such models are applied post-facto to numerical dark matter only (DMO) simulations, by assigning galaxies to nodes of halo merger trees. The formation and evolution of galaxies is then modelled using a set of equations describing key mechanisms such as cosmological accretion, gas cooling and star formation, black hole growth, and stellar and AGN feedback. The merger history of a galaxy and whether it is a central or satellite are traced using the underlying dark matter and inform its physical treatment \citep{Somerville_2015_rev}. We refer the reader to the highlighted papers for details of these treatments, but summarise the forward modelling and catalogue details below.

The full star formation and metal enrichment histories of the galaxies are known and can be used to self-consistently model their intrinsic emission. \scsam\ lightcones are based on the $250 \ \rm{Mpc/h}$ Bolshoi-Planck \mbox{MultiDark} DMO simulation, with $1.55\times10^{8} \ \mathrm{M_{\odot}/h}$ mass resolution \citep{Klypin_2016, Rodriguez-Puebla_2016}. The GOODS-S lightcone covers $1000 \ \rm{arcmin}^{2}$ and includes all $0<z<10$ galaxies with $M_{\ast} > 10^{7} \ \mathrm{M_{\odot}}$ and $M_{\mathrm{DM}} > 10^{10} \ \mathrm{M_{\odot}}$ \citep{Somerville_2021, Yung_2022}. Photometry is based on the BC03 stellar population synthesis (SPS) model \citep{Bruzual_2003}, assuming a \cite{Chabrier_2003} initial mass function (IMF). Nebular emission arising from birth clouds and post-AGB stars is modelled following the \cite{Hirschmann_2017, Hirschmann_2019, Hirschmann_2023} approach. Dust attenuation is modelled by a 'slab' intervening the line of sight \citep{Devriendt_1999, Somerville_2012}, which adopts a \cite{Calzetti_2000} dust attenuation curve with a metallicity and cold gas density dependent normalisation. Additional attenuation is applied to account for stellar birth clouds and the intervening intergalactic medium \citep[IGM; ][]{Madau_1996}.

We use the Theoretical Astrophysical Observatory \citep[TAO; ][]{Bernyk_2016} to generate a \sage\ lightcone overlapping the same field. We remain consistent with \scsam\ by adopting the same underlying DMO simulation, SPS model and IMF. The dust model is a similar metallicity-based slab, this time adopting a \cite{Mathis_1983} attenuation curve and additional redshift and albedo-dependent scaling factors. IGM absorption is modelled following \cite{Fan_2006}, which we do not expect to cause a significant difference and only affects galaxies at $z\gtrsim4$. In contrast, the TAO does not support nebular emission modelling, which will significantly alter the photometric colours at all redshifts. The catalogue spans the same redshift range as \scsam, but includes galaxies down to $M_{\ast} = 10^{5} \ \rm{M_{\odot}}$. While \scsam\ and \sage\ both implement AGN in their physical models, they do not contribute to the photometry. 

\subsubsection{Semi-Empirical Models}

Semi-empirical models are constructed from observed distribution functions and scaling relations, so the underlying relationships are descriptive rather than causal. They therefore offer limited insight into the processes driving galaxy evolution, but can be used to demonstrate the statistical distinguishing power of DR regardless.

The JAdes extraGalactic Ultradeep Artificial Realizations \citep[\jaguar;][]{Williams_2018} describe the evolution of the galaxy number counts and spectral energy distributions across cosmic time. Number density and mass distributions at $z\lesssim4$ are derived from empirical stellar mass functions, measured by \hst\ and \emph{Spitzer}/IRAC. At higher redshifts, they are instead forward modelled from the UV-luminosity function, which can be directly probed by these observatories. Galaxies are assigned spectra dependent on their redshift and stellar mass, reflecting their observability. Massive galaxies at $z<4$ are matched to real sources from the 3D-HST catalogue \citep{Skelton_2014}, adopting the spectrum of the source with the most comparable stellar mass and redshift. These quantities are inferred from the observations by SED fitting with \beagle\ \citep{Chevallard_2016}. This assumes a delayed exponential star formation history, the BC03 stellar spectra and \cite{Chabrier_2003} IMF. Photoionisation is modelled by \texttt{Cloudy} \citep[version 13.3;][]{Ferland_2013}, and dust attenuation is accounted for by a two-component \cite{Charlot_2000} model. The spectra of low mass galaxies and those at $z>4$ are selected from a grid generated by varying the parameters underpinning this model. As well as the stellar mass and redshift, the UV-luminosity and continuum slope are derived from empirical relations and used to determine the most appropriate synthetic spectrum. A $121 \ \rm{arcmin}^{2}$ realisation of includes galaxies with stellar masses $M_{\ast} > 10^{6} \ \rm{M_{\odot}}$ and spans $0.2 < z < 15$. Galaxies from GOODS-S constitute a significant fraction of the 3D-HST catalogue, which should be considered when evaluating \jaguar's agreement with observations.

Spectro-Photometric Realisations of IR-Selected Targets at all-Z \citep[\spritz;][]{Bisigello_2021b} also adopts a semi-empirical approach. The number density and redshift distributions of seven separate galaxy populations are derived from dedicated luminosity or galaxy stellar mass function measurements, primarily measured by \emph{Herschel} \citep{Gruppioni_2013}. These could not be observed at $z>3$ for spirals, starbursts and two composite AGN populations, so the relations in this regime are extrapolated from lower redshift. This assumes a constant characteristic luminosity and a decreasing number density at the knee $\propto(1 + z)^{k_{\Phi}}$. Galaxies are assigned one of 72 empirical rest-frame SED templates, after applying dust attenuation \citep{Charlot_2000}, nebular emission based on empirical relations, and narrow line AGN emission \citep{Feltre_2016}. Physical properties are derived from these templates using further empirical relations. We use two $46 \ \rm{arcmin}^{2}$ realisations of GOODS-S, adopting $k_{\Phi} = -1$ and $-4$, which we refer to as \spritz$_{1}$ and \spritz$_{4}$ respectively. Both catalogues contain galaxies with $M_{\ast} > 10^{4} \ \rm{M_{\odot}}$ and span $0.01 < z < 10$.

\subsubsection{Synthetic Noise}

Observed fluxes are affected by background noise and contamination that cause the measured value to deviate from the truth. These can arise from detector effects or background objects, neither of which affect simulated galaxies. We account for any potential biases by adding synthetic noise to the model fluxes. To get an estimate of the appropriate level, we first fit the relationship between the flux observed in each filter and its uncertainty. We adopt a linear function in logarithmic space, which well recovers the overall trend. Each flux distribution is then split into 25 logarithmically spaced bins, ensuring reliable statistics in each bin and that the noise-sensitive faint end is well sampled. The standard deviation of the residual between true and fit uncertainty is calculated within each bin, and interpolated to get an estimate of the flux-dependent scatter $\sigma_{\mathrm{noise}}$. The final uncertainty on each simulated flux is the sum of the fit extracted value and a Gaussian term $\mathcal{N}
(0, \sigma_{\mathrm{noise}})$. Each simulated flux is then resampled from a Gaussian centred on the true value with a width defined by this combined uncertainty. The resampled fluxes are used exclusively herein.

\subsection{Sample Selection}\label{sec:sample}

While unconstrained, the \scsam, \sage\ and \jaguar\ lightcones replicate the geometry of GOODS-S, so it is straightforward to select galaxies that fall within the selection mask described in Section \ref{sec:obs}. \spritz\ does not provide field-specific RA and DEC values, so we apply an RA cut that recovers an equivalent sky area. To guard against the inclusion of observed spurious sources, we require each galaxy to be detected with $S/N>3$ in at least five filters. This ensures that we select objects that appear consistently, without biasing against high-redshift galaxies that may `drop out' in short wavelength filters probing below the Lyman break. The synthetic noise estimates allow this same selection to be applied to the models.

\begin{figure}
 	\includegraphics[width=\columnwidth]{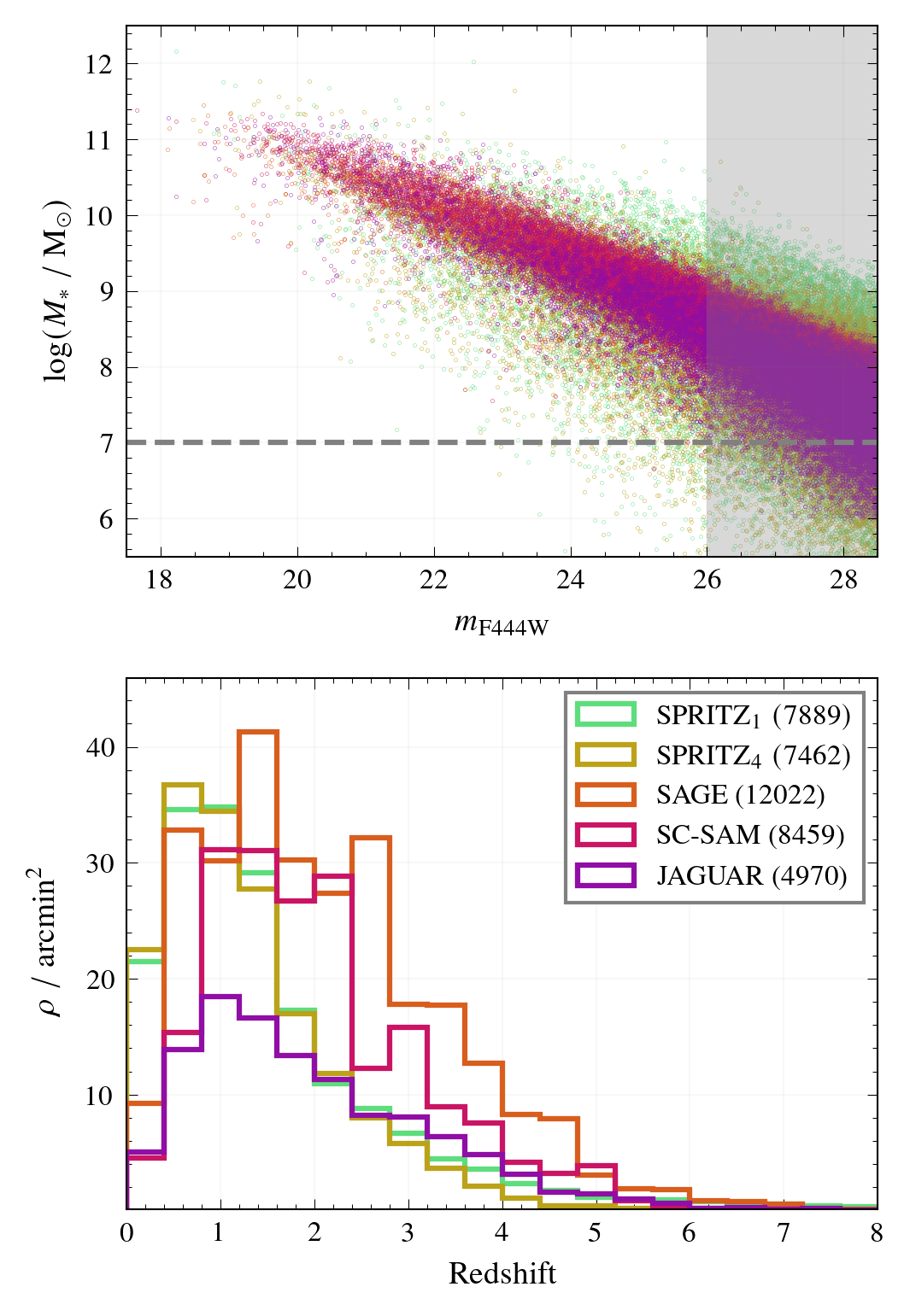}
	\caption{\emph{Top}: The relation between stellar mass and apparent F444W magnitude as predicted by the five models. The grey dashed line denotes the mass resolution of \scsam, the lowest of the five. Galaxies in the grey shaded region are discarded by requiring $m_{\mathrm{F444W}} < 26$. \emph{Bottom}: The sky area density of remaining model sources in 0.5 width redshift bins. The total number of sources predicted by each model is indicated in the legend.
	\label{fig:selection}}
\end{figure}

The minimum mass limits of the lightcones currently span \mbox{$10^{4} < M_{\ast} \ / \ \mathrm{M_{\odot}} < 10^{7}$}, which introduces another likely source of bias that can be mitigated by ensuring all galaxies exceed the upper limit. As this approach is intended as an alternative to SED fitting, one cannot simply apply a direct stellar mass cut to the observations. We instead plot the relationship between stellar mass and F444W apparent magnitude in the top panel of Figure \ref{fig:selection}, where the relationship is tightest and most simulation independent. We apply a pseudo mass cut using this magnitude, producing a high-purity sample while preserving statistical significance. By selecting all sources with $m_{\mathrm{F444W}} < 26$, we obtain 40802 simulated galaxies of which $>99\%$ have $M_{\ast} > 10^{7} \ \rm{M_{\odot}}$. The same cut preserves 4590 observed sources for a combined total of 45392. \flagsi\ showed that a comparable source extraction approach results in $<1\%$ incompleteness at this magnitude, so our results should not be impacted by missing sources. The redshift distribution of each dataset is shown in the bottom panel of Figure \ref{fig:selection}, in terms of the number of sources per unit sky area per redshift interval. Differences between the models are already clear, with \sage\ producing more than twice the number of galaxies predicted by \jaguar. These galaxies are also observed at different redshifts. For example, the \spritz\ distributions peak at $z\sim0.75$, whereas \sage\ peaks at $z\sim1.5$ and still produces many galaxies at $z=2.5$. While the sample does include galaxies as far out as $z=8$, the strict magnitude cut causes the numbers to fall off quickly at $z \gtrsim 3$, so our analysis will be most sensitive to differences at low redshift.

%% file: sections/3_analysis.tex
\section{Analysis}\label{sec:analysis}

\subsection{Comparing Binned Distributions}

Having selected a consistent sample of galaxies, we can begin evaluating the agreement between observed and modelled distributions using all thirteen filters. Perhaps the most obvious approach is to apply nearest point binning on a grid where each dimension corresponds to one of these filters. While conceptually simple, this approach preserves all the information we want to evaluate. The redshift and internal mechanisms of a galaxy determine its SED shape and position on the grid, while the number counts determine the overall normalisation. We adopt a simple bin-wise score to quantify the agreement between distributions \begin{equation}
    s_{i} = \frac{q_{\mathrm{obs},i} - q_{\mathrm{sim},i}}{\sqrt{\sigma_{\mathrm{obs},i}^{2}+\sigma_{\mathrm{sim},i}^{2}}}\ ,
\label{eq:metric}\end{equation} where $q$ is a preferably normalised quantity describing the occupation of bin $i$ and $\sigma$ is the associated uncertainty. While other metrics were explored, this is the most informative when visualised, with positive and negative values indicating more or fewer observed galaxies than predicted by the model respectively. The absolute value depends on the statistical significance of this difference and scales predictably with binning resolution. The quantity $q$ can be adjusted depending on the science target. Using the density of sources per unit sky area $\rho$ includes the number counts, but its contribution can be ignored by using the fraction of a dataset $f$ in a bin. Both quantities will be utilised in this work. We only include the Poisson contribution to the uncertainty in both cases, as while a cosmic variance \citep{Driver_2010} term could easily be included based on different realisations, these are not currently available for \spritz. The overall agreement of each model can be summarised with a single score \begin{equation}
    \mathcal{S} = \sum_{i} s^{2}_{i} \ ,
\label{eq:summary}\end{equation} the sum of squares of the metric over all bins. This is essentially the $\chi^{2}$, but we refer to it as $\mathcal{S}$ for consistency with equation \ref{eq:metric} and to free up the superscript. As with most quantities derived from bins, the absolute value depends on the binning resolution. Intuitively, lower resolution results in improved apparent agreement and vice versa, but the relative uncertainty also decreases at lower resolution. Any comparison between binning schemes must be made with this in mind, so we will remain consistent throughout this work.

\subsection{Addressing the Curse of Dimensionality}

\begin{figure}
 	\includegraphics[width=\columnwidth]{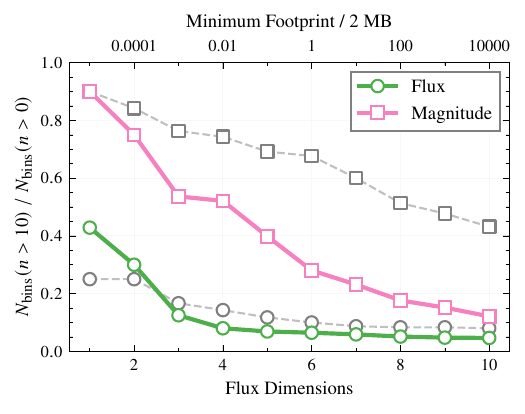}
	\caption{The fraction of occupied bins containing at least 10 observed galaxies when splitting each dimension of an increasingly high-dimensional flux space into deciles. The green and pink lines show the results when binning in flux and magnitude space respectively. The top axis labels show the minimum memory footprint required for a grid of each size. The dashed grey lines show the result when using $>10^{5.5}$ \scsam\ galaxies.
	\label{fig:occupation}}
\end{figure}

The use of such a metric does not inherently demand DR, and we are free to attempt an analysis in the native space. However, it quickly becomes apparent that this is impractical. Figure \ref{fig:occupation} shows the fraction of occupied bins that contain at least ten observed galaxies when constructing grids of increasing dimensionality. This is a loose measure of statistical significance, requiring a relative Poisson uncertainty of $\lesssim 30\%$. Each dimension is split into deciles, possibly the lowest resolution from which we could hope to draw informative conclusions. Despite this favourable setup, the distribution is already sparsely sampled in a single dimension, with only three of the seven occupied bins considered robust when using the raw fluxes. We would expect $>400$ objects per bin if they were uniformly distributed, but the order of magnitude differences between the brightest galaxies and the general population mean this is not the case. Converting to magnitudes increases the fraction to $>90\%$, as this is equivalent to using logarithmically spaced bins, but this requires omitting galaxies that are undetected in any filter. Even in this suboptimal scenario, the space again becomes sparsely sampled beyond two dimensions, making statistically significant conclusions unattainable. This is not merely a consequence of our limited sample size, as the grey dashed lines show the same quantity calculated using $\sim 10^{5.5}$ \scsam\ galaxies selected from additional lightcones. This $\sim 2 \ \mathrm{dex}$ increase does little to improve the statistics when using raw fluxes, demonstrating how intrinsically sparse the space is. The number of dimensions could be pushed to $\sim7$ when using magnitudes, but we are still unable to reliably include all thirteen.

Bins of varying size can overcome this issue by enforcing a minimum number of sources in each bin, but this comes at the cost of resolution in sparse regions. However, this cannot surmount the exponentially increasing memory footprint. Splitting all thirteen dimensions into ten bins of any size requires at least $20 \ \mathrm{TB}$ of memory at 16-bit integer resolution, without accounting for additional processing overheads. The high-resolution floats required for normalisation and further analysis demand twice this amount, and subsequently doubling the number of bins in each dimension raises the requirement to $>100 \ \mathrm{PB}$, making such an analysis entirely unfeasible. Dimensionality reduction is not only desirable, but a necessity if we want to fully exploit multi-wavelength datasets in the observer frame.

Each of these issues can be alleviated by the UMAP algorithm, details of which are provided by \citet{McInnes_2018}. In short, UMAP works by assuming that the input data is uniformly distributed on a locally connected Riemannian manifold. It constructs a graph of nearest neighbours in this high-dimensional space, where each object must be connected to at least $k$ neighbours. To maintain this assumption of a uniform distribution, an object-dependent distance metric $D$ is defined. This leads to a `fuzzy' structure, where each vertex between points on the graph is weighted by the probability that they are connected based on their different distance definitions. UMAP then approaches dimensionality reduction as an optimisation problem, searching for the lower-dimensional embedding that best preserves the fuzzy structure. This search is stochastic and conducted over a given number of epochs $t_{e}$. The enforced minimum distance between points embedded in euclidean output space is $d$. This space may have any number of dimensions, which can be set to control sparsity and the memory demands of subsequent binning. 

Classical dimensionality reduction approaches such as PCA are linear, meaning the reduced features are simple combinations of the inputs. The transform can therefore be written as a computationally inexpensive matrix multiplication. While slower, the UMAP approach is non-linear, allowing it to preserve more of the complex structure at both global and local scales. This is an important quality in the context of galaxy SEDs, where we know there is not only a correlation between emission in adjacent filters, but also across broad wavelength ranges. For example, while not probed by our filter set, dust attenuation and re-emission cause an energy balancing connection between rest-frame optical and far-infrared emission. Dust emission can often be treated as a greybody, meaning that all the photons attenuated at one wavelength may not necessarily be re-emitted in the same long wavelength filter \citep{Salim_2020}. The connection between AGN and young stellar emission, arising from AGN feedback, is an example more likely to be probed by \mbox{NIRCam} \citep{Fabian_2012}. UMAP is not the only algorithm adopting such an approach, with t-distributed Stochastic Neighbour Embedding working similarly. However, UMAP's superior scaling with both dataset size and dimensionality \citep{McInnes_2018} make it ideal for processing large extragalactic surveys and minimising stochasticity.

\subsection{Fitting UMAP Models}\label{sec:umap_fitting}

UMAP hyperparameters can be tuned according to the nature of the input data and the science goal. Unlike a classification problem, there is no ground truth that we are trying to recover, so we should be able to determine consistent model constraints regardless of the specific values. Reasonable choices are still required to preserve as much of the underlying information as possible and draw meaningful conclusions. Requiring fewer neighbours preserves more local structure, whereas higher values produce a more global view of the manifold. The latter may therefore `wash out' subpopulations, making datasets appear more consistent, whereas the former makes results more sensitive to the treatment of noise and cosmic variance, which could cause spurious points on the graph. We choose $k=20$ in an attempt to balance these effects. 

The metric used to compute distances in the high-dimensional space can be chosen to target different physics or minimise the effect of outliers. We fit UMAP models adopting both the `euclidean' and `cosine' metrics, with scores computed from embeddings constructed with each denoted by superscript $e$ and $c$ respectively. As one might expect, the euclidean metric considers both the magnitude and angle between each feature when determining the distance between inputs. We pair this metric with the sky area density of sources in each bin $\rho$ when computing summary scores. This allows us to probe whether models correctly reproduce the total number of galaxies, their apparent brightness distribution, and shapes, the latter of which probes both the redshift distribution and underlying astrophysics. While there is no inherent requirement to adopt $q\equiv\rho$ when using this metric, we refer to scores computed in this way as euclidean and density-based interchangeably in this work ($s^{e}$, $\mathcal{S}^{e}$). A cosine metric ignores the amplitude of features, which in this context means only the SED shape of each galaxy is considered when positioning it in embedding space. For this reason, it pairs well with a summary score computed with $q\equiv f$, the fraction of a given dataset within each bin. This allows us to quantify how well each model replicates the observed distribution of SED shapes, probing the evolutionary mechanisms of galaxies more closely. We refer to scores computed in this way as cosine and fraction-based interchangeably in this work ($s^{c}$, $\mathcal{S}^{c}$).

We choose a high number of training epochs $t_{e}=1000$ to minimise the impact of UMAP's stochastic elements. No minimum distance is imposed on the space between points in embedding space $d=0$, to preserve the high-dimensional structure as closely as possible. Section \ref{sec:systematics} quantifies the impact of these choices on the measured scores. We generate a 2-dimensional embedding, which allows the space to be visualised straightforwardly and binned at sufficiently high resolution.

\begin{figure}
 	\includegraphics[width=\columnwidth]{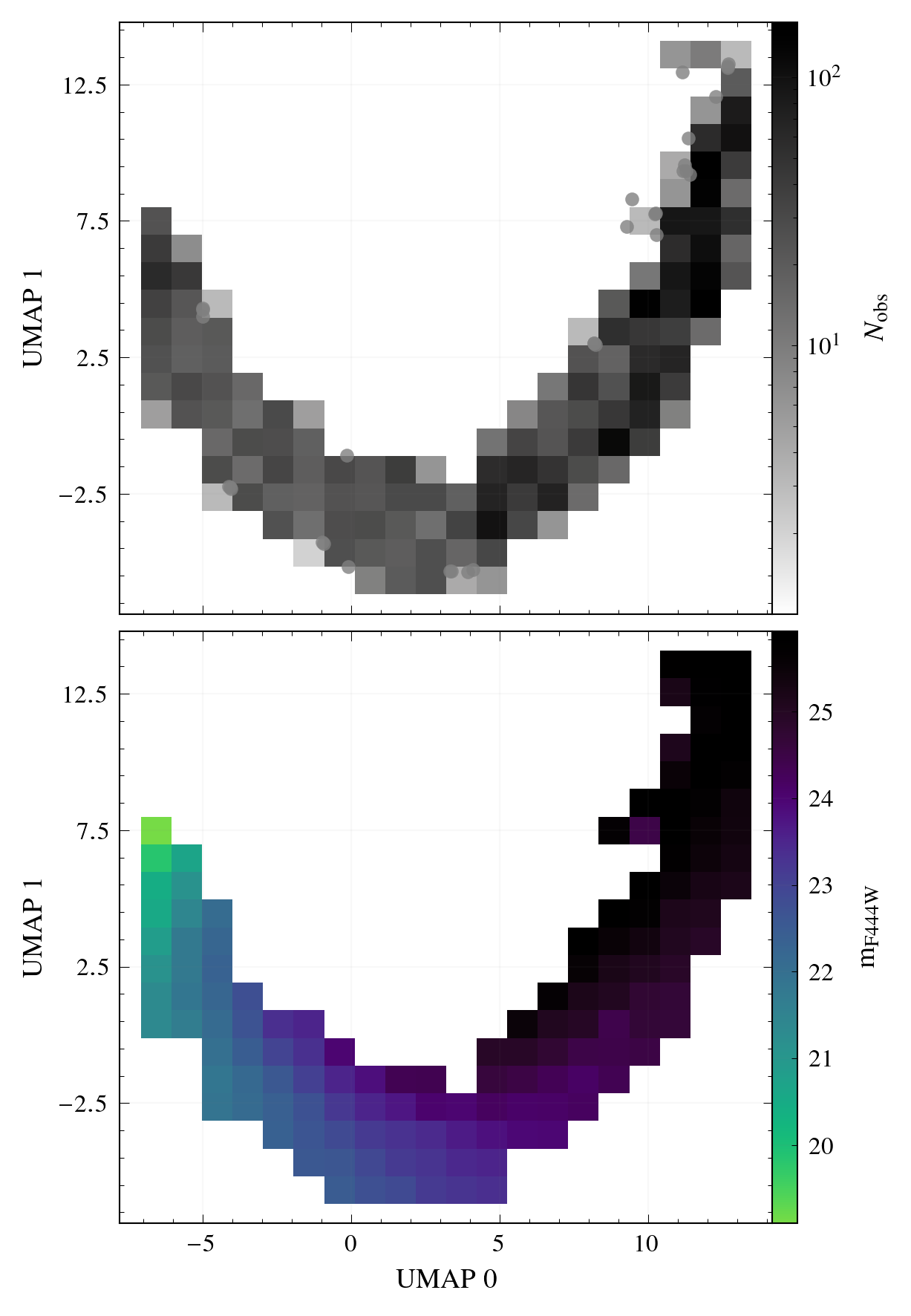}
	\caption{\emph{Top}: The binned distribution of observed sources in UMAP embedding space generated with a euclidean metric. Each source is positioned using the median of 500 UMAP iterations. Both dimensions are split into 20 bins of equal width, and individual galaxies are overlain as grey points if $N_{\mathrm{obs}} < 5$. \emph{Bottom:} Same as above, but with the bins coloured by the mean apparent magnitude of the constituent galaxies.
	\label{fig:umap1}}
\end{figure}

As discussed previously, using magnitudes necessitates the omission of galaxies with even a single non-detection, which introduces potential bias, and limits the sample to $z<4$ as the Lyman break otherwise appears in F435W. We instead fit the UMAP models using raw fluxes, including negative values. These features are not scaled in any way, as they are clearly not independent of one another, so scaling would remove the meaningful connections we aim to capture. Scaling is often applied to guard against a single feature dominating the learned relationship between samples, simply because it spans far higher orders of magnitude. This is not of concern in this case, as the fluxes in each filter span comparable ranges and the cosine metric will be insensitive regardless. We repeat the fitting of each UMAP model 500 times, with each iteration adopting a different random seed. By recording the position of each galaxy in each of the generated embeddings, we can account for the effect of UMAP's stochastic elements when drawing conclusions. While the relative positions of galaxies do not change significantly across these iterations, their orientation within the UMAP space can. Rather than assigning an uncertainty to the position of each galaxy, we use their median positions when plotting distributions and reporting $s_{i}$, and calculate the $16^{\mathrm{th}}$, $50^{\mathrm{th}}$ and $84^{\mathrm{th}}$ percentiles of $\mathcal{S}$ over all 500 iterations. 

\begin{figure*}[t]
        \centering
 	\includegraphics[width=\textwidth]{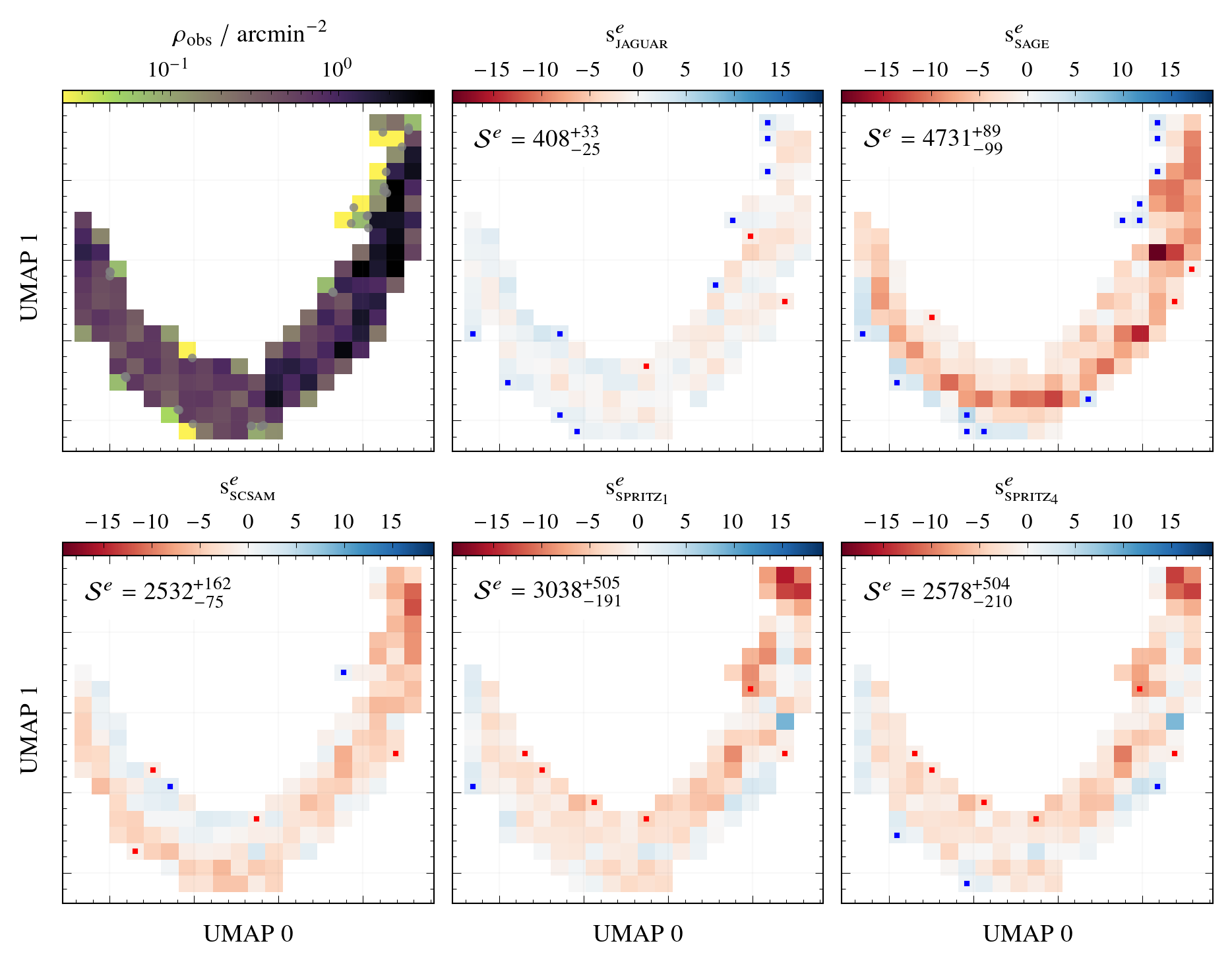}
	\caption{\emph{Top left}: The density of observed sources per unit sky-area in the 2D UMAP embedding constructed using a euclidean metric. Individual galaxies are overlain as grey points if $\rho_{\mathrm{obs}} < 0.1 \ \mathrm{arcmin^{-2}}$. \emph{Other panels}: Maps showing how accurately each model reproduces the observed distribution, quantified using the metric $s^{e}$. From left to right and top to bottom, \jaguar, \sage, \scsam, \spritz$_{1}$ and \spritz$_{4}$. The corresponding summary score is shown in the top left of each panel. Overlain blue and red squares indicate that a bin contains only observed or model galaxies respectively.
	\label{fig:metric_e}}
\end{figure*}

These fits are performed across 20 CPU cores, which complete each iteration per object in $\sim0.007 \ \rm{seconds}$ of CPU time, which is equivalent to $\sim3.5 \ \rm{seconds}$ per object per model fit. No more than $25 \ \mathrm{GB}$ of memory is required, demonstrating a significant improvement upon N-dimensional binning. As a comparison, we process the same sample with the Bayesian SED fitting code \texttt{Bagpipes} \citep{Carnall_2018}, adopting a simple literature-informed model consisting of only six free parameters \citep{Harvey_2025}. These are the redshift, total mass and metallicity of stars, peak age and FWHM of a log-normal star formation history, and the V-band dust attenuation assuming a \citep{Calzetti_2000} dust curve. On average, it takes $> 400 \ \rm{seconds}$ of CPU time per object fit when using all thirteen filters, which is $>100$ times slower than finding the embedded position with UMAP. While the inferred physical properties may allow the physics of galaxies to be interpreted more directly, assuming the associated biases are well understood, DR offers a more efficient alternative for performing statistical comparisons between the potentially millions of galaxies observed by large surveys and produced by model variations. \citet{McInnes_2018} provide further details of UMAP's scaling with dataset size and ambient dimension.
 
The top panel of Figure \ref{fig:umap1} shows the distribution of observed galaxies in the embedding space generated with a euclidean metric. The absolute values attributed to each UMAP dimension are not physically meaningful, so while we include them here for completeness, they are omitted herein. Separating both dimensions into twenty bins ensures that the average fraction of robust bins across all iterations is $80\%$. This already demonstrates a clear motivation for working in the embedding, rather than native space, as reduced sparsity improves the reliability of any conclusions drawn \citep{Ashmead_2025}.

The galaxies lie on a connected horseshoe structure,  with no significant separated clusters. This is not particularly surprising, as apparent brightness, which we expect to vary smoothly, is the main differentiating characteristic of galaxies in the observer frame. The bottom panel shows the median $m_{\mathrm{F444W}}$ of objects within each bin, demonstrating a clear trend, with the brightest galaxies at one end and the faintest at the other. There is the expected smooth transition, which is reproduced by the models, but also a clear substructure, with fainter galaxies tending to lie on the inside edge. When coupled with the non-linear shape and extended nature of the distribution, this tells us that additional information from the shape of the SED is being preserved by UMAP.

%% file: sections/4_results.tex
\section{Model Evaluation \& Differentiation}\label{sec:results}

\subsection{Euclidean metric ($q\equiv\rho$)}\label{sec:euclidean}

The top left panel of Figure \ref{fig:metric_e} shows this same distribution of observed galaxies in an embedding space constructed with a euclidean metric, now normalised by the total sky area. The remaining panels show how well each of the five models replicates the distribution, using maps of the $s^{e}$ score computed when $q\equiv\rho$. This metric contains all the available information, so it quantifies the overall performance of each model. The model galaxies generally occupy the same region of the embedding space, replicating a horseshoe structure with the same extent. Bins overlaid with blue or red squares contain only observed or model galaxies respectively. These bins are on the outskirts of the distribution, and the top left panel shows that these contain few observed galaxies, which lie on the edge of more occupied bins. These also contain very few model galaxies, so are likely a result of noise, which is reflected in the low $s^{e}$ scores. These bins are not considered an indicator of model divergence. 

The top left panel does show two observed galaxies that are separated from well-populated bins by $>1$ bin width. While both \spritz\ models produce analogues, the remaining three do not, which could be an indicator of unique physics, or a consequence of the specific noise realisation. The most separated of these objects is the galaxy JADES-GS-53.08745-27.81492, which was observed with the \mbox{NIRCam} grism as part of the FRESCO programme \citep{Hainline_2024b}. \cite{Xiao_2024} measured a spectroscopic redshift $z_{\mathrm{spec}}=7.846$, and through the process of SED fitting identified it as a dusty star-forming galaxy, with no AGN component. This is surprising, as the inclusion of AGN emission is a significant difference between \spritz\ and the other model photometry. The \spritz\ galaxies in this bin are instead dusty $0.06 < A_{V}<3.2$ dwarfs at $z\sim0.7$, which highlights the well-documented issue of nebular emission and dust reddening combining to mimic the photometry of higher redshift galaxies \citep{Arrabal-Haro_2023, Zavala_2023, Gandolfi_2026}. The \texttt{Bagpipes} inferred redshifts of these dusty \spritz$_{4}$ galaxies are systematically overestimated, ranging from $1<z<8$. Even considering the full posteriors, the significance of the differences does not fall below $20\sigma$. While studies based on SED fitting inferred properties may be substantially biased by such objects, the direct observable approach is unaffected by nature, which is a key benefit of constraining models this way.

The second of these separated objects was fit by \citet{Hainline_2024} using the template fitting code \texttt{EAZY}, which returned a photometric redshift $z_{\mathrm{phot}}=7.51$. This is consistent with previous studies \citep{Schenker_2014, Finkelstein_2015}, which appear to place it in the same easily contaminated population of high-redshift galaxies. However, it was also fit with a brown dwarf template, which returned an order of magnitude lower $\chi^{2}$. Proper motion between WFC3 and NIRCam imaging all but confirms it as a brown dwarf. This serves as an important reminder that while the direct observable approach is insensitive to many potential biases, those introduced by spurious galactic sources cannot be easily avoided, although we expect only a small fraction of these to avoid GAIA-informed masking.
 
Returning now to the comparison panels of Figure \ref{fig:metric_e}, which demonstrate the statistical distinguishing power of the DR approach. \scsam\ and both \spritz\ models are discrepant with the observations at the $\sim5\sigma$ level across the distribution, which rises to $\sim10\sigma$ in bins hosting the faintest galaxies. A similar, albeit amplified trend is produced by \sage, exceeding $15\sigma$ in several of the most densely populated bins. These trends are reflected in the summary scores, with all four unable to predict the observed distribution as well as \jaguar. No \jaguar\ bin is discrepant at a level $>5\sigma$, and it achieves $\mathcal{S}^{e} = 408^{+33}_{-25}$ overall. Normalising by the number of occupied bins yields a quantity equivalent to $\chi^{2}_{\nu}$, and \jaguar's value of $\sim3$ confirms excellent replication of the galaxy population in GOODS-S. This is not particularly surprising, given that semi-empirical models are naturally constructed from observations, and a subset of the \jaguar\ SEDs are taken directly from the field. The best performing causal model is \scsam, with its score $\mathcal{S}^{e}=2532^{+162}_{-75}$ indicating that it performs almost twice as well as \sage\ with $\mathcal{S}^{e}=4731^{+89}_{-99}$. It is unclear what causes the high percentage uncertainties in the \spritz\ scores, but this makes them statistically consistent with one another. A more abundant sample of galaxies at $z>3$ might allow us to distinguish between these models, as this is the only regime where model-driven differences between the two are expected. These results demonstrate that models can be differentiated and therefore constrained in UMAP embedding space.

\begin{figure}
 	\includegraphics[width=\columnwidth]{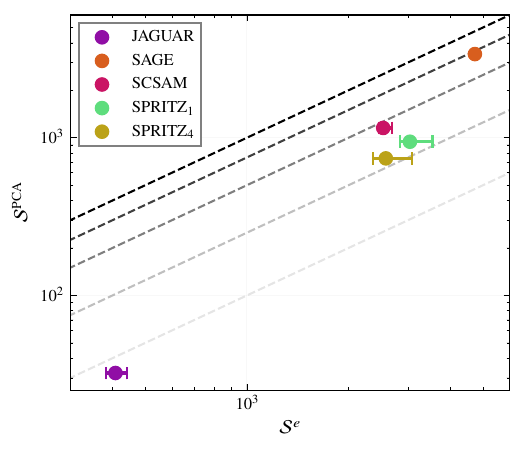}
	\caption{The difference in summary scores measured for each of the five models in 2D spaces constructed with PCA ($\mathcal{S}^{\mathrm{PCA}}$), and with UMAP using a euclidean metric. Regions below the increasingly faint dashed lines enclose models with $\mathcal{S}^{\mathrm{PCA}}$ at least 1, 25, 50, 75 or $90\%$ lower than $\mathcal{S}^{e}$.
	\label{fig:pca}}
\end{figure}

Figure~\ref{fig:pca} demonstrates the benefit of preserving non-linear information with UMAP. It compares the $\mathcal{S}^{e}$ scores to a second set, also calculated using equation \ref{eq:metric}, but within a 2D space generated by linear PCA. The latter are found to be systematically lower, showing that ignoring non-linear components oversimplifies the distributions, making model predictions appear artificially more accurate. This difference is significant, rising from $\sim25\%$ for the worst performing model \sage, to $>90\%$ for \jaguar. We can therefore expect non-linear approaches to become increasingly important over time, as modelling inevitably improves. Perhaps most concerningly, the relative performance of the models is also different in the PCA space. UMAP shows that \scsam\ performs better than \spritz$_{1}$, but this flips when using PCA. If these were both causal models, we might favour the physics implemented in \spritz\ rather than \scsam. This issue will also become more prevalent as models converge, or when analysing those that explore astrophysical and cosmological parameter variations, which may only introduce small changes. Preserving non-linear information is vital if models are to be constrained accurately in the observer frame.

\begin{figure*}
        \centering
 	\includegraphics[width=\textwidth]{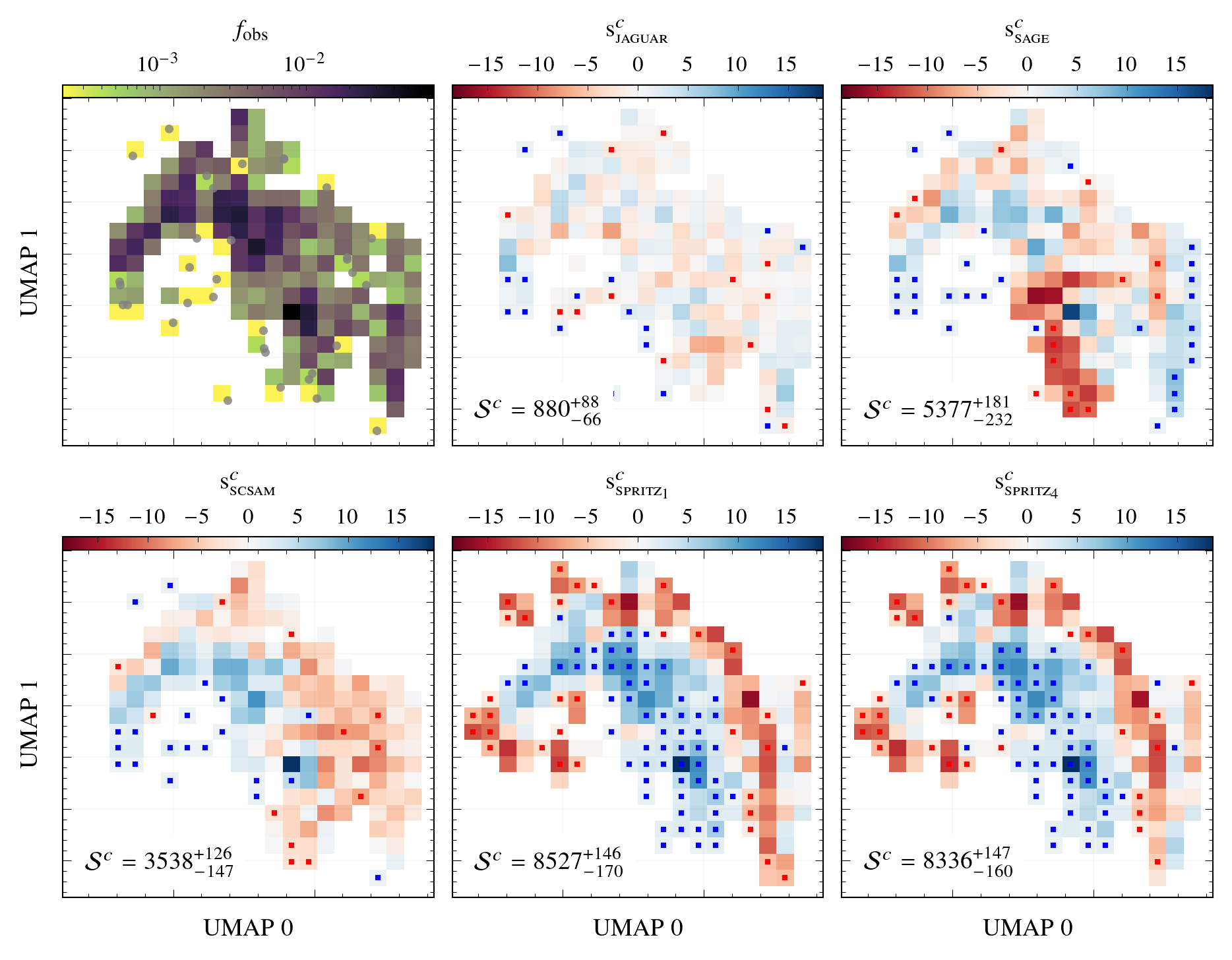}
	\caption{\emph{Top left}: The fraction of observed sources falling within each bin of a 2D UMAP embedding constructed using a cosine metric. Individual galaxies are overlain as grey points if $f_{\mathrm{obs}} < 5\times10^{-4}$. \emph{Other panels}: Maps showing how accurately each model reproduces the observed distribution, quantified using the metric $s^{c}$. From left to right and top to bottom, \jaguar, \sage, \scsam, \spritz$_{1}$ and \spritz$_{4}$. The corresponding summary score is shown in the bottom left of each panel. Overlain blue and red squares indicate that a bin contains only observed or model galaxies respectively.
	\label{fig:metric_c}}
\end{figure*}

While flux information is clearly valuable when encoded within the summary statistic, the largest disparities that can be discerned from Figure \ref{fig:metric_e} are the same as those discussed in \flagsi: many semi-empirical and semi-analytic models produce too many galaxies across the magnitude distribution. \flagsi\ showed that inefficient SNe feedback is the primary driver of this issue in \sage, by comparing its implementation to that of \scsam. The mass loading factors are significantly lower in low mass \sage\ halos, which accelerates the growth of the constituent galaxies and inflates the number counts across a broad range of magnitudes. \spritz\ does not directly implement this physics, so there may be an issue with its calibration. The superior, albeit not statistically significant, performance of \spritz$_{4}$ may at first motivate a more negative value of $k_{\phi}$, particularly given improved performance in bins at the faint end. However, Figure \ref{fig:selection} shows that \spritz$_{1}$ is already under-predicting the number of $z>3$ galaxies relative to the best performing model. An offset between the near and far-infrared (FIR) luminosities of the template SEDs could explain the overprediction observed across a broad range of redshifts and throughout the horseshoe distribution. If the NIR component is overestimated, the FIR luminosity function can still be well recovered, but the number of galaxies which are bright in F444W will be inflated. These SED shapes can be probed more closely with a cosine metric.

\subsection{Cosine metric ($q\equiv f$)}\label{sec:cosine}

The approach can be adjusted to emphasise different physics. The top left panel of Figure \ref{fig:metric_c} is now showing the fraction of observed galaxies within each bin of a different embedding space. We now use this fraction $f$ when evaluating equation \ref{eq:metric}, thereby removing the contribution from the number counts. The space is constructed in an almost identical way, but adopting a cosine distance metric, which further omits any contribution from the apparent brightness. The score maps therefore demonstrate how well the models replicate the observed distribution of SED shapes. These shapes are still measured in the observer frame, so are not independent of the predicted redshift distributions. 

As within the first embedding, the majority of galaxies fall on a main connected cluster, with only a small number constituting more separated bins. These contain the faintest galaxies, so they are likely a consequence of noise, but there is no clear apparent magnitude trend in the main cluster. The models again occupy the same broad region of the space, and their distributions within can still vary greatly. In fact, the models perform more poorly in this space, regardless of whether $f$ or $\rho$ is used to calculate the score. Bin-wise discrepancies again exceed $15\sigma$ in \sage, but now also in \scsam\ and \spritz. While the most significant discrepancies were previously a result of models overpredicting the number of galaxies, there are now many bins containing observed galaxies that the models predict to be almost, if not entirely empty. \jaguar\ is least affected by this issue, and those bins left unoccupied are primarily in the noise-dominated region. It is again the standout performer with a score $\mathcal{S}^{c}=880^{+88}_{-66}$, albeit with the same caveat regarding the SEDs sourced from GOODS-S. \scsam\ remains the best causal model and is now the outright second best overall with $\mathcal{S}^{c}=3538^{+126}_{-147}$, despite a systematic offset from the observations in the main cluster. \sage\ fails to occupy an extended region of the space, while also producing $>10\sigma$ discrepancies in a region partly unoccupied by the observations, and its performance worsens to $\mathcal{S}^{c}=5377^{+181}_{-232}$ as a result. The \spritz\ models are still consistent with one another, but perform considerably worse in this space, such that $\mathcal{S}^{c}>8000$. This again confirms the efficacy of UMAP for statistically differentiating between models, even when the number counts are ignored. We will now investigate how the underlying physics can be probed and used to suggest future model improvements.

Large regions of the UMAP space are entirely unoccupied by \spritz, with the majority of galaxies instead lying on a more narrow manifold. This is likely due to the template-based approach to generating photometry, which limits \spritz\ to the 39 rest-frame shapes included in the lightcone catalogue. Clearly, more are needed to recover the observed variation, supporting the idea that the shapes are not representative, and highlighting the need for self-consistent modelling where possible. \spritz\ may not be best placed for applications where the NIR SED shapes are important, such as when testing photometric redshift accuracy in this regime, but it has already been shown to perform equally well, if not better, than other models overall (\S\ref{sec:euclidean}). In addition, \spritz\ recovers the more separated galaxies, albeit in overabundance, where the other models often do not. This appears to be due to its dedicated inclusion of irregular dwarf galaxies, with $\sim95\%$ of the galaxies in these regions having an SED of that type.

\begin{figure*}
        \centering
 	\includegraphics[width=\textwidth]{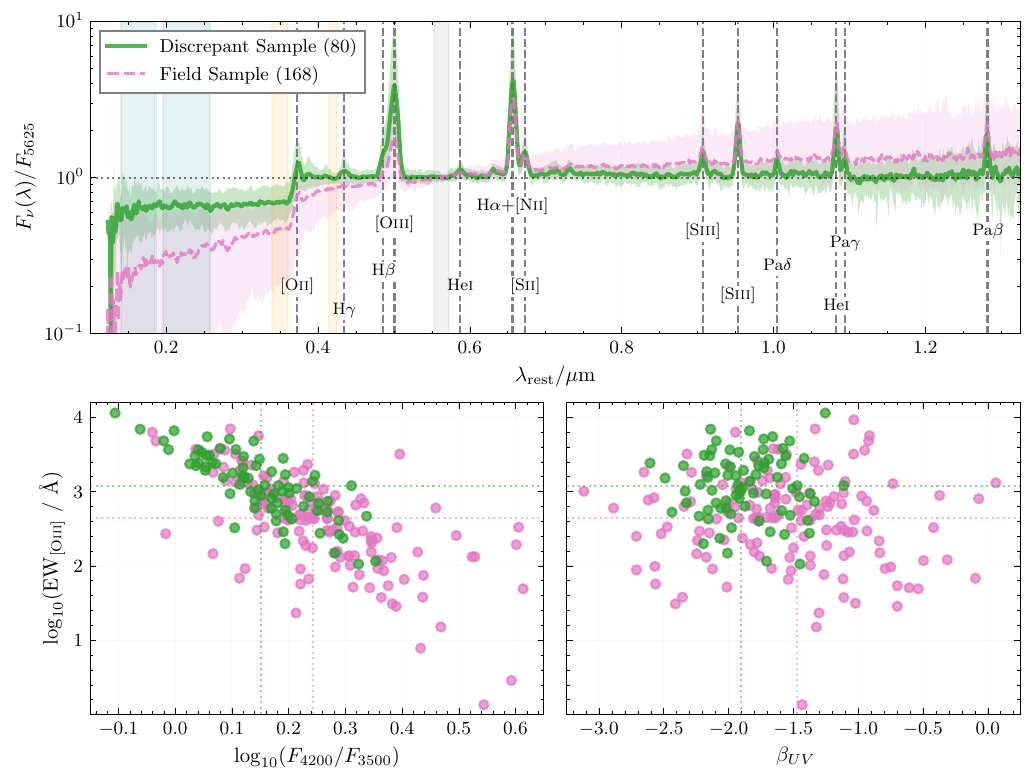}
	\caption{\emph{Top}: The median spectra of observed galaxies in the most model-discrepant bin of Figure \ref{fig:metric_c} (green) and field galaxies over a comparable redshift range $2.5<z<3.5$ (pink), after normalising by the median flux in a [5525, 5725]\AA\ window shown by a grey band. Shaded regions indicate the 16th and 84th percentile ranges, while the teal and orange bands show the wavelength range over which $\beta_{\mathrm{UV}}$ and D4000 are estimated. Dashed lines indicate the position of notable spectral features. \emph{Bottom}: The relationship between $\mathrm{EW_{[O\textsc{iii}]}}$, D4000 and $\beta_{\mathrm{UV}}$ for the same samples. Dotted lines denote the median of each quantity.
	\label{fig:stacks}}
\end{figure*}

We note one bin in particular demonstrates a consistent discrepancy, and we choose to explore this in detail. This contains the largest fraction of the observations, but is underpredicted at a level $>15\sigma$ by \scsam\ and \sage. \spritz\ predicts no galaxies in the bin at all, but this is likely due to the same template SED issue. Visual inspection of the observed galaxies confirms that they are well-resolved, and there is no indication of preferential noise or bright source contamination, so this could represent a significant failure of the models. To aid interpretation of the physical drivers, we retrieve NIRSpec/PRISM spectra from the DJA archive \citep{Heintz_2024, Heintz_2025}, after matching to the galaxies in our sample with a $0\farcs4$ tolerance. This results in 734 galaxies, each with a robust spectroscopic redshift measured by \texttt{msaexp} \citep{brammer_2023_msaexp}. Only $25\%$ of the galaxies in the discrepant bin have a matching spectrum, which while not entirely representative, will give an initial idea of where the models could be improved. Only two of these galaxies reside at $z>4$, while the remainder cluster strongly at $2.5<z<3.5$. We use this range to select 168 comparable galaxies from the field. The median rest-frame spectra of these populations are shown in the top panel of Figure~\ref{fig:stacks}, after normalising the individual spectra by the median flux in a [5525, 5725]\AA\ window. The galaxies in the discrepant bin have surprisingly flat spectra, with the only gradient arising from the Balmer/4000\AA\ break (D4000), whereas the brightness of the field galaxies steadily increases with wavelength. Both samples demonstrate clear line emission, but this appears to be stronger in the discrepant sample.

We run \texttt{msaexp} to quantify the strength of these features. The D4000 break is defined as the ratio of median continuum fluxes in a red window [4150,4250]\AA\ and a blue window [3400,3600]\AA, which avoids strong nebular line emission \citep{Binggeli_2019}. The reddening is quantified by fitting a power-law \mbox{$f_{\lambda} \propto \lambda^{\beta_{\mathrm{UV}}}$} to the UV continuum over the wavelength range [1400,2580]\AA, ignoring the [1860,1955]\AA\ window to minimise contamination from the C\textsc{ii}]$\lambda1909$ line. 

The bottom panel of Figure \ref{fig:stacks} compares the strength of these features to the combined equivalent width (EW) of [O\textsc{iii}]$\lambda4959, \lambda5007$, for all galaxies with fits satisfying $\chi^{2}_{\nu}<5$. The discrepant population does not appear to be unique in this space, so they are unlikely to represent new physics, but they are clearly extreme. All exceed $\mathrm{EW_{[O\textsc{iii}]}}=100 \ \rm{\AA}$, and $68\%$ satisfy the $\mathrm{EW_{[O\textsc{iii}]}}>750 \ \rm{\AA}$ condition imposed by \citet{Tang_2019} to define extreme emission line galaxies (EELGs). The field sample demonstrates far greater scatter, with only $31\%$ considered EELGs. Such strong emission is consistent with an excess of ionising photons, which can in principle arise either from a large fraction of massive young O and B-type stars, or from very low metallicity stars which are capable of maintaining elevated ionisation at ages well beyond $10 \ \mathrm{Myr}$ \citep{Stanway_2016, Kewley_2019}. The D4000 break strength has regularly been used as an indicator of stellar age \citep{Bruzual_1983, Poggianti_1997}, with which \citet{Wilkins_2024} demonstrated a strong correlation and an inverse correlation with specific star formation rate. The combination of weak breaks and high EWs in the discrepant sample suggests that they are undergoing intense bursts of star formation, with durations sufficiently short ($\sim10 \ \mathrm{Myr}$) that young massive stars still dominate the emission and can ionise their surrounding gas to produce strong nebular emission lines. The UV continua of these galaxies also evolve steeply, consistent with limited dust attenuation \citep{Calzetti_1994, Meurer_1999}.

Combined, these features reveal these galaxies to be dust-poor starbursts residing at $2.5<z<3.5$, shortly before the peak in the cosmic star formation rate density \citep{Madau_1996}. Their intense line emission alters the photometric colours relative to the field, causing them to cluster together in the UMAP space. Ionisation strength decreases rapidly as stellar populations exceed ages of $10 \ \rm{Myr}$ \citep{Xiao_2018}, so these colours can also be expected to vary greatly over short timescales. The failure of both \scsam\ and \sage\ to produce comparable galaxies may suggest an issue with the implemented galaxy-evolution physics, but other explanations should be tested before revising the model. 

Firstly, the TAO \sage\ photometry does not include nebular emission, which must be rectified to produce the strong emission lines. Secondly, the BC03 stellar models do not include binary systems, which have been shown to make a significant contribution to the ionising photon budget of galaxies \citep{Stanway_2016, Ma_2016_bin}. \citet{Wilkins_2024} showed that BPASS \citep{Stanway_2018}, which includes binaries, produced breaks that were $\sim0.1 \ \mathrm{dex}$ weaker than BC03 and other commonly used single star models. Given access to the raw model data, it would be trivial to investigate the impact of BPASS on UMAP space predictions. Thirdly, the resolution of SFHs produced by SAMs is dependent on the snapshot cadence of the underlying DMO simulation. The short-lived spectral features characteristic of brief intense starbursts can easily be washed out when averaged over long timescales \citep{Benson_2012}. The MultiDark DMO saved only 178 snapshots, with a scale factor dependent cadence that decreases from $10 \ \mathrm{Myr}$ at $z=15$, to $100 \ \mathrm{Myr}$ at $z=0$. While early starbursts may be identifiable, the SFHs at $2.5<z<3.5$ will be averaged out, so phases dominated by young stars are highly unlikely. The associated weak breaks and strong lines will not manifest, and the corresponding photometric colours will appear artificially uniform. The impact on UMAP space predictions could be tested by running \scsam\ on a higher cadence DMO, or by comparing to hydrodynamical model predictions, where the exact age of each star particle is known. The star formation and feedback prescriptions should only be revised after considering these effects.

\begin{figure}
 	\includegraphics[width=\columnwidth]{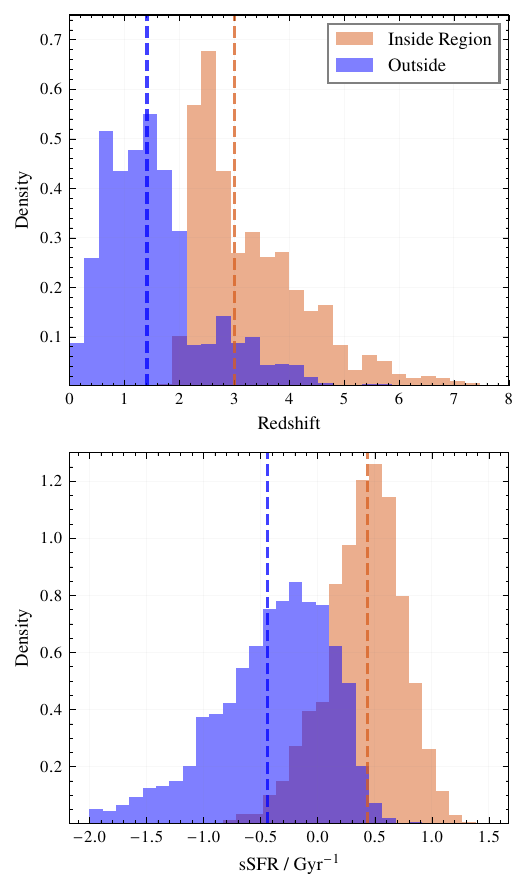}
	\caption{The count normalised distribution of physical properties of galaxies within (orange) and outside (blue) the most disparate extended region of the \sage\ map in Figure~\ref{fig:metric_c}. We construct this region from all connected bins in the bottom-centre region with $s^{c} <-5$. The top and bottom panels show the redshift and instantaneous sSFR respectively. Dashed lines show the median of each distribution.
	\label{fig:sage_props}}
\end{figure}

\begin{figure*}
        \centering
 	\includegraphics[width=\textwidth]{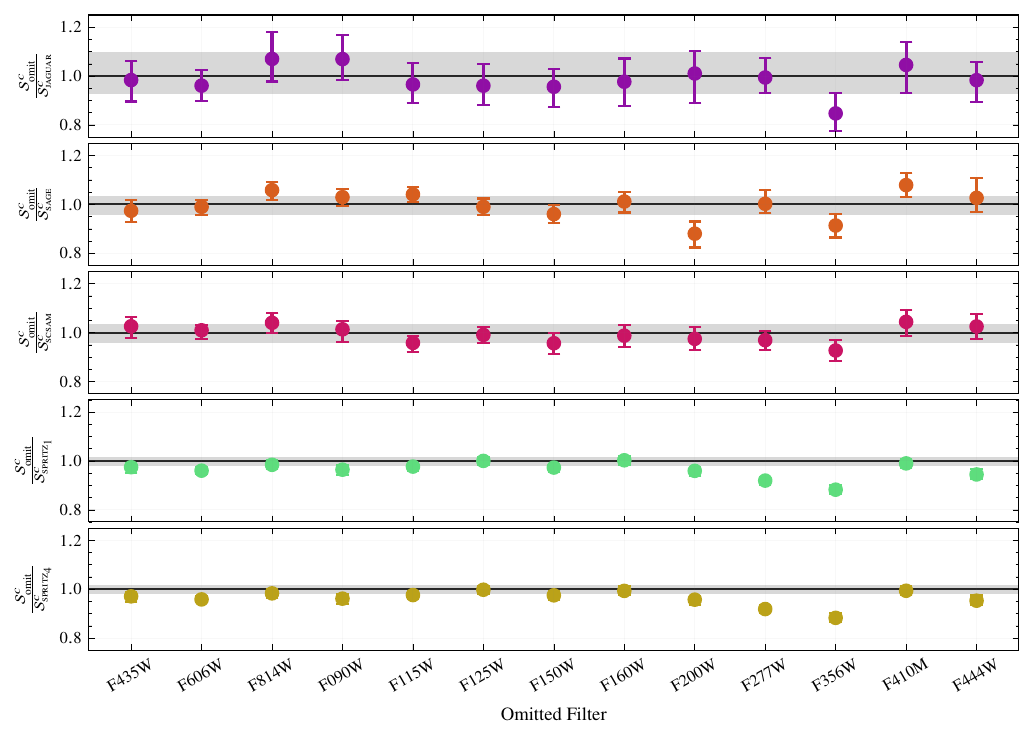}
	\caption{The results of applying a degradation layer before analysing models in an embedding generated by a cosine metric. The panels show the fractional shift from the fiducially measured $\mathcal{S}^{c}$ when removing a different photometric filter. An increase or decrease in score suggests emission in that filter is modelled better or worse than average. Each row shows the result for one of the five models. The shaded grey region shows the uncertainty on the fiducial value. 
	\label{fig:deg_layer}}
\end{figure*}

While the UMAP approach is intended to avoid inferred physical properties, those of the model galaxies, which are known confidently, can be used to inform areas of improvement. For example, an extended region of the \sage\ distribution, located at the bottom-centre, contains a far larger fraction of galaxies than expected from observations. We define this region as the 27 connected bins with $s^{c}<-5$, which contain a third of the total \sage\ sample. Six of these bins are discrepant at a level $>10\sigma$ and are unoccupied by the observations. The top and bottom panels of Figure~\ref{fig:sage_props} compare the distribution of redshifts and specific star formation rates (sSFR) respectively, for \sage\ galaxies inside and outside this 27 bin region. Those inside are the most star-forming, with the median sSFR $>0.8 \ \rm{dex}$ higher than the remaining population. These galaxies also have the highest redshifts, with the difference in medians again $0.8 \ \rm{dex}$, and a long tail including the earliest galaxies. This region therefore contains the most star-forming galaxies in the earliest stages of their evolution. 

\flagsi\ showed that the \sage\ number counts are inflated by galaxies in low mass halos growing too quickly, due to inefficient SNe feedback. The number of star-forming galaxies at high redshift will therefore also be inflated, causing the associated region of the UMAP parameter space to be misrepresented. This will eventually lead to an excess of SEDs reddened by old and metal-enriched stellar populations at intermediate redshift, which will introduce discrepancies in other regions of the UMAP space. Combined, these effects likely drive the inferior performance of \sage\ relative to \scsam. While the effects of photoionisation modelling may be somewhat washed out at lower redshifts, it is still important to produce realistic synthetic observables of highly star-forming galaxies at high redshift. Its omission from the TAO photometry will also contribute to this discrepancy.

A degradation layer offers a more generalisable approach to interpreting the physical drivers of identified discrepancies. We implement this by removing each filter in turn and rerunning the same UMAP setup on the degraded datasets. A decrease in the measured $\mathcal{S}^{c}$ suggests that the omitted filter was driving a discrepancy, and some element of the physics that it probes is poorly modelled, whereas an increase in the score indicates that the physics is particularly well modelled. This is unlikely to be very informative in the observer frame, as the redshifting of the underlying spectra will both wash out potential signatures and make it more difficult to pinpoint the physics responsible. We employ this approach as a means of investigating the minimum expected signal from more appropriate datasets (\S\ref{sec:applications}).

\begin{figure*}
        \centering
 	\includegraphics[width=\textwidth]{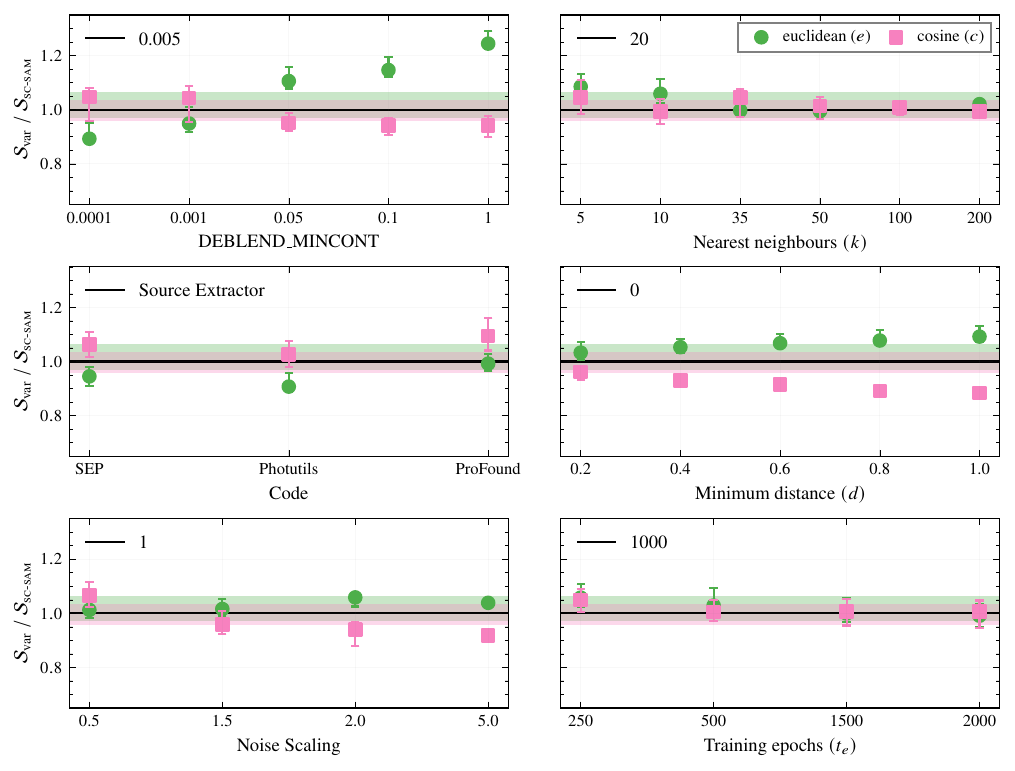}
	\caption{The effect of systematic uncertainties on the $\mathcal{S}_{e}$ (green) and $\mathcal{S}_{c}$ (pink) summary scores measured for \scsam. Each panel shows a different systematic, with the y-axis value indicating the fractional shift from the fiducial approach. The left column shows the minimum deblending contrast, source extraction code and noise level applied to the models. The right column shows the effect of key UMAP hyperparameter choices. The fiducial choices are shown by the solid black lines, with green and pink shaded regions indicating the associated uncertainty.
	\label{fig:systematics}}
\end{figure*}

Figure~\ref{fig:deg_layer} shows the fractional shift in $\mathcal{S}^{c}$ relative to the fiducial run when omitting each filter from each model. There are no statistically significant differences at $\lambda_{\mathrm{obs}} <2 \ \mu\rm{m}$, but longer wavelengths appear to be more poorly modelled. For example, removing F200W reduces the \sage\ score by $10\%$. This does not extend to neighbouring filters, suggesting that it may be due to a strong nebular feature appearing in the filter, rather than inaccurate modelling of a large population. In fact, this filter aligns with the aforementioned [O\textsc{iii}] emission at $z\sim3$. Both \spritz\ models steadily decline between F200W, F277W and F356W, which is unlikely to be caused by a nebular feature as \sage\ is unaffected. While very uncertain, a rough comparison between this trend and the redshift distribution shown in Figure \ref{fig:selection} appears to indicate an issue with rest-frame emission in the range $1<\lambda_{\mathrm{rest}} <2 \ \mathrm{\mu m}$, where the emission of old stars is expected to peak. This could suggest an issue with the treatment of these populations in \spritz, which would again be a consequence of the template SEDs. The fact that these signals can be identified in the observer frame, albeit weakly, suggests that this approach could be highly informative if performed in the rest frame, or when applied to model variations where the underlying physics is varied systematically. While the UMAP approach is intended for statistical evaluation, it is promising that the score maps and degradation layer can be used to identify potential areas of improvement.

\subsection{Understanding Systematics}\label{sec:systematics}

It is important to determine whether differing scores are not only significant with respect to UMAP's stochasticity, but also systematic uncertainties. Forward modelling is a potential source, but these cannot be quantified with lightcone catalogues alone. We instead defer this to the future work of Vijayan et al. (\emph{in prep}), but also refer the reader to \flagsi. This showed that while the predicted galaxy number counts were largely insensitive to dust modelling, the assumed IMF and SPS model can cause shifts of up to 0.3 and $0.9 \ \mathrm{dex}$ respectively. 

We instead focus on source extraction, by varying SE parameters in turn to produce 45 catalogue variations. Identical selection criteria and fitting approaches are then applied to these catalogues to recover comparable summary scores using the euclidean and cosine approaches. The left column of Figure~\ref{fig:systematics} shows the ratio of these scores to the fiducial measured for \scsam. While each model is affected differently, depending on the nature, degree and direction of its discrepancies, we find that no justifiable variation causes the ordering of model performance to change. This alone is a desirable result, demonstrating that models can confidently be differentiated from each other. It is therefore appropriate to limit further comparison to a single model, such that the sensitivity of the absolute score can be presented clearly. We choose \scsam\ based on its moderate performance. 

Most of the parameters, such as the detection filter size, threshold and minimum pixels, cause no change in the absolute value of either score. The broad consistency is likely due to the magnitude selection cut being far brighter than 30th magnitude, which is the average $5\sigma$ point source depth of the NIRCam imaging. The majority of sources in our sample are very robustly detected and therefore insensitive to these changes. Varying the scaling factor of the Kron aperture between $1-3$ induces no significant change. This demonstrates the efficacy of the aperture correction, as we would otherwise expect this to have a significant effect by changing the number of bright galaxies and their colours. The top left panel of Figure~\ref{fig:systematics} shows the scores obtained when using different minimum deblending contrasts, which control how aggressively segmented objects are split into separate galaxies. While over-deblending could affect the measured colours, $\mathcal{S}^{c}$ is insensitive. The density-based score can be decreased by reducing the minimum contrast, which is unsurprising given that \scsam\ overpredicts the number of galaxies. Visual inspection makes it clear that values below 0.001 tear apart well-defined bright galaxies, and those above 0.05 do not sufficiently deblend those of intermediate brightness. Any reasonable potential uncertainty is limited to $<10\%$. 

The middle panel shows the results of using different source extraction codes, namely \texttt{SEP} \citep{Barbary_2016}, \texttt{Photutils} \citep{Bradley_2025} and \texttt{ProFound} \citep{Robotham_2018}. We match parameter values exactly when in common with SE, and are otherwise informed by the literature and visual inspection. ProFound does not implement Kron photometry and instead iteratively dilates segments until the curve of growth has converged. We do not apply any aperture correction to these fluxes, but do apply them to the Kron fluxes produced by \texttt{SEP} and \texttt{Photutils}. Reassuringly, the uncertainties are again limited to $<10\%$, with the differing photometry approach of \texttt{ProFound} causing surprisingly little change in the SED shapes. 

The bottom panel demonstrates the sensitivity of the results to the level of noise applied to the models. By resampling each flux from a Gaussian defined by the observational uncertainties, we assume that these uncertainties are an accurate representation of the underlying noise levels. We test the impact of this assumption by performing new UMAP runs on the same sample of galaxies after resampling the model fluxes from Gaussians of different widths. No change is observed when scaling the width within a factor of two, and while the cosine score can be reduced by $<10\%$ with a factor of five, such a significant underestimation of the noise is unlikely given the brightness of these galaxies relative to the background level. Data processing appears to contribute an uncertainty of no more than $0.1 \ \mathrm{dex}$ to the measured scores. However, analysis of more shallow imaging, or higher resolution models requiring a less restrictive magnitude cut, could be more sensitive. These uncertainties could be removed almost entirely with full image-based forward modelling if required.

The right column of Figure~\ref{fig:systematics} shows the impact of changing the UMAP hyperparameters discussed in Section \ref{sec:umap_fitting}. Like the observational parameters, these do not change the ordering of model performance. Varying the number of neighbours has little effect on the absolute score, with all values consistent with the fiducial. The minimum embedded distance is more impactful, with the shift exceeding $10\%$ in the cosine space when $d=1.0$. This is intuitive, as larger values wash out the more detailed topological structure, making the models and observations appear more consistent. The tentative opposite trend in the euclidean space is less intuitive, and could indicate increased noise sensitivity. Noise can cause galaxies within the same population to be assigned to different nodes, but if $d>0$, these nodes cannot be placed at the same point in embedding space. We therefore consider $d=0$ to be the most viable option. The final panel shows that while the number of training epochs has little impact on the median $\mathcal{S}^{e}$ score, the percentile ranges can narrow when using larger values. When $t_{e} = 2000$, the \spritz\ scores are no longer statistically consistent, and \spritz$_{4}$ can be claimed to perform best, albeit at a level $<2\sigma$.

Overall, our results are largely insensitive to potential systematic uncertainties. There are no differences in performance rankings, indicating that we can reliably differentiate between models. The absolute agreement varies at the $<0.1 \ \mathrm{dex}$ level, but this could be minimised significantly with image-based forward modelling.

\subsection{Ideal Applications \& Future Work}\label{sec:applications}

While we have demonstrated that UMAP can be used to statistically differentiate between models, this does not represent the ideal use case of the DR approach. Only two of the models are based on physical prescriptions, which limits what can be learned about the physics of galaxy formation. The models are also quite dissimilar, so it would have been straightforward to identify \jaguar\ as the best performing model from the source counts alone. However, this will not always be the case when analysing detailed hydrodynamical models, particularly those that vary the implemented physics, where we can expect the predictions to be far more consistent. Where models may appear consistent within the uncertainties on the number counts or colour-redshift relations, these degeneracies may be broken by including the additional non-linear information highlighted by Figure~\ref{fig:pca}. For example, the \textsc{camels} simulations \citep{Villaescusa-Navarro_2021} produce thousands of galaxy catalogues exploring both cosmological and astrophysical parameter variation, which are being used for simulation-based inference approaches \citep{Lovell_2025}. While the simulated volumes are currently too small to produce reliable lightcones, this will not be the case for the third generation, which will be the ideal dataset with which to implement the DR approach.

This approach will also be enhanced by the next generation of survey data. The grism capabilities of both \emph{JWST}/NIRCam and \emph{Euclid}/NISP allow the identification of large and unbiased spectroscopic samples. These spectra can be rebinned into any number of faux filters, allowing the same analysis to be performed with higher wavelength resolution. Spectroscopic redshifts can also be derived from the grism data, allowing this analysis to be performed in the rest frame. \citet{Katz_2025} showed that UMAP analysis of rest-frame spectra from the \mbox{\textsc{megatron}} simulation produced clusters corresponding to distinct populations. We can therefore expect rest-frame studies to more straightforwardly derive constraints on the underlying physics from the metric maps and degradation layer.

%% file: sections/5_conclusion.tex
\section{Conclusions}\label{sec:conclusion}

We have presented the second instalment of the \flags\ series, which expands on the direct observable approach by investigating methods for analysing high-dimensional datasets. \jwst\ and \hst\ imaging of GOODS-S is consistently processed to produce a robust catalogue of 4590 bright galaxies ($m_{\mathrm{AB}}<26$) observed across thirteen optical-NIR filters. By applying observationally motivated noise to the predicted fluxes of galaxies from semi-empirical and semi-analytic models, we produce a dataset that can be compared directly to real galaxies in the observer frame. Comparing the flux distributions in thirteen dimensions is found to be unfeasible due to sparse sampling and memory demands, which we seek to mitigate using UMAP dimensionality reduction.

UMAP determines the position of each galaxy in a 2D embedding space constructed with a euclidean distance metric. The space is divided into 400 equally spaced bins, within which we compute a $\chi^{2}$-like metric $s$, which describes the statistical significance of differences between the observed and model-predicted sky area densities. We perform 500 UMAP fitting iterations to quantify the effect of its stochastic elements on a summary statistic $\mathcal{S}^{e}$, which describes how well each model predicts the observed distribution in this space. We find that \jaguar\ performs best overall ($\mathcal{S}^{e} = 408^{+33}_{-25}$), and \scsam\ ($\mathcal{S}^{e}=2578^{+504}_{-210}$) performs best of the semi-analytic models. UMAP retains valuable non-linear connections between the fluxes in different filters, producing more representative statistics than those generated by PCA, which can differ by up to $90\%$. However, discrepancies in this space are primarily driven by the models overpredicting the number of galaxies which pass the selection cuts.

We generate a second embedding using a cosine metric, which considers only the shape of the SEDs when positioning them in the 2D space. The bin-wise score is recomputed, now comparing the fraction of observed and model datasets residing in each region. The models perform more poorly, with \spritz' ($\mathcal{S}^{c}>8000$) inability to occupy large regions of the space highlighting limitations of template-based forward modelling. A lack of nebular emission modelling in \sage\ contributes to a significant discrepancy in the positioning of star-forming galaxies at high redshift, but the most significant individual discrepancy may be due to limited SAM snapshot cadence. The relative and absolute summary scores computed by both approaches are largely insensitive to systematic uncertainties introduced by source extraction and the UMAP reduction.

We demonstrate that a degradation layer can be used to determine which filters drive identified discrepancies, a technique that will be greatly enhanced by spectroscopic datasets. Each 500-iteration UMAP fit can be performed $>100$ times more quickly than Bayesian SED fitting of the same filters, demonstrating that dimensionality reduction provides a viable alternative to facilitate reliable and unbiased statistical comparisons between large datasets generated by \emph{Euclid}, LSST and simulation suites exploring astrophysical and cosmological parameter variation.

%% file: flags_umap.bib
@ARTICLE{Adams_2024,
       author = {{Adams}, Nathan J. and {Conselice}, Christopher J. and {Austin}, Duncan and {Harvey}, Thomas and {Ferreira}, Leonardo and {Trussler}, James and {Juod{\v{z}}balis}, Ignas and {Li}, Qiong and {Windhorst}, Rogier and {Cohen}, Seth H. and {Jansen}, Rolf A. and {Summers}, Jake and {Tompkins}, Scott and {Driver}, Simon P. and {Robotham}, Aaron and {D'Silva}, Jordan C.~J. and {Yan}, Haojing and {Coe}, Dan and {Frye}, Brenda and {Grogin}, Norman A. and {Koekemoer}, Anton M. and {Marshall}, Madeline A. and {Pirzkal}, Nor and {Ryan}, Russell E. and {Maksym}, W. Peter and {Rutkowski}, Michael J. and {Willmer}, Christopher N.~A. and {Hammel}, Heidi B. and {Nonino}, Mario and {Bhatawdekar}, Rachana and {Wilkins}, Stephen M. and {Bradley}, Larry D. and {Broadhurst}, Tom and {Cheng}, Cheng and {Dole}, Herv{\'e} and {Hathi}, Nimish P. and {Zitrin}, Adi},
        title = "{EPOCHS. II. The Ultraviolet Luminosity Function from 7.5 < z < 13.5 Using 180 arcmin$^{2}$ of Deep, Blank Fields from the PEARLS Survey and Public JWST Data}",
      journal = {\apj},
         year = 2024,
        month = apr,
       volume = {965},
       number = {2},
          eid = {169},
        pages = {169},
          doi = {10.3847/1538-4357/ad2a7b},
archivePrefix = {arXiv},
       eprint = {2304.13721},
 primaryClass = {astro-ph.GA},
       adsurl = {https://ui.adsabs.harvard.edu/abs/2024ApJ...965..169A}
}

@ARTICLE{Allen_2025,
       author = {{Euclid Collaboration} and {Allen}, N. and {Oesch}, P.~A. and {Bowler}, R.~A.~A. and {Toft}, S. and {Matharu}, J. and {Weaver}, J.~R. and {McPartland}, C.~J.~R. and {Shuntov}, M. and {Sanders}, D.~B. and {Mobasher}, B. and {McCracken}, H.~J. and {Atek}, H. and {Ba{\~n}ados}, E. and {Barrow}, S.~W.~J. and {Belladitta}, S. and {Carollo}, D. and {Castellano}, M. and {Conselice}, C.~J. and {Eisenhardt}, P.~R.~M. and {Harikane}, Y. and {Murphree}, G. and {Stefanon}, M. and {Wilkins}, S.~M. and {Amara}, A. and {Andreon}, S. and {Auricchio}, N. and {Baccigalupi}, C. and {Baldi}, M. and {Balestra}, A. and {Bardelli}, S. and {Battaglia}, P. and {Bender}, R. and {Biviano}, A. and {Branchini}, E. and {Brescia}, M. and {Brinchmann}, J. and {Camera}, S. and {Ca{\~n}as-Herrera}, G. and {Capobianco}, V. and {Carbone}, C. and {Carretero}, J. and {Castignani}, G. and {Cavuoti}, S. and {Chambers}, K.~C. and {Cimatti}, A. and {Colodro-Conde}, C. and {Congedo}, G. and {Conversi}, L. and {Copin}, Y. and {Courbin}, F. and {Courtois}, H.~M. and {Cropper}, M. and {Da Silva}, A. and {Degaudenzi}, H. and {De Lucia}, G. and {Dole}, H. and {Dubath}, F. and {Duncan}, C.~A.~J. and {Dupac}, X. and {Dusini}, S. and {Escoffier}, S. and {Farina}, M. and {Farinelli}, R. and {Faustini}, F. and {Ferriol}, S. and {Finelli}, F. and {Fourmanoit}, N. and {Frailis}, M. and {Franceschi}, E. and {Fumana}, M. and {Galeotta}, S. and {George}, K. and {Gillis}, B. and {Giocoli}, C. and {Gracia-Carpio}, J. and {Grazian}, A. and {Grupp}, F. and {Haugan}, S.~V.~H. and {Hoekstra}, H. and {Holmes}, W. and {Hook}, I.~M. and {Hormuth}, F. and {Hornstrup}, A. and {Jahnke}, K. and {Jhabvala}, M. and {Joachimi}, B. and {Keih{\"a}nen}, E. and {Kermiche}, S. and {Kiessling}, A. and {Kubik}, B. and {Kuijken}, K. and {K{\"u}mmel}, M. and {Kunz}, M. and {Kurki-Suonio}, H. and {Le Brun}, A.~M.~C. and {Le Mignant}, D. and {Ligori}, S. and {Lilje}, P.~B. and {Lindholm}, V. and {Lloro}, I. and {Mainetti}, G. and {Maino}, D. and {Maiorano}, E. and {Mansutti}, O. and {Marggraf}, O. and {Martinelli}, M. and {Martinet}, N. and {Marulli}, F. and {Massey}, R.~J. and {Medinaceli}, E. and {Mei}, S. and {Mellier}, Y. and {Meneghetti}, M. and {Merlin}, E. and {Meylan}, G. and {Mora}, A. and {Moresco}, M. and {Moscardini}, L. and {Nakajima}, R. and {Neissner}, C. and {Niemi}, S.-M. and {Padilla}, C. and {Paltani}, S. and {Pasian}, F. and {Pedersen}, K. and {Percival}, W.~J. and {Pettorino}, V. and {Pires}, S. and {Polenta}, G. and {Poncet}, M. and {Popa}, L.~A. and {Pozzetti}, L. and {Raison}, F. and {Renzi}, A. and {Rhodes}, J. and {Riccio}, G. and {Romelli}, E. and {Roncarelli}, M. and {Saglia}, R. and {Sakr}, Z. and {Sapone}, D. and {Sartoris}, B. and {Schirmer}, M. and {Schneider}, P. and {Schrabback}, T. and {Secroun}, A. and {Sefusatti}, E. and {Seidel}, G. and {Serrano}, S. and {Simon}, P. and {Sirignano}, C. and {Sirri}, G. and {Stanco}, L. and {Steinwagner}, J. and {Tallada-Cresp{\'\i}}, P. and {Taylor}, A.~N. and {Teplitz}, H.~I. and {Tereno}, I. and {Tessore}, N. and {Toledo-Moreo}, R. and {Torradeflot}, F. and {Tutusaus}, I. and {Valenziano}, L. and {Valiviita}, J. and {Vassallo}, T. and {Wang}, Y. and {Weller}, J. and {Zamorani}, G. and {Zerbi}, F.~M. and {Zucca}, E. and {Allevato}, V. and {Ballardini}, M. and {Bolzonella}, M. and {Bozzo}, E. and {Burigana}, C. and {Cabanac}, R. and {Calabrese}, M. and {Cappi}, A. and {Di Ferdinando}, D. and {Escartin Vigo}, J.~A. and {Gabarra}, L. and {Hartley}, W.~G. and {Maoli}, R. and {Mart{\'\i}n-Fleitas}, J. and {Matthew}, S. and {Maturi}, M. and {Mauri}, N. and {Metcalf}, R.~B. and {Pezzotta}, A. and {P{\"o}ntinen}, M. and {Porciani}, C. and {Risso}, I. and {Scottez}, V. and {Sereno}, M. and {Tenti}, M. and {Viel}, M. and {Wiesmann}, M. and {Akrami}, Y. and {Andika}, I.~T.},
        title = "{Euclid Quick Data Release (Q1): XXV. Hunting for luminous z > 6 galaxies in the Euclid Deep Fields ─ Forecasts and the first bright detections}",
      journal = {\aap},
         year = 2026,
        month = jun,
       volume = {711},
          eid = {A25},
        pages = {A25},
          doi = {10.1051/0004-6361/202557785},
archivePrefix = {arXiv},
       eprint = {2511.02926},
 primaryClass = {astro-ph.GA},
       adsurl = {https://ui.adsabs.harvard.edu/abs/2026A&A...711A..25E}
}

@ARTICLE{Arrabal-Haro_2023,
       author = {{Arrabal Haro}, Pablo and {Dickinson}, Mark and {Finkelstein}, Steven L. and {Kartaltepe}, Jeyhan S. and {Donnan}, Callum T. and {Burgarella}, Denis and {Carnall}, Adam C. and {Cullen}, Fergus and {Dunlop}, James S. and {Fern{\'a}ndez}, Vital and {Fujimoto}, Seiji and {Jung}, Intae and {Krips}, Melanie and {Larson}, Rebecca L. and {Papovich}, Casey and {P{\'e}rez-Gonz{\'a}lez}, Pablo G. and {Amor{\'\i}n}, Ricardo O. and {Bagley}, Micaela B. and {Buat}, V{\'e}ronique and {Casey}, Caitlin M. and {Chworowsky}, Katherine and {Cohen}, Seth H. and {Ferguson}, Henry C. and {Giavalisco}, Mauro and {Huertas-Company}, Marc and {Hutchison}, Taylor A. and {Kocevski}, Dale D. and {Koekemoer}, Anton M. and {Lucas}, Ray A. and {McLeod}, Derek J. and {McLure}, Ross J. and {Pirzkal}, Norbert and {Seill{\'e}}, Lise-Marie and {Trump}, Jonathan R. and {Weiner}, Benjamin J. and {Wilkins}, Stephen M. and {Zavala}, Jorge A.},
        title = "{Confirmation and refutation of very luminous galaxies in the early Universe}",
      journal = {\nat},
         year = 2023,
        month = oct,
       volume = {622},
       number = {7984},
        pages = {707-711},
          doi = {10.1038/s41586-023-06521-7},
archivePrefix = {arXiv},
       eprint = {2303.15431},
 primaryClass = {astro-ph.GA},
       adsurl = {https://ui.adsabs.harvard.edu/abs/2023Natur.622..707A}
}

@ARTICLE{Asadi_2025,
       author = {{Asadi}, Vahid and {Chartab}, Nima and {Zonoozi}, Akram Hasani and {Haghi}, Hosein and {Gozaliasl}, Ghassem and {Haghjoo}, Aryana and {Mobasher}, Bahram},
        title = "{Machine Learning Classification of COSMOS2020 Galaxies: Quiescent versus Star-forming}",
      journal = {\apj},
         year = 2025,
        month = nov,
       volume = {993},
       number = {1},
          eid = {123},
        pages = {123},
          doi = {10.3847/1538-4357/ae0a2c},
archivePrefix = {arXiv},
       eprint = {2509.03039},
 primaryClass = {astro-ph.GA},
       adsurl = {https://ui.adsabs.harvard.edu/abs/2025ApJ...993..123A}
}

@ARTICLE{Ashmead_2025,
       author = {{Ashmead}, Finian and {Newman}, Jeffrey A. and {Andrews}, Brett H. and {Bezanson}, Rachel and {Dey}, Biprateep and {Masters}, Daniel C. and {Stanford}, S.~A.},
        title = "{Optimizing Photometric Redshift Training Sets I: Efficient Compression of the Galaxy Color-Redshift Relation with UMAP}",
      journal = {arXiv e-prints},
         year = 2025,
        month = dec,
          eid = {arXiv:2512.09032},
        pages = {arXiv:2512.09032},
          doi = {10.48550/arXiv.2512.09032},
archivePrefix = {arXiv},
       eprint = {2512.09032},
 primaryClass = {astro-ph.GA},
       adsurl = {https://ui.adsabs.harvard.edu/abs/2025arXiv251209032A}
}

@ARTICLE{Astropy_2022,
       author = {{Astropy Collaboration} and {Price-Whelan}, Adrian M. and {Lim}, Pey Lian and {Earl}, Nicholas and {Starkman}, Nathaniel and {Bradley}, Larry and {Shupe}, David L. and {Patil}, Aarya A. and {Corrales}, Lia and {Brasseur}, C.~E. and {N{\"o}the}, Maximilian and {Donath}, Axel and {Tollerud}, Erik and {Morris}, Brett M. and {Ginsburg}, Adam and {Vaher}, Eero and {Weaver}, Benjamin A. and {Tocknell}, James and {Jamieson}, William and {van Kerkwijk}, Marten H. and {Robitaille}, Thomas P. and {Merry}, Bruce and {Bachetti}, Matteo and {G{\"u}nther}, H. Moritz and {Aldcroft}, Thomas L. and {Alvarado-Montes}, Jaime A. and {Archibald}, Anne M. and {B{\'o}di}, Attila and {Bapat}, Shreyas and {Barentsen}, Geert and {Baz{\'a}n}, Juanjo and {Biswas}, Manish and {Boquien}, M{\'e}d{\'e}ric and {Burke}, D.~J. and {Cara}, Daria and {Cara}, Mihai and {Conroy}, Kyle E. and {Conseil}, Simon and {Craig}, Matthew W. and {Cross}, Robert M. and {Cruz}, Kelle L. and {D'Eugenio}, Francesco and {Dencheva}, Nadia and {Devillepoix}, Hadrien A.~R. and {Dietrich}, J{\"o}rg P. and {Eigenbrot}, Arthur Davis and {Erben}, Thomas and {Ferreira}, Leonardo and {Foreman-Mackey}, Daniel and {Fox}, Ryan and {Freij}, Nabil and {Garg}, Suyog and {Geda}, Robel and {Glattly}, Lauren and {Gondhalekar}, Yash and {Gordon}, Karl D. and {Grant}, David and {Greenfield}, Perry and {Groener}, Austen M. and {Guest}, Steve and {Gurovich}, Sebastian and {Handberg}, Rasmus and {Hart}, Akeem and {Hatfield-Dodds}, Zac and {Homeier}, Derek and {Hosseinzadeh}, Griffin and {Jenness}, Tim and {Jones}, Craig K. and {Joseph}, Prajwel and {Kalmbach}, J. Bryce and {Karamehmetoglu}, Emir and {Ka{\l}uszy{\'n}ski}, Miko{\l}aj and {Kelley}, Michael S.~P. and {Kern}, Nicholas and {Kerzendorf}, Wolfgang E. and {Koch}, Eric W. and {Kulumani}, Shankar and {Lee}, Antony and {Ly}, Chun and {Ma}, Zhiyuan and {MacBride}, Conor and {Maljaars}, Jakob M. and {Muna}, Demitri and {Murphy}, N.~A. and {Norman}, Henrik and {O'Steen}, Richard and {Oman}, Kyle A. and {Pacifici}, Camilla and {Pascual}, Sergio and {Pascual-Granado}, J. and {Patil}, Rohit R. and {Perren}, Gabriel I. and {Pickering}, Timothy E. and {Rastogi}, Tanuj and {Roulston}, Benjamin R. and {Ryan}, Daniel F. and {Rykoff}, Eli S. and {Sabater}, Jose and {Sakurikar}, Parikshit and {Salgado}, Jes{\'u}s and {Sanghi}, Aniket and {Saunders}, Nicholas and {Savchenko}, Volodymyr and {Schwardt}, Ludwig and {Seifert-Eckert}, Michael and {Shih}, Albert Y. and {Jain}, Anany Shrey and {Shukla}, Gyanendra and {Sick}, Jonathan and {Simpson}, Chris and {Singanamalla}, Sudheesh and {Singer}, Leo P. and {Singhal}, Jaladh and {Sinha}, Manodeep and {Sip{\H{o}}cz}, Brigitta M. and {Spitler}, Lee R. and {Stansby}, David and {Streicher}, Ole and {{\v{S}}umak}, Jani and {Swinbank}, John D. and {Taranu}, Dan S. and {Tewary}, Nikita and {Tremblay}, Grant R. and {de Val-Borro}, Miguel and {Van Kooten}, Samuel J. and {Vasovi{\'c}}, Zlatan and {Verma}, Shresth and {de Miranda Cardoso}, Jos{\'e} Vin{\'\i}cius and {Williams}, Peter K.~G. and {Wilson}, Tom J. and {Winkel}, Benjamin and {Wood-Vasey}, W.~M. and {Xue}, Rui and {Yoachim}, Peter and {Zhang}, Chen and {Zonca}, Andrea and {Astropy Project Contributors}},
        title = "{The Astropy Project: Sustaining and Growing a Community-oriented Open-source Project and the Latest Major Release (v5.0) of the Core Package}",
      journal = {\apj},
         year = 2022,
        month = aug,
       volume = {935},
       number = {2},
          eid = {167},
        pages = {167},
          doi = {10.3847/1538-4357/ac7c74},
archivePrefix = {arXiv},
       eprint = {2206.14220},
 primaryClass = {astro-ph.IM},
       adsurl = {https://ui.adsabs.harvard.edu/abs/2022ApJ...935..167A}
}

@ARTICLE{Bagley_2023,
       author = {{Bagley}, Micaela B. and {Finkelstein}, Steven L. and {Koekemoer}, Anton M. and {Ferguson}, Henry C. and {Arrabal Haro}, Pablo and {Dickinson}, Mark and {Kartaltepe}, Jeyhan S. and {Papovich}, Casey and {P{\'e}rez-Gonz{\'a}lez}, Pablo G. and {Pirzkal}, Nor and {Somerville}, Rachel S. and {Willmer}, Christopher N.~A. and {Yang}, Guang and {Yung}, L.~Y. Aaron and {Fontana}, Adriano and {Grazian}, Andrea and {Grogin}, Norman A. and {Hirschmann}, Michaela and {Kewley}, Lisa J. and {Kirkpatrick}, Allison and {Kocevski}, Dale D. and {Lotz}, Jennifer M. and {Medrano}, Aubrey and {Morales}, Alexa M. and {Pentericci}, Laura and {Ravindranath}, Swara and {Trump}, Jonathan R. and {Wilkins}, Stephen M. and {Calabr{\`o}}, Antonello and {Cooper}, M.~C. and {Costantin}, Luca and {de la Vega}, Alexander and {Hilbert}, Bryan and {Hutchison}, Taylor A. and {Larson}, Rebecca L. and {Lucas}, Ray A. and {McGrath}, Elizabeth J. and {Ryan}, Russell and {Wang}, Xin and {Wuyts}, Stijn},
        title = "{CEERS Epoch 1 NIRCam Imaging: Reduction Methods and Simulations Enabling Early JWST Science Results}",
      journal = {\apjl},
         year = 2023,
        month = mar,
       volume = {946},
       number = {1},
          eid = {L12},
        pages = {L12},
          doi = {10.3847/2041-8213/acbb08},
archivePrefix = {arXiv},
       eprint = {2211.02495},
 primaryClass = {astro-ph.IM},
       adsurl = {https://ui.adsabs.harvard.edu/abs/2023ApJ...946L..12B}
}

@ARTICLE{Baker_2025,
       author = {{Baker}, William M. and {Valentino}, Francesco and {Lagos}, Claudia del P. and {Ito}, Kei and {Jespersen}, Christian Kragh and {Gottumukkala}, Rashmi and {Hjorth}, Jens and {Langeroodi}, Danial and {Sedgewick}, Aidan},
        title = "{Exploring over 700 massive quiescent galaxies at z = 2─7: Demographics and stellar mass functions}",
      journal = {\aap},
         year = 2025,
        month = oct,
       volume = {702},
          eid = {A270},
        pages = {A270},
          doi = {10.1051/0004-6361/202555829},
archivePrefix = {arXiv},
       eprint = {2506.04119},
 primaryClass = {astro-ph.GA},
       adsurl = {https://ui.adsabs.harvard.edu/abs/2025A&A...702A.270B}
}

@ARTICLE{Bakx_2023,
       author = {{Bakx}, Tom J.~L.~C. and {Zavala}, Jorge A. and {Mitsuhashi}, Ikki and {Treu}, Tommaso and {Fontana}, Adriano and {Tadaki}, Ken-ichi and {Casey}, Caitlin M. and {Castellano}, Marco and {Glazebrook}, Karl and {Hagimoto}, Masato and {Ikeda}, Ryota and {Jones}, Tucker and {Leethochawalit}, Nicha and {Mason}, Charlotte and {Morishita}, Takahiro and {Nanayakkara}, Themiya and {Pentericci}, Laura and {Roberts-Borsani}, Guido and {Santini}, Paola and {Serjeant}, Stephen and {Tamura}, Yoichi and {Trenti}, Michele and {Vanzella}, Eros},
        title = "{Deep ALMA redshift search of a z {\ensuremath{\sim}} 12 GLASS-JWST galaxy candidate}",
      journal = {\mnras},
         year = 2023,
        month = mar,
       volume = {519},
       number = {4},
        pages = {5076-5085},
          doi = {10.1093/mnras/stac3723},
archivePrefix = {arXiv},
       eprint = {2208.13642},
 primaryClass = {astro-ph.GA},
       adsurl = {https://ui.adsabs.harvard.edu/abs/2023MNRAS.519.5076B}
}

@article{Barbary_2016, doi = {10.21105/joss.00058}, url = {https://doi.org/10.21105/joss.00058}, year = {2016}, publisher = {The Open Journal}, volume = {1}, number = {6}, pages = {58}, author = {Kyle Barbary}, title = {SEP: Source Extractor as a library}, journal = {Journal of Open Source Software} }

@ARTICLE{Benson_2012,
       author = {{Benson}, Andrew J. and {Borgani}, Stefano and {De Lucia}, Gabriella and {Boylan-Kolchin}, Michael and {Monaco}, Pierluigi},
        title = "{Convergence of galaxy properties with merger tree temporal resolution}",
      journal = {\mnras},
         year = 2012,
        month = feb,
       volume = {419},
       number = {4},
        pages = {3590-3603},
          doi = {10.1111/j.1365-2966.2011.20002.x},
archivePrefix = {arXiv},
       eprint = {1107.4098},
 primaryClass = {astro-ph.CO},
       adsurl = {https://ui.adsabs.harvard.edu/abs/2012MNRAS.419.3590B}
}

@ARTICLE{Bernyk_2016,
       author = {{Bernyk}, Maksym and {Croton}, Darren J. and {Tonini}, Chiara and {Hodkinson}, Luke and {Hassan}, Amr H. and {Garel}, Thibault and {Duffy}, Alan R. and {Mutch}, Simon J. and {Poole}, Gregory B. and {Hegarty}, Sarah},
        title = "{The Theoretical Astrophysical Observatory: Cloud-based Mock Galaxy Catalogs}",
      journal = {\apjs},
         year = 2016,
        month = mar,
       volume = {223},
       number = {1},
          eid = {9},
        pages = {9},
          doi = {10.3847/0067-0049/223/1/9},
archivePrefix = {arXiv},
       eprint = {1403.5270},
 primaryClass = {astro-ph.GA},
       adsurl = {https://ui.adsabs.harvard.edu/abs/2016ApJS..223....9B}
}

@ARTICLE{Bertin_1996,
       author = {{Bertin}, E. and {Arnouts}, S.},
        title = "{SExtractor: Software for source extraction.}",
      journal = {\aaps},
         year = 1996,
        month = jun,
       volume = {117},
        pages = {393-404},
          doi = {10.1051/aas:1996164},
       adsurl = {https://ui.adsabs.harvard.edu/abs/1996A&AS..117..393B}
}

@ARTICLE{Binggeli_2019,
       author = {{Binggeli}, Christian and {Zackrisson}, Erik and {Ma}, Xiangcheng and {Inoue}, Akio K. and {Vikaeus}, Anton and {Hashimoto}, Takuya and {Mawatari}, Ken and {Shimizu}, Ikkoh and {Ceverino}, Daniel},
        title = "{Balmer breaks in simulated galaxies at z > 6}",
      journal = {\mnras},
         year = 2019,
        month = nov,
       volume = {489},
       number = {3},
        pages = {3827-3835},
          doi = {10.1093/mnras/stz2387},
archivePrefix = {arXiv},
       eprint = {1908.11393},
 primaryClass = {astro-ph.GA},
       adsurl = {https://ui.adsabs.harvard.edu/abs/2019MNRAS.489.3827B}
}

@ARTICLE{Bisigello_2021b,
       author = {{Bisigello}, L. and {Gruppioni}, C. and {Calura}, F. and {Feltre}, A. and {Pozzi}, F. and {Vignali}, C. and {Barchiesi}, L. and {Rodighiero}, G. and {Negrello}, M. and {Carrera}, F.~J. and {Dasyra}, K.~M. and {Fern{\'a}ndez-Ontiveros}, J.~A. and {Giard}, M. and {Hatziminaoglou}, E. and {Kaneda}, H. and {Lusso}, E. and {Pereira-Santaella}, M. and {P{\'e}rez Gonz{\'a}lez}, P.~G. and {Ricci}, C. and {Schaerer}, D. and {Spinoglio}, L. and {Wang}, L.},
        title = "{Simulating infrared spectro-photometric surveys with a SPRITZ}",
      journal = {\pasa},
         year = 2021,
        month = dec,
       volume = {38},
          eid = {e064},
        pages = {e064},
          doi = {10.1017/pasa.2021.57},
archivePrefix = {arXiv},
       eprint = {2111.11453},
 primaryClass = {astro-ph.GA},
       adsurl = {https://ui.adsabs.harvard.edu/abs/2021PASA...38...64B}
}

@ARTICLE{Boucaud_2016,
       author = {{Boucaud}, A. and {Bocchio}, M. and {Abergel}, A. and {Orieux}, F. and {Dole}, H. and {Hadj-Youcef}, M.~A.},
        title = "{Convolution kernels for multi-wavelength imaging}",
      journal = {\aap},
         year = 2016,
        month = dec,
       volume = {596},
          eid = {A63},
        pages = {A63},
          doi = {10.1051/0004-6361/201629080},
archivePrefix = {arXiv},
       eprint = {1609.02006},
 primaryClass = {astro-ph.IM},
       adsurl = {https://ui.adsabs.harvard.edu/abs/2016A&A...596A..63B}
}

@software{Bradley_2025,
  author       = {Larry Bradley and
                  Brigitta Sipőcz and
                  Thomas Robitaille and
                  Erik Tollerud and
                  Zé Vinícius and
                  Christoph Deil and
                  Kyle Barbary and
                  Tom J Wilson and
                  Ivo Busko and
                  Axel Donath and
                  Hans Moritz Günther and
                  Mihai Cara and
                  P. L. Lim and
                  Sebastian Meßlinger and
                  Simon Conseil and
                  Michael Droettboom and
                  Azalee Bostroem and
                  E. M. Bray and
                  Lars Andersen Bratholm and
                  Zach Burnett and
                  William Jamieson and
                  Adam Ginsburg and
                  Dan Taranu and
                  Geert Barentsen and
                  Matt Craig and
                  Brett M. Morris and
                  Marshall Perrin and
                  Shivangee Rathi},
  title        = {astropy/photutils: 2.3.0},
  month        = sep,
  year         = 2025,
  publisher    = {Zenodo},
  version      = {2.3.0},
  doi          = {10.5281/zenodo.17129028},
  url          = {https://doi.org/10.5281/zenodo.17129028},
  swhid        = {swh:1:dir:dd51869167d76d722ba87e3f80f9f4199ec08c3f;origin=https://doi.org/10.5281/zenodo.596036;visit=swh:1:snp:30a5f50b0586911dc674668853d9abc352a2bc22;anchor=swh:1:rel:e97861da904cf010c499a4211cd8a612373e912a;path=astropy-photutils-2294e35},
}

@software{brammer_2023_msaexp,
  author       = {Brammer, Gabriel},
  title        = {msaexp: NIRSpec analyis tools},
  month        = sep,
  year         = 2023,
  publisher    = {Zenodo},
  version      = {0.6.17},
  doi          = {10.5281/zenodo.8319596},
  url          = {https://doi.org/10.5281/zenodo.8319596},
}

@article{Bruzual_1983,
author = {Bruzual, Gustavo},
year = {1983},
month = {09},
pages = {105-127},
title = {Spectral evolution of galaxies. I - Early-type systems},
volume = {273},
journal = {\apj},
doi = {10.1086/161352}
}

@ARTICLE{Bruzual_2003,
       author = {{Bruzual}, G. and {Charlot}, S.},
        title = "{Stellar population synthesis at the resolution of 2003}",
      journal = {\mnras},
         year = 2003,
        month = oct,
       volume = {344},
       number = {4},
        pages = {1000-1028},
          doi = {10.1046/j.1365-8711.2003.06897.x},
archivePrefix = {arXiv},
       eprint = {astro-ph/0309134},
 primaryClass = {astro-ph},
       adsurl = {https://ui.adsabs.harvard.edu/abs/2003MNRAS.344.1000B}
}

@ARTICLE{Bulbul_2024,
       author = {{Bulbul}, E. and {Liu}, A. and {Kluge}, M. and {Zhang}, X. and {Sanders}, J.~S. and {Bahar}, Y.~E. and {Ghirardini}, V. and {Artis}, E. and {Seppi}, R. and {Garrel}, C. and {Ramos-Ceja}, M.~E. and {Comparat}, J. and {Balzer}, F. and {B{\"o}ckmann}, K. and {Br{\"u}ggen}, M. and {Clerc}, N. and {Dennerl}, K. and {Dolag}, K. and {Freyberg}, M. and {Grandis}, S. and {Gruen}, D. and {Kleinebreil}, F. and {Krippendorf}, S. and {Lamer}, G. and {Merloni}, A. and {Migkas}, K. and {Nandra}, K. and {Pacaud}, F. and {Predehl}, P. and {Reiprich}, T.~H. and {Schrabback}, T. and {Veronica}, A. and {Weller}, J. and {Zelmer}, S.},
        title = "{The SRG/eROSITA All-Sky Survey. The first catalog of galaxy clusters and groups in the Western Galactic Hemisphere}",
      journal = {\aap},
         year = 2024,
        month = may,
       volume = {685},
          eid = {A106},
        pages = {A106},
          doi = {10.1051/0004-6361/202348264},
archivePrefix = {arXiv},
       eprint = {2402.08452},
 primaryClass = {astro-ph.CO},
       adsurl = {https://ui.adsabs.harvard.edu/abs/2024A&A...685A.106B}
}

@ARTICLE{Camps_2015,
       author = {{Camps}, P. and {Baes}, M.},
        title = "{SKIRT: An advanced dust radiative transfer code with a user-friendly architecture}",
      journal = {Astronomy and Computing},
         year = 2015,
        month = mar,
       volume = {9},
        pages = {20-33},
          doi = {10.1016/j.ascom.2014.10.004},
archivePrefix = {arXiv},
       eprint = {1410.1629},
 primaryClass = {astro-ph.IM},
       adsurl = {https://ui.adsabs.harvard.edu/abs/2015A&C.....9...20C}
}

@ARTICLE{Calzetti_1994,
       author = {{Calzetti}, Daniela and {Kinney}, Anne L. and {Storchi-Bergmann}, Thaisa},
        title = "{Dust Extinction of the Stellar Continua in Starburst Galaxies: The Ultraviolet and Optical Extinction Law}",
      journal = {\apj},
         year = 1994,
        month = jul,
       volume = {429},
        pages = {582},
          doi = {10.1086/174346},
       adsurl = {https://ui.adsabs.harvard.edu/abs/1994ApJ...429..582C}
}

@ARTICLE{Calzetti_2000,
       author = {{Calzetti}, Daniela and {Armus}, Lee and {Bohlin}, Ralph C. and {Kinney}, Anne L. and {Koornneef}, Jan and {Storchi-Bergmann}, Thaisa},
        title = "{The Dust Content and Opacity of Actively Star-forming Galaxies}",
      journal = {\apj},
         year = 2000,
        month = apr,
       volume = {533},
       number = {2},
        pages = {682-695},
          doi = {10.1086/308692},
archivePrefix = {arXiv},
       eprint = {astro-ph/9911459},
 primaryClass = {astro-ph},
       adsurl = {https://ui.adsabs.harvard.edu/abs/2000ApJ...533..682C}
}

@ARTICLE{Carnall_2018,
       author = {{Carnall}, A.~C. and {McLure}, R.~J. and {Dunlop}, J.~S. and {Dav{\'e}}, R.},
        title = "{Inferring the star formation histories of massive quiescent galaxies with BAGPIPES: evidence for multiple quenching mechanisms}",
      journal = {\mnras},
         year = 2018,
        month = nov,
       volume = {480},
       number = {4},
        pages = {4379-4401},
          doi = {10.1093/mnras/sty2169},
archivePrefix = {arXiv},
       eprint = {1712.04452},
 primaryClass = {astro-ph.GA},
       adsurl = {https://ui.adsabs.harvard.edu/abs/2018MNRAS.480.4379C}
}

@ARTICLE{Carniani_2024,
       author = {{Carniani}, Stefano and {Hainline}, Kevin and {D'Eugenio}, Francesco and {Eisenstein}, Daniel J. and {Jakobsen}, Peter and {Witstok}, Joris and {Johnson}, Benjamin D. and {Chevallard}, Jacopo and {Maiolino}, Roberto and {Helton}, Jakob M. and {Willott}, Chris and {Robertson}, Brant and {Alberts}, Stacey and {Arribas}, Santiago and {Baker}, William M. and {Bhatawdekar}, Rachana and {Boyett}, Kristan and {Bunker}, Andrew J. and {Cameron}, Alex J. and {Cargile}, Phillip A. and {Charlot}, St{\'e}phane and {Curti}, Mirko and {Curtis-Lake}, Emma and {Egami}, Eiichi and {Giardino}, Giovanna and {Isaak}, Kate and {Ji}, Zhiyuan and {Jones}, Gareth C. and {Kumari}, Nimisha and {Maseda}, Michael V. and {Parlanti}, Eleonora and {P{\'e}rez-Gonz{\'a}lez}, Pablo G. and {Rawle}, Tim and {Rieke}, George and {Rieke}, Marcia and {Del Pino}, Bruno Rodr{\'\i}guez and {Saxena}, Aayush and {Scholtz}, Jan and {Smit}, Renske and {Sun}, Fengwu and {Tacchella}, Sandro and {{\"U}bler}, Hannah and {Venturi}, Giacomo and {Williams}, Christina C. and {Willmer}, Christopher N.~A.},
        title = "{Spectroscopic confirmation of two luminous galaxies at a redshift of 14}",
      journal = {\nat},
         year = 2024,
        month = sep,
       volume = {633},
       number = {8029},
        pages = {318-322},
          doi = {10.1038/s41586-024-07860-9},
archivePrefix = {arXiv},
       eprint = {2405.18485},
 primaryClass = {astro-ph.GA},
       adsurl = {https://ui.adsabs.harvard.edu/abs/2024Natur.633..318C}
}

@ARTICLE{Carrasco_2018,
       author = {{Carrasco}, D. and {Trenti}, M. and {Mutch}, S. and {Oesch}, P.~A.},
        title = "{GLACiAR, an Open-Source Python Tool for Simulations of Source Recovery and Completeness in Galaxy Surveys}",
      journal = {\pasa},
         year = 2018,
        month = jun,
       volume = {35},
          eid = {e022},
        pages = {e022},
          doi = {10.1017/pasa.2018.17},
archivePrefix = {arXiv},
       eprint = {1805.08985},
 primaryClass = {astro-ph.GA},
       adsurl = {https://ui.adsabs.harvard.edu/abs/2018PASA...35...22C}
}

@ARTICLE{Chabrier_2003,
       author = {{Chabrier}, Gilles},
        title = "{Galactic Stellar and Substellar Initial Mass Function}",
      journal = {\pasp},
         year = 2003,
        month = jul,
       volume = {115},
       number = {809},
        pages = {763-795},
          doi = {10.1086/376392},
archivePrefix = {arXiv},
       eprint = {astro-ph/0304382},
 primaryClass = {astro-ph},
       adsurl = {https://ui.adsabs.harvard.edu/abs/2003PASP..115..763C}
}

@ARTICLE{Charlot_2000,
       author = {{Charlot}, St{\'e}phane and {Fall}, S. Michael},
        title = "{A Simple Model for the Absorption of Starlight by Dust in Galaxies}",
      journal = {\apj},
         year = 2000,
        month = aug,
       volume = {539},
       number = {2},
        pages = {718-731},
          doi = {10.1086/309250},
archivePrefix = {arXiv},
       eprint = {astro-ph/0003128},
 primaryClass = {astro-ph},
       adsurl = {https://ui.adsabs.harvard.edu/abs/2000ApJ...539..718C}
}

@ARTICLE{Chevallard_2016,
       author = {{Chevallard}, Jacopo and {Charlot}, St{\'e}phane},
        title = "{Modelling and interpreting spectral energy distributions of galaxies with BEAGLE}",
      journal = {\mnras},
         year = 2016,
        month = oct,
       volume = {462},
       number = {2},
        pages = {1415-1443},
          doi = {10.1093/mnras/stw1756},
archivePrefix = {arXiv},
       eprint = {1603.03037},
 primaryClass = {astro-ph.GA},
       adsurl = {https://ui.adsabs.harvard.edu/abs/2016MNRAS.462.1415C}
}

@article{Conroy_2013,
	author = {Conroy, Charlie},
	doi = {https://doi.org/10.1146/annurev-astro-082812-141017},
	issn = {1545-4282},
	journal = {\araa},
	number = {Volume 51, 2013},
	pages = {393-455},
	publisher = {Annual Reviews},
	title = {Modeling the Panchromatic Spectral Energy Distributions of Galaxies},
	type = {Journal Article},
	url = {https://www.annualreviews.org/content/journals/10.1146/annurev-astro-082812-141017},
	volume = {51},
	year = {2013}}

@ARTICLE{Cook_2024,
       author = {{Cook}, Todd L. and {Bandi}, Behnood and {Philipsborn}, Sam and {Loveday}, Jon and {Bellstedt}, Sabine and {Driver}, Simon P. and {Robotham}, Aaron S.~G. and {Bilicki}, Maciej and {Kaur}, Gursharanjit and {Tempel}, Elmo and {Baldry}, Ivan and {Gruen}, Daniel and {Longhetti}, Marcella and {Iovino}, Angela and {Holwerda}, Benne W. and {Demarco}, Ricardo},
        title = "{Wide Area VISTA Extra-galactic Survey (WAVES): unsupervised star-galaxy separation on the WAVES-Wide photometric input catalogue using UMAP and HDBSCAN}",
      journal = {\mnras},
         year = 2024,
        month = dec,
       volume = {535},
       number = {3},
        pages = {2129-2148},
          doi = {10.1093/mnras/stae2389},
archivePrefix = {arXiv},
       eprint = {2406.11611},
 primaryClass = {astro-ph.GA},
       adsurl = {https://ui.adsabs.harvard.edu/abs/2024MNRAS.535.2129C}
}

@ARTICLE{Crain_2023,
       author = {{Crain}, Robert A. and {van de Voort}, Freeke},
        title = "{Hydrodynamical Simulations of the Galaxy Population: Enduring Successes and Outstanding Challenges}",
      journal = {\araa},
         year = 2023,
        month = aug,
       volume = {61},
        pages = {473-515},
          doi = {10.1146/annurev-astro-041923-043618},
archivePrefix = {arXiv},
       eprint = {2309.17075},
 primaryClass = {astro-ph.GA},
       adsurl = {https://ui.adsabs.harvard.edu/abs/2023ARA&A..61..473C}
}

@ARTICLE{Croton_2006,
       author = {{Croton}, Darren J. and {Springel}, Volker and {White}, Simon D.~M. and {De Lucia}, G. and {Frenk}, C.~S. and {Gao}, L. and {Jenkins}, A. and {Kauffmann}, G. and {Navarro}, J.~F. and {Yoshida}, N.},
        title = "{The many lives of active galactic nuclei: cooling flows, black holes and the luminosities and colours of galaxies}",
      journal = {\mnras},
         year = 2006,
        month = jan,
       volume = {365},
       number = {1},
        pages = {11-28},
          doi = {10.1111/j.1365-2966.2005.09675.x},
archivePrefix = {arXiv},
       eprint = {astro-ph/0508046},
 primaryClass = {astro-ph},
       adsurl = {https://ui.adsabs.harvard.edu/abs/2006MNRAS.365...11C}
}

@ARTICLE{Croton_2016,
       author = {{Croton}, Darren J. and {Stevens}, Adam R.~H. and {Tonini}, Chiara and {Garel}, Thibault and {Bernyk}, Maksym and {Bibiano}, Antonio and {Hodkinson}, Luke and {Mutch}, Simon J. and {Poole}, Gregory B. and {Shattow}, Genevieve M.},
        title = "{Semi-Analytic Galaxy Evolution (SAGE): Model Calibration and Basic Results}",
      journal = {\apjs},
         year = 2016,
        month = feb,
       volume = {222},
       number = {2},
          eid = {22},
        pages = {22},
          doi = {10.3847/0067-0049/222/2/22},
archivePrefix = {arXiv},
       eprint = {1601.04709},
 primaryClass = {astro-ph.GA},
       adsurl = {https://ui.adsabs.harvard.edu/abs/2016ApJS..222...22C}
}

@ARTICLE{Dahlen_2013,
       author = {{Dahlen}, Tomas and {Mobasher}, Bahram and {Faber}, Sandra M. and {Ferguson}, Henry C. and {Barro}, Guillermo and {Finkelstein}, Steven L. and {Finlator}, Kristian and {Fontana}, Adriano and {Gruetzbauch}, Ruth and {Johnson}, Seth and {Pforr}, Janine and {Salvato}, Mara and {Wiklind}, Tommy and {Wuyts}, Stijn and {Acquaviva}, Viviana and {Dickinson}, Mark E. and {Guo}, Yicheng and {Huang}, Jiasheng and {Huang}, Kuang-Han and {Newman}, Jeffrey A. and {Bell}, Eric F. and {Conselice}, Christopher J. and {Galametz}, Audrey and {Gawiser}, Eric and {Giavalisco}, Mauro and {Grogin}, Norman A. and {Hathi}, Nimish and {Kocevski}, Dale and {Koekemoer}, Anton M. and {Koo}, David C. and {Lee}, Kyoung-Soo and {McGrath}, Elizabeth J. and {Papovich}, Casey and {Peth}, Michael and {Ryan}, Russell and {Somerville}, Rachel and {Weiner}, Benjamin and {Wilson}, Grant},
        title = "{A Critical Assessment of Photometric Redshift Methods: A CANDELS Investigation}",
      journal = {\apj},
         year = 2013,
        month = oct,
       volume = {775},
       number = {2},
          eid = {93},
        pages = {93},
          doi = {10.1088/0004-637X/775/2/93},
archivePrefix = {arXiv},
       eprint = {1308.5353},
 primaryClass = {astro-ph.CO},
       adsurl = {https://ui.adsabs.harvard.edu/abs/2013ApJ...775...93D}
}

@ARTICLE{Dai_2023,
       author = {{Dai}, Yao and {Xu}, Jun and {Song}, Jie and {Fang}, Guanwen and {Zhou}, Chichun and {Ba}, Shuo and {Gu}, Yizhou and {Lin}, Zesen and {Kong}, Xu},
        title = "{The Classification of Galaxy Morphology in the H Band of the COSMOS-DASH Field: A Combination-based Machine-learning Clustering Model}",
      journal = {\apjs},
         year = 2023,
        month = sep,
       volume = {268},
       number = {1},
          eid = {34},
        pages = {34},
          doi = {10.3847/1538-4365/ace69e},
archivePrefix = {arXiv},
       eprint = {2307.02335},
 primaryClass = {astro-ph.GA},
       adsurl = {https://ui.adsabs.harvard.edu/abs/2023ApJS..268...34D}
}

@ARTICLE{Deeming_1964,
       author = {{Deeming}, T.~J.},
        title = "{Stellar spectral classification, I.}",
      journal = {\mnras},
         year = 1964,
        month = jan,
       volume = {127},
        pages = {493},
          doi = {10.1093/mnras/127.6.493},
       adsurl = {https://ui.adsabs.harvard.edu/abs/1964MNRAS.127..493D}
}

@ARTICLE{Devriendt_1999,
       author = {{Devriendt}, J.~E.~G. and {Guiderdoni}, B. and {Sadat}, R.},
        title = "{Galaxy modelling. I. Spectral energy distributions from far-UV to sub-mm wavelengths}",
      journal = {\aap},
         year = 1999,
        month = oct,
       volume = {350},
        pages = {381-398},
          doi = {10.48550/arXiv.astro-ph/9906332},
archivePrefix = {arXiv},
       eprint = {astro-ph/9906332},
 primaryClass = {astro-ph},
       adsurl = {https://ui.adsabs.harvard.edu/abs/1999A&A...350..381D}
}

@ARTICLE{Dong_2018,
       author = {{Zhang}, Dong},
        title = "{A Review of the Theory of Galactic Winds Driven by Stellar Feedback}",
      journal = {Galaxies},
         year = 2018,
        month = nov,
       volume = {6},
       number = {4},
          eid = {114},
        pages = {114},
          doi = {10.3390/galaxies6040114},
archivePrefix = {arXiv},
       eprint = {1811.00558},
 primaryClass = {astro-ph.GA},
       adsurl = {https://ui.adsabs.harvard.edu/abs/2018Galax...6..114Z}
}

@ARTICLE{Driver_2010,
       author = {{Driver}, Simon P. and {Robotham}, Aaron S.~G.},
        title = "{Quantifying cosmic variance}",
      journal = {\mnras},
         year = 2010,
        month = oct,
       volume = {407},
       number = {4},
        pages = {2131-2140},
          doi = {10.1111/j.1365-2966.2010.17028.x},
archivePrefix = {arXiv},
       eprint = {1005.2538},
 primaryClass = {astro-ph.CO},
       adsurl = {https://ui.adsabs.harvard.edu/abs/2010MNRAS.407.2131D}
}

@ARTICLE{Eisenstein_2026,
       author = {{Eisenstein}, Daniel J. and {Willott}, Chris and {Alberts}, Stacey and {Arribas}, Santiago and {Bonaventura}, Nina and {Bunker}, Andrew J. and {Cameron}, Alex J. and {Carniani}, Stefano and {Charlot}, Stephane and {Curtis-Lake}, Emma and {D'Eugenio}, Francesco and {Ferruit}, Pierre and {Giardino}, Giovanna and {Hainline}, Kevin and {Hausen}, Ryan and {Jakobsen}, Peter and {Johnson}, Benjamin D. and {Maiolino}, Roberto and {Rauscher}, Bernard J. and {Rieke}, Marcia and {Rieke}, George and {Rix}, Hans-Walter and {Robertson}, Brant and {Stark}, Daniel P. and {Tacchella}, Sandro and {Williams}, Christina C. and {Willmer}, Christopher N.~A. and {Baker}, William M. and {Baum}, Stefi and {Bhatawdekar}, Rachana and {Boyett}, Kristan and {Chen}, Zuyi and {Chevallard}, Jacopo and {Circosta}, Chiara and {Curti}, Mirko and {Danhaive}, A. Lola and {DeCoursey}, Christa and {Endsley}, Ryan and {de Graaff}, Anna and {Dressler}, Alan and {Egami}, Eiichi and {Helton}, Jakob M. and {Hviding}, Raphael E. and {Ji}, Zhiyuan and {Jones}, Gareth C. and {Kumari}, Nimisha and {L{\"u}tzgendorf}, Nora and {Laseter}, Isaac and {Looser}, Tobias J. and {Lyu}, Jianwei and {Maseda}, Michael V. and {Nelson}, Erica and {Parlanti}, Eleonora and {Perna}, Michele and {Pusk{\'a}s}, D{\'a}vid and {Rawle}, Tim and {Rodr{\'\i}guez Del Pino}, Bruno and {Rujopakarn}, Wiphu and {Sandles}, Lester and {Saxena}, Aayush and {Scholtz}, Jan and {Sharpe}, Katherine and {Shivaei}, Irene and {Silcock}, Maddie S. and {Simmonds}, Charlotte and {Skarbinski}, Maya and {Smit}, Renske and {Stone}, Meredith and {Suess}, Katherine A. and {Sun}, Fengwu and {Tang}, Mengtao and {Topping}, Michael W. and {{\"U}bler}, Hannah and {Villanueva}, Natalia C. and {Wallace}, Imaan E.~B. and {Whitler}, Lily and {Witstok}, Joris and {Woodrum}, Charity},
        title = "{Overview of the JWST Advanced Deep Extragalactic Survey (JADES)}",
      journal = {\apjs},
         year = 2026,
        month = mar,
       volume = {283},
       number = {1},
          eid = {6},
        pages = {6},
          doi = {10.3847/1538-4365/ae3163},
archivePrefix = {arXiv},
       eprint = {2306.02465},
 primaryClass = {astro-ph.GA},
       adsurl = {https://ui.adsabs.harvard.edu/abs/2026ApJS..283....6E}
}

@ARTICLE{Fabian_2012,
       author = {{Fabian}, A.~C.},
        title = "{Observational Evidence of Active Galactic Nuclei Feedback}",
      journal = {\araa},
         year = 2012,
        month = sep,
       volume = {50},
        pages = {455-489},
          doi = {10.1146/annurev-astro-081811-125521},
archivePrefix = {arXiv},
       eprint = {1204.4114},
 primaryClass = {astro-ph.CO},
       adsurl = {https://ui.adsabs.harvard.edu/abs/2012ARA&A..50..455F}
}

@ARTICLE{Fan_2006,
       author = {{Fan}, Xiaohui and {Strauss}, Michael A. and {Becker}, Robert H. and {White}, Richard L. and {Gunn}, James E. and {Knapp}, Gillian R. and {Richards}, Gordon T. and {Schneider}, Donald P. and {Brinkmann}, J. and {Fukugita}, Masataka},
        title = "{Constraining the Evolution of the Ionizing Background and the Epoch of Reionization with z\raisebox{-0.5ex}\textasciitilde6 Quasars. II. A Sample of 19 Quasars}",
      journal = {\aj},
         year = 2006,
        month = jul,
       volume = {132},
       number = {1},
        pages = {117-136},
          doi = {10.1086/504836},
archivePrefix = {arXiv},
       eprint = {astro-ph/0512082},
 primaryClass = {astro-ph},
       adsurl = {https://ui.adsabs.harvard.edu/abs/2006AJ....132..117F}
}

@ARTICLE{Feltre_2016,
       author = {{Feltre}, A. and {Charlot}, S. and {Gutkin}, J.},
        title = "{Nuclear activity versus star formation: emission-line diagnostics at ultraviolet and optical wavelengths}",
      journal = {\mnras},
         year = 2016,
        month = mar,
       volume = {456},
       number = {3},
        pages = {3354-3374},
          doi = {10.1093/mnras/stv2794},
archivePrefix = {arXiv},
       eprint = {1511.08217},
 primaryClass = {astro-ph.GA},
       adsurl = {https://ui.adsabs.harvard.edu/abs/2016MNRAS.456.3354F}
}

@ARTICLE{Ferland_2013,
       author = {{Ferland}, G.~J. and {Porter}, R.~L. and {van Hoof}, P.~A.~M. and {Williams}, R.~J.~R. and {Abel}, N.~P. and {Lykins}, M.~L. and {Shaw}, G. and {Henney}, W.~J. and {Stancil}, P.~C.},
        title = "{The 2013 Release of Cloudy}",
      journal = {\rmxaa},
         year = 2013,
        month = apr,
       volume = {49},
        pages = {137-163},
          doi = {10.48550/arXiv.1302.4485},
archivePrefix = {arXiv},
       eprint = {1302.4485},
 primaryClass = {astro-ph.GA},
       adsurl = {https://ui.adsabs.harvard.edu/abs/2013RMxAA..49..137F}
}

@ARTICLE{Finkelstein_2015,
       author = {{Finkelstein}, Steven L. and {Ryan}, Jr., Russell E. and {Papovich}, Casey and {Dickinson}, Mark and {Song}, Mimi and {Somerville}, Rachel S. and {Ferguson}, Henry C. and {Salmon}, Brett and {Giavalisco}, Mauro and {Koekemoer}, Anton M. and {Ashby}, Matthew L.~N. and {Behroozi}, Peter and {Castellano}, Marco and {Dunlop}, James S. and {Faber}, Sandy M. and {Fazio}, Giovanni G. and {Fontana}, Adriano and {Grogin}, Norman A. and {Hathi}, Nimish and {Jaacks}, Jason and {Kocevski}, Dale D. and {Livermore}, Rachael and {McLure}, Ross J. and {Merlin}, Emiliano and {Mobasher}, Bahram and {Newman}, Jeffrey A. and {Rafelski}, Marc and {Tilvi}, Vithal and {Willner}, S.~P.},
        title = "{The Evolution of the Galaxy Rest-frame Ultraviolet Luminosity Function over the First Two Billion Years}",
      journal = {\apj},
         year = 2015,
        month = sep,
       volume = {810},
       number = {1},
          eid = {71},
        pages = {71},
          doi = {10.1088/0004-637X/810/1/71},
archivePrefix = {arXiv},
       eprint = {1410.5439},
 primaryClass = {astro-ph.GA},
       adsurl = {https://ui.adsabs.harvard.edu/abs/2015ApJ...810...71F}
}

@ARTICLE{Fortuni_2023,
       author = {{Fortuni}, Flaminia and {Merlin}, Emiliano and {Fontana}, Adriano and {Giocoli}, Carlo and {Romelli}, Erik and {Graziani}, Luca and {Santini}, Paola and {Castellano}, Marco and {Charlot}, St{\'e}phane and {Chevallard}, Jacopo},
        title = "{FORECAST: A flexible software to forward model cosmological hydrodynamical simulations mimicking real observations}",
      journal = {\aap},
         year = 2023,
        month = sep,
       volume = {677},
          eid = {A102},
        pages = {A102},
          doi = {10.1051/0004-6361/202346725},
archivePrefix = {arXiv},
       eprint = {2305.19166},
 primaryClass = {astro-ph.IM},
       adsurl = {https://ui.adsabs.harvard.edu/abs/2023A&A...677A.102F}
}

@ARTICLE{GAIA_2016,
       author = {{Gaia Collaboration} and {Prusti}, T. and {de Bruijne}, J.~H.~J. and {Brown}, A.~G.~A. and {Vallenari}, A. and {Babusiaux}, C. and {Bailer-Jones}, C.~A.~L. and {Bastian}, U. and {Biermann}, M. and {Evans}, D.~W. and {Eyer}, L. and {Jansen}, F. and {Jordi}, C. and {Klioner}, S.~A. and {Lammers}, U. and {Lindegren}, L. and {Luri}, X. and {Mignard}, F. and {Milligan}, D.~J. and {Panem}, C. and {Poinsignon}, V. and {Pourbaix}, D. and {Randich}, S. and {Sarri}, G. and {Sartoretti}, P. and {Siddiqui}, H.~I. and {Soubiran}, C. and {Valette}, V. and {van Leeuwen}, F. and {Walton}, N.~A. and {Aerts}, C. and {Arenou}, F. and {Cropper}, M. and {Drimmel}, R. and {H{\o}g}, E. and {Katz}, D. and {Lattanzi}, M.~G. and {O'Mullane}, W. and {Grebel}, E.~K. and {Holland}, A.~D. and {Huc}, C. and {Passot}, X. and {Bramante}, L. and {Cacciari}, C. and {Casta{\~n}eda}, J. and {Chaoul}, L. and {Cheek}, N. and {De Angeli}, F. and {Fabricius}, C. and {Guerra}, R. and {Hern{\'a}ndez}, J. and {Jean-Antoine-Piccolo}, A. and {Masana}, E. and {Messineo}, R. and {Mowlavi}, N. and {Nienartowicz}, K. and {Ord{\'o}{\~n}ez-Blanco}, D. and {Panuzzo}, P. and {Portell}, J. and {Richards}, P.~J. and {Riello}, M. and {Seabroke}, G.~M. and {Tanga}, P. and {Th{\'e}venin}, F. and {Torra}, J. and {Els}, S.~G. and {Gracia-Abril}, G. and {Comoretto}, G. and {Garcia-Reinaldos}, M. and {Lock}, T. and {Mercier}, E. and {Altmann}, M. and {Andrae}, R. and {Astraatmadja}, T.~L. and {Bellas-Velidis}, I. and {Benson}, K. and {Berthier}, J. and {Blomme}, R. and {Busso}, G. and {Carry}, B. and {Cellino}, A. and {Clementini}, G. and {Cowell}, S. and {Creevey}, O. and {Cuypers}, J. and {Davidson}, M. and {De Ridder}, J. and {de Torres}, A. and {Delchambre}, L. and {Dell'Oro}, A. and {Ducourant}, C. and {Fr{\'e}mat}, Y. and {Garc{\'\i}a-Torres}, M. and {Gosset}, E. and {Halbwachs}, J.-L. and {Hambly}, N.~C. and {Harrison}, D.~L. and {Hauser}, M. and {Hestroffer}, D. and {Hodgkin}, S.~T. and {Huckle}, H.~E. and {Hutton}, A. and {Jasniewicz}, G. and {Jordan}, S. and {Kontizas}, M. and {Korn}, A.~J. and {Lanzafame}, A.~C. and {Manteiga}, M. and {Moitinho}, A. and {Muinonen}, K. and {Osinde}, J. and {Pancino}, E. and {Pauwels}, T. and {Petit}, J.-M. and {Recio-Blanco}, A. and {Robin}, A.~C. and {Sarro}, L.~M. and {Siopis}, C. and {Smith}, M. and {Smith}, K.~W. and {Sozzetti}, A. and {Thuillot}, W. and {van Reeven}, W. and {Viala}, Y. and {Abbas}, U. and {Abreu Aramburu}, A. and {Accart}, S. and {Aguado}, J.~J. and {Allan}, P.~M. and {Allasia}, W. and {Altavilla}, G. and {{\'A}lvarez}, M.~A. and {Alves}, J. and {Anderson}, R.~I. and {Andrei}, A.~H. and {Anglada Varela}, E. and {Antiche}, E. and {Antoja}, T. and {Ant{\'o}n}, S. and {Arcay}, B. and {Atzei}, A. and {Ayache}, L. and {Bach}, N. and {Baker}, S.~G. and {Balaguer-N{\'u}{\~n}ez}, L. and {Barache}, C. and {Barata}, C. and {Barbier}, A. and {Barblan}, F. and {Baroni}, M. and {Barrado y Navascu{\'e}s}, D. and {Barros}, M. and {Barstow}, M.~A. and {Becciani}, U. and {Bellazzini}, M. and {Bellei}, G. and {Bello Garc{\'\i}a}, A. and {Belokurov}, V. and {Bendjoya}, P. and {Berihuete}, A. and {Bianchi}, L. and {Bienaym{\'e}}, O. and {Billebaud}, F. and {Blagorodnova}, N. and {Blanco-Cuaresma}, S. and {Boch}, T. and {Bombrun}, A. and {Borrachero}, R. and {Bouquillon}, S. and {Bourda}, G. and {Bouy}, H. and {Bragaglia}, A. and {Breddels}, M.~A. and {Brouillet}, N. and {Br{\"u}semeister}, T. and {Bucciarelli}, B. and {Budnik}, F. and {Burgess}, P. and {Burgon}, R. and {Burlacu}, A. and {Busonero}, D. and {Buzzi}, R. and {Caffau}, E. and {Cambras}, J. and {Campbell}, H. and {Cancelliere}, R. and {Cantat-Gaudin}, T. and {Carlucci}, T. and {Carrasco}, J.~M. and {Castellani}, M. and {Charlot}, P. and {Charnas}, J. and {Charvet}, P. and {Chassat}, F. and {Chiavassa}, A. and {Clotet}, M. and {Cocozza}, G. and {Collins}, R.~S. and {Collins}, P. and {Costigan}, G.},
        title = "{The Gaia mission}",
      journal = {\aap},
         year = 2016,
        month = nov,
       volume = {595},
          eid = {A1},
        pages = {A1},
          doi = {10.1051/0004-6361/201629272},
archivePrefix = {arXiv},
       eprint = {1609.04153},
 primaryClass = {astro-ph.IM},
       adsurl = {https://ui.adsabs.harvard.edu/abs/2016A&A...595A...1G}
}

@ARTICLE{GAIA_2023,
       author = {{Gaia Collaboration} and {Vallenari}, A. and {Brown}, A.~G.~A. and {Prusti}, T. and {de Bruijne}, J.~H.~J. and {Arenou}, F. and {Babusiaux}, C. and {Biermann}, M. and {Creevey}, O.~L. and {Ducourant}, C. and {Evans}, D.~W. and {Eyer}, L. and {Guerra}, R. and {Hutton}, A. and {Jordi}, C. and {Klioner}, S.~A. and {Lammers}, U.~L. and {Lindegren}, L. and {Luri}, X. and {Mignard}, F. and {Panem}, C. and {Pourbaix}, D. and {Randich}, S. and {Sartoretti}, P. and {Soubiran}, C. and {Tanga}, P. and {Walton}, N.~A. and {Bailer-Jones}, C.~A.~L. and {Bastian}, U. and {Drimmel}, R. and {Jansen}, F. and {Katz}, D. and {Lattanzi}, M.~G. and {van Leeuwen}, F. and {Bakker}, J. and {Cacciari}, C. and {Casta{\~n}eda}, J. and {De Angeli}, F. and {Fabricius}, C. and {Fouesneau}, M. and {Fr{\'e}mat}, Y. and {Galluccio}, L. and {Guerrier}, A. and {Heiter}, U. and {Masana}, E. and {Messineo}, R. and {Mowlavi}, N. and {Nicolas}, C. and {Nienartowicz}, K. and {Pailler}, F. and {Panuzzo}, P. and {Riclet}, F. and {Roux}, W. and {Seabroke}, G.~M. and {Sordo}, R. and {Th{\'e}venin}, F. and {Gracia-Abril}, G. and {Portell}, J. and {Teyssier}, D. and {Altmann}, M. and {Andrae}, R. and {Audard}, M. and {Bellas-Velidis}, I. and {Benson}, K. and {Berthier}, J. and {Blomme}, R. and {Burgess}, P.~W. and {Busonero}, D. and {Busso}, G. and {C{\'a}novas}, H. and {Carry}, B. and {Cellino}, A. and {Cheek}, N. and {Clementini}, G. and {Damerdji}, Y. and {Davidson}, M. and {de Teodoro}, P. and {Nu{\~n}ez Campos}, M. and {Delchambre}, L. and {Dell'Oro}, A. and {Esquej}, P. and {Fern{\'a}ndez-Hern{\'a}ndez}, J. and {Fraile}, E. and {Garabato}, D. and {Garc{\'\i}a-Lario}, P. and {Gosset}, E. and {Haigron}, R. and {Halbwachs}, J.-L. and {Hambly}, N.~C. and {Harrison}, D.~L. and {Hern{\'a}ndez}, J. and {Hestroffer}, D. and {Hodgkin}, S.~T. and {Holl}, B. and {Jan{\ss}en}, K. and {Jevardat de Fombelle}, G. and {Jordan}, S. and {Krone-Martins}, A. and {Lanzafame}, A.~C. and {L{\"o}ffler}, W. and {Marchal}, O. and {Marrese}, P.~M. and {Moitinho}, A. and {Muinonen}, K. and {Osborne}, P. and {Pancino}, E. and {Pauwels}, T. and {Recio-Blanco}, A. and {Reyl{\'e}}, C. and {Riello}, M. and {Rimoldini}, L. and {Roegiers}, T. and {Rybizki}, J. and {Sarro}, L.~M. and {Siopis}, C. and {Smith}, M. and {Sozzetti}, A. and {Utrilla}, E. and {van Leeuwen}, M. and {Abbas}, U. and {{\'A}brah{\'a}m}, P. and {Abreu Aramburu}, A. and {Aerts}, C. and {Aguado}, J.~J. and {Ajaj}, M. and {Aldea-Montero}, F. and {Altavilla}, G. and {{\'A}lvarez}, M.~A. and {Alves}, J. and {Anders}, F. and {Anderson}, R.~I. and {Anglada Varela}, E. and {Antoja}, T. and {Baines}, D. and {Baker}, S.~G. and {Balaguer-N{\'u}{\~n}ez}, L. and {Balbinot}, E. and {Balog}, Z. and {Barache}, C. and {Barbato}, D. and {Barros}, M. and {Barstow}, M.~A. and {Bartolom{\'e}}, S. and {Bassilana}, J.-L. and {Bauchet}, N. and {Becciani}, U. and {Bellazzini}, M. and {Berihuete}, A. and {Bernet}, M. and {Bertone}, S. and {Bianchi}, L. and {Binnenfeld}, A. and {Blanco-Cuaresma}, S. and {Blazere}, A. and {Boch}, T. and {Bombrun}, A. and {Bossini}, D. and {Bouquillon}, S. and {Bragaglia}, A. and {Bramante}, L. and {Breedt}, E. and {Bressan}, A. and {Brouillet}, N. and {Brugaletta}, E. and {Bucciarelli}, B. and {Burlacu}, A. and {Butkevich}, A.~G. and {Buzzi}, R. and {Caffau}, E. and {Cancelliere}, R. and {Cantat-Gaudin}, T. and {Carballo}, R. and {Carlucci}, T. and {Carnerero}, M.~I. and {Carrasco}, J.~M. and {Casamiquela}, L. and {Castellani}, M. and {Castro-Ginard}, A. and {Chaoul}, L. and {Charlot}, P. and {Chemin}, L. and {Chiaramida}, V. and {Chiavassa}, A. and {Chornay}, N. and {Comoretto}, G. and {Contursi}, G. and {Cooper}, W.~J. and {Cornez}, T. and {Cowell}, S. and {Crifo}, F. and {Cropper}, M. and {Crosta}, M. and {Crowley}, C. and {Dafonte}, C. and {Dapergolas}, A. and {David}, M. and {David}, P. and {de Laverny}, P. and {De Luise}, F. and {De March}, R.},
        title = "{Gaia Data Release 3. Summary of the content and survey properties}",
      journal = {\aap},
         year = 2023,
        month = jun,
       volume = {674},
          eid = {A1},
        pages = {A1},
          doi = {10.1051/0004-6361/202243940},
archivePrefix = {arXiv},
       eprint = {2208.00211},
 primaryClass = {astro-ph.GA},
       adsurl = {https://ui.adsabs.harvard.edu/abs/2023A&A...674A...1G}
}

@ARTICLE{Galametz_2013,
       author = {{Galametz}, Audrey and {Grazian}, Andrea and {Fontana}, Adriano and {Ferguson}, Henry C. and {Ashby}, M.~L.~N. and {Barro}, Guillermo and {Castellano}, Marco and {Dahlen}, Tomas and {Donley}, Jennifer L. and {Faber}, Sandy M. and {Grogin}, Norman and {Guo}, Yicheng and {Huang}, Kuang-Han and {Kocevski}, Dale D. and {Koekemoer}, Anton M. and {Lee}, Kyoung-Soo and {McGrath}, Elizabeth J. and {Peth}, Michael and {Willner}, S.~P. and {Almaini}, Omar and {Cooper}, Michael and {Cooray}, Asantha and {Conselice}, Christopher J. and {Dickinson}, Mark and {Dunlop}, James S. and {Fazio}, G.~G. and {Foucaud}, Sebastien and {Gardner}, Jonathan P. and {Giavalisco}, Mauro and {Hathi}, N.~P. and {Hartley}, Will G. and {Koo}, David C. and {Lai}, Kamson and {de Mello}, Duilia F. and {McLure}, Ross J. and {Lucas}, Ray A. and {Paris}, Diego and {Pentericci}, Laura and {Santini}, Paola and {Simpson}, Chris and {Sommariva}, Veronica and {Targett}, Thomas and {Weiner}, Benjamin J. and {Wuyts}, Stijn and {CANDELS Team}},
        title = "{CANDELS Multiwavelength Catalogs: Source Identification and Photometry in the CANDELS UKIDSS Ultra-deep Survey Field}",
      journal = {\apjs},
         year = 2013,
        month = jun,
       volume = {206},
       number = {2},
          eid = {10},
        pages = {10},
          doi = {10.1088/0067-0049/206/2/10},
archivePrefix = {arXiv},
       eprint = {1305.1823},
 primaryClass = {astro-ph.CO},
       adsurl = {https://ui.adsabs.harvard.edu/abs/2013ApJS..206...10G}
}

@ARTICLE{Galaz_1998,
       author = {{Galaz}, Gaspar and {de Lapparent}, Valerie},
        title = "{The ESO-Sculptor Survey: spectral classification of galaxies with Z < 0.5}",
      journal = {\aap},
         year = 1998,
        month = apr,
       volume = {332},
        pages = {459-478},
          doi = {10.48550/arXiv.astro-ph/9711093},
archivePrefix = {arXiv},
       eprint = {astro-ph/9711093},
 primaryClass = {astro-ph},
       adsurl = {https://ui.adsabs.harvard.edu/abs/1998A&A...332..459G}
}

@ARTICLE{Gandolfi_2026,
       author = {{Gandolfi}, G. and {Rodighiero}, G. and {Bisigello}, L. and {Grazian}, A. and {Finkelstein}, S.~L. and {Dickinson}, M. and {Castellano}, M. and {Merlin}, E. and {Calabr{\`o}}, A. and {Papovich}, C. and {Bianchetti}, A. and {Ba{\~n}ados}, E. and {Benotto}, P. and {Catone}, M. and {Buitrago}, F. and {Daddi}, E. and {Girardi}, G. and {Giulietti}, M. and {Hirschmann}, M. and {Holwerda}, B.~W. and {Arrabal Haro}, P. and {Lapi}, A. and {Lucas}, R.~A. and {Lyu}, Y. and {Massardi}, M. and {Pacucci}, F. and {P{\'e}rez-Gonz{\'a}lez}, P.~G. and {Ronconi}, T. and {Tarrasse}, M. and {Wilkins}, S. and {Vulcani}, B. and {Yung}, L.~Y.~A. and {Zavala}, J.~A. and {Backhaus}, B. and {Bagley}, M. and {Buat}, V. and {Burgarella}, D. and {Kartaltepe}, J. and {Khusanova}, Y. and {Kirkpatrick}, A. and {Kocevski}, D. and {Koekemoer}, A.~M. and {Lambrides}, E. and {Pirzkal}, N. and {Yang}, G.},
        title = "{Ultrahigh-redshift or closer-by, dust-obscured galaxies?: Deciphering the nature of faint, previously missed F200W dropouts in CEERS}",
      journal = {\aap},
         year = 2026,
        month = apr,
       volume = {708},
          eid = {A195},
        pages = {A195},
          doi = {10.1051/0004-6361/202554009},
archivePrefix = {arXiv},
       eprint = {2502.02637},
 primaryClass = {astro-ph.GA},
       adsurl = {https://ui.adsabs.harvard.edu/abs/2026A&A...708A.195G}
}

@ARTICLE{Graham_2005,
       author = {{Graham}, Alister W. and {Driver}, Simon P.},
        title = "{A Concise Reference to (Projected) S{\'e}rsic R$^{1/n}$ Quantities, Including Concentration, Profile Slopes, Petrosian Indices, and Kron Magnitudes}",
      journal = {\pasa},
         year = 2005,
        month = jan,
       volume = {22},
       number = {2},
        pages = {118-127},
          doi = {10.1071/AS05001},
archivePrefix = {arXiv},
       eprint = {astro-ph/0503176},
 primaryClass = {astro-ph},
       adsurl = {https://ui.adsabs.harvard.edu/abs/2005PASA...22..118G}
}

@ARTICLE{Gruppioni_2013,
       author = {{Gruppioni}, C. and {Pozzi}, F. and {Rodighiero}, G. and {Delvecchio}, I. and {Berta}, S. and {Pozzetti}, L. and {Zamorani}, G. and {Andreani}, P. and {Cimatti}, A. and {Ilbert}, O. and {Le Floc'h}, E. and {Lutz}, D. and {Magnelli}, B. and {Marchetti}, L. and {Monaco}, P. and {Nordon}, R. and {Oliver}, S. and {Popesso}, P. and {Riguccini}, L. and {Roseboom}, I. and {Rosario}, D.~J. and {Sargent}, M. and {Vaccari}, M. and {Altieri}, B. and {Aussel}, H. and {Bongiovanni}, A. and {Cepa}, J. and {Daddi}, E. and {Dom{\'\i}nguez-S{\'a}nchez}, H. and {Elbaz}, D. and {F{\"o}rster Schreiber}, N. and {Genzel}, R. and {Iribarrem}, A. and {Magliocchetti}, M. and {Maiolino}, R. and {Poglitsch}, A. and {P{\'e}rez Garc{\'\i}a}, A. and {Sanchez-Portal}, M. and {Sturm}, E. and {Tacconi}, L. and {Valtchanov}, I. and {Amblard}, A. and {Arumugam}, V. and {Bethermin}, M. and {Bock}, J. and {Boselli}, A. and {Buat}, V. and {Burgarella}, D. and {Castro-Rodr{\'\i}guez}, N. and {Cava}, A. and {Chanial}, P. and {Clements}, D.~L. and {Conley}, A. and {Cooray}, A. and {Dowell}, C.~D. and {Dwek}, E. and {Eales}, S. and {Franceschini}, A. and {Glenn}, J. and {Griffin}, M. and {Hatziminaoglou}, E. and {Ibar}, E. and {Isaak}, K. and {Ivison}, R.~J. and {Lagache}, G. and {Levenson}, L. and {Lu}, N. and {Madden}, S. and {Maffei}, B. and {Mainetti}, G. and {Nguyen}, H.~T. and {O'Halloran}, B. and {Page}, M.~J. and {Panuzzo}, P. and {Papageorgiou}, A. and {Pearson}, C.~P. and {P{\'e}rez-Fournon}, I. and {Pohlen}, M. and {Rigopoulou}, D. and {Rowan-Robinson}, M. and {Schulz}, B. and {Scott}, D. and {Seymour}, N. and {Shupe}, D.~L. and {Smith}, A.~J. and {Stevens}, J.~A. and {Symeonidis}, M. and {Trichas}, M. and {Tugwell}, K.~E. and {Vigroux}, L. and {Wang}, L. and {Wright}, G. and {Xu}, C.~K. and {Zemcov}, M. and {Bardelli}, S. and {Carollo}, M. and {Contini}, T. and {Le F{\'e}vre}, O. and {Lilly}, S. and {Mainieri}, V. and {Renzini}, A. and {Scodeggio}, M. and {Zucca}, E.},
        title = "{The Herschel PEP/HerMES luminosity function - I. Probing the evolution of PACS selected Galaxies to z ≃ 4}",
      journal = {\mnras},
         year = 2013,
        month = jun,
       volume = {432},
       number = {1},
        pages = {23-52},
          doi = {10.1093/mnras/stt308},
archivePrefix = {arXiv},
       eprint = {1302.5209},
 primaryClass = {astro-ph.CO},
       adsurl = {https://ui.adsabs.harvard.edu/abs/2013MNRAS.432...23G}
}

@ARTICLE{Hainline_2024,
       author = {{Hainline}, Kevin N. and {Helton}, Jakob M. and {Johnson}, Benjamin D. and {Sun}, Fengwu and {Topping}, Michael W. and {Leisenring}, Jarron M. and {Baker}, William M. and {Eisenstein}, Daniel J. and {Hausen}, Ryan and {Hviding}, Raphael E. and {Lyu}, Jianwei and {Robertson}, Brant and {Tacchella}, Sandro and {Williams}, Christina C. and {Willmer}, Christopher N.~A. and {Roellig}, Thomas L.},
        title = "{Brown Dwarf Candidates in the JADES and CEERS Extragalactic Surveys}",
      journal = {\apj},
         year = 2024,
        month = mar,
       volume = {964},
       number = {1},
          eid = {66},
        pages = {66},
          doi = {10.3847/1538-4357/ad20d1},
archivePrefix = {arXiv},
       eprint = {2309.03250},
 primaryClass = {astro-ph.SR},
       adsurl = {https://ui.adsabs.harvard.edu/abs/2024ApJ...964...66H}
}

@ARTICLE{Hainline_2024b,
       author = {{Hainline}, Kevin N. and {Johnson}, Benjamin D. and {Robertson}, Brant and {Tacchella}, Sandro and {Helton}, Jakob M. and {Sun}, Fengwu and {Eisenstein}, Daniel J. and {Simmonds}, Charlotte and {Topping}, Michael W. and {Whitler}, Lily and {Willmer}, Christopher N.~A. and {Rieke}, Marcia and {Suess}, Katherine A. and {Hviding}, Raphael E. and {Cameron}, Alex J. and {Alberts}, Stacey and {Baker}, William M. and {Baum}, Stefi and {Bhatawdekar}, Rachana and {Bonaventura}, Nina and {Boyett}, Kristan and {Bunker}, Andrew J. and {Carniani}, Stefano and {Charlot}, Stephane and {Chevallard}, Jacopo and {Chen}, Zuyi and {Curti}, Mirko and {Curtis-Lake}, Emma and {D'Eugenio}, Francesco and {Egami}, Eiichi and {Endsley}, Ryan and {Hausen}, Ryan and {Ji}, Zhiyuan and {Looser}, Tobias J. and {Lyu}, Jianwei and {Maiolino}, Roberto and {Nelson}, Erica and {Pusk{\'a}s}, D{\'a}vid and {Rawle}, Tim and {Sandles}, Lester and {Saxena}, Aayush and {Smit}, Renske and {Stark}, Daniel P. and {Williams}, Christina C. and {Willott}, Chris and {Witstok}, Joris},
        title = "{The Cosmos in Its Infancy: JADES Galaxy Candidates at z > 8 in GOODS-S and GOODS-N}",
      journal = {\apj},
         year = 2024,
        month = mar,
       volume = {964},
       number = {1},
          eid = {71},
        pages = {71},
          doi = {10.3847/1538-4357/ad1ee4},
archivePrefix = {arXiv},
       eprint = {2306.02468},
 primaryClass = {astro-ph.GA},
       adsurl = {https://ui.adsabs.harvard.edu/abs/2024ApJ...964...71H}
}

@Article{Harris_2020,
 title         = {Array programming with {NumPy}},
 author        = {Charles R. Harris and K. Jarrod Millman and St{\'{e}}fan J.
                 van der Walt and Ralf Gommers and Pauli Virtanen and David
                 Cournapeau and Eric Wieser and Julian Taylor and Sebastian
                 Berg and Nathaniel J. Smith and Robert Kern and Matti Picus
                 and Stephan Hoyer and Marten H. van Kerkwijk and Matthew
                 Brett and Allan Haldane and Jaime Fern{\'{a}}ndez del
                 R{\'{i}}o and Mark Wiebe and Pearu Peterson and Pierre
                 G{\'{e}}rard-Marchant and Kevin Sheppard and Tyler Reddy and
                 Warren Weckesser and Hameer Abbasi and Christoph Gohlke and
                 Travis E. Oliphant},
 year          = {2020},
 month         = sep,
 journal       = {Nature},
 volume        = {585},
 number        = {7825},
 pages         = {357--362},
 doi           = {10.1038/s41586-020-2649-2},
 publisher     = {Springer Science and Business Media {LLC}},
 url           = {https://doi.org/10.1038/s41586-020-2649-2}
}

@ARTICLE{Harvey_2025,
       author = {{Harvey}, Thomas and {Conselice}, Christopher J. and {Adams}, Nathan J. and {Austin}, Duncan and {Juod{\v{z}}balis}, Ignas and {Trussler}, James and {Li}, Qiong and {Ormerod}, Katherine and {Ferreira}, Leonardo and {Lovell}, Christopher C. and {Duan}, Qiao and {Westcott}, Lewi and {Harris}, Honor and {Bhatawdekar}, Rachana and {Coe}, Dan and {Cohen}, Seth H. and {Caruana}, Joseph and {Cheng}, Cheng and {Driver}, Simon P. and {Frye}, Brenda and {Furtak}, Lukas J. and {Grogin}, Norman A. and {Hathi}, Nimish P. and {Holwerda}, Benne W. and {Jansen}, Rolf A. and {Koekemoer}, Anton M. and {Marshall}, Madeline A. and {Nonino}, Mario and {Vijayan}, Aswin P. and {Wilkins}, Stephen M. and {Windhorst}, Rogier and {Willmer}, Christopher N.~A. and {Yan}, Haojing and {Zitrin}, Adi},
        title = "{EPOCHS. IV. SED Modeling Assumptions and Their Impact on the Stellar Mass Function at 6.5 {\ensuremath{\leq}} z {\ensuremath{\leq}} 13.5 Using PEARLS and Public JWST Observations}",
      journal = {\apj},
         year = 2025,
        month = jan,
       volume = {978},
       number = {1},
          eid = {89},
        pages = {89},
          doi = {10.3847/1538-4357/ad8c29},
archivePrefix = {arXiv},
       eprint = {2403.03908},
 primaryClass = {astro-ph.GA},
       adsurl = {https://ui.adsabs.harvard.edu/abs/2025ApJ...978...89H}
}

@ARTICLE{Harvey_2025_syn,
       author = {{Harvey}, Thomas and {Lovell}, Christopher C. and {Newman}, Sophie and {Conselice}, Christopher J. and {Austin}, Duncan and {Roper}, William J. and {Vijayan}, Aswin P. and {Wilkins}, Stephen M. and {Iglesias-Navarro}, Patricia and {Rusakov}, Vadim and {Li}, Qiong and {Adams}, Nathan and {Magdwick}, Kai and {Goolsby}, Caio M. and {Huertas-Company}, Marc and {Ho}, Matthew},
        title = "{Flexible simulation-based inference for galaxy photometric fitting with synthesizer}",
      journal = {\mnras},
         year = 2026,
        month = mar,
       volume = {547},
       number = {1},
          eid = {stag282},
        pages = {stag282},
          doi = {10.1093/mnras/stag282},
archivePrefix = {arXiv},
       eprint = {2511.10640},
 primaryClass = {astro-ph.GA},
       adsurl = {https://ui.adsabs.harvard.edu/abs/2026MNRAS.547ag282H}
}

@ARTICLE{Heintz_2023,
       author = {{Heintz}, K.~E. and {Gim{\'e}nez-Arteaga}, C. and {Fujimoto}, S. and {Brammer}, G. and {Espada}, D. and {Gillman}, S. and {Gonz{\'a}lez-L{\'o}pez}, J. and {Greve}, T.~R. and {Harikane}, Y. and {Hatsukade}, B. and {Knudsen}, K.~K. and {Koekemoer}, A.~M. and {Kohno}, K. and {Kokorev}, V. and {Lee}, M.~M. and {Magdis}, G.~E. and {Nelson}, E.~J. and {Rizzo}, F. and {Sanders}, R.~L. and {Schaerer}, D. and {Shapley}, A.~E. and {Strait}, V.~B. and {Toft}, S. and {Valentino}, F. and {van der Wel}, A. and {Vijayan}, A.~P. and {Watson}, D. and {Bauer}, F.~E. and {Christiansen}, C.~R. and {Wilson}, S.~N.},
        title = "{The Gas and Stellar Content of a Metal-poor Galaxy at z = 8.496 as Revealed by JWST and ALMA}",
      journal = {\apjl},
         year = 2023,
        month = feb,
       volume = {944},
       number = {2},
          eid = {L30},
        pages = {L30},
          doi = {10.3847/2041-8213/acb2cf},
archivePrefix = {arXiv},
       eprint = {2212.06877},
 primaryClass = {astro-ph.GA},
       adsurl = {https://ui.adsabs.harvard.edu/abs/2023ApJ...944L..30H}
}

@ARTICLE{Heintz_2024,
       author = {{Heintz}, Kasper E. and {Watson}, Darach and {Brammer}, Gabriel and {Vejlgaard}, Simone and {Hutter}, Anne and {Strait}, Victoria B. and {Matthee}, Jorryt and {Oesch}, Pascal A. and {Jakobsson}, P{\'a}ll and {Tanvir}, Nial R. and {Laursen}, Peter and {Naidu}, Rohan P. and {Mason}, Charlotte A. and {Killi}, Meghana and {Jung}, Intae and {Hsiao}, Tiger Yu-Yang and {Abdurro'uf} and {Coe}, Dan and {Arrabal Haro}, Pablo and {Finkelstein}, Steven L. and {Toft}, Sune},
        title = "{Strong damped Lyman-{\ensuremath{\alpha}} absorption in young star-forming galaxies at redshifts 9 to 11}",
      journal = {Science},
         year = 2024,
        month = may,
       volume = {384},
       number = {6698},
        pages = {890-894},
          doi = {10.1126/science.adj0343},
archivePrefix = {arXiv},
       eprint = {2306.00647},
 primaryClass = {astro-ph.GA},
       adsurl = {https://ui.adsabs.harvard.edu/abs/2024Sci...384..890H}
}

@ARTICLE{Heintz_2025,
       author = {{Heintz}, K.~E. and {Brammer}, G.~B. and {Watson}, D. and {Oesch}, P.~A. and {Keating}, L.~C. and {Hayes}, M.~J. and {Abdurro'uf} and {Arellano-C{\'o}rdova}, K.~Z. and {Carnall}, A.~C. and {Christiansen}, C.~R. and {Cullen}, F. and {Dav{\'e}}, R. and {Dayal}, P. and {Ferrara}, A. and {Finlator}, K. and {Fynbo}, J.~P.~U. and {Flury}, S.~R. and {Gelli}, V. and {Gillman}, S. and {Gottumukkala}, R. and {Gould}, K. and {Greve}, T.~R. and {Hardin}, S.~E. and {Hsiao}, T.~Y.-Y. and {Hutter}, A. and {Jakobsson}, P. and {Killi}, M. and {Khosravaninezhad}, N. and {Laursen}, P. and {Lee}, M.~M. and {Magdis}, G.~E. and {Matthee}, J. and {Naidu}, R.~P. and {Narayanan}, D. and {Pollock}, C. and {Prescott}, M.~K.~M. and {Rusakov}, V. and {Shuntov}, M. and {Sneppen}, A. and {Smit}, R. and {Tanvir}, N.~R. and {Terp}, C. and {Toft}, S. and {Valentino}, F. and {Vijayan}, A.~P. and {Weaver}, J.~R. and {Wise}, J.~H. and {Witstok}, J.},
        title = "{The JWST-PRIMAL archival survey: A JWST/NIRSpec reference sample for the physical properties and Lyman-{\ensuremath{\alpha}} absorption and emission of {\ensuremath{\sim}}600 galaxies at z = 5.0 {\ensuremath{-}} 13.4}",
      journal = {\aap},
         year = 2025,
        month = jan,
       volume = {693},
          eid = {A60},
        pages = {A60},
          doi = {10.1051/0004-6361/202450243},
archivePrefix = {arXiv},
       eprint = {2404.02211},
 primaryClass = {astro-ph.GA},
       adsurl = {https://ui.adsabs.harvard.edu/abs/2025A&A...693A..60H}
}

@ARTICLE{Hirschmann_2017,
       author = {{Hirschmann}, Michaela and {Charlot}, Stephane and {Feltre}, Anna and {Naab}, Thorsten and {Choi}, Ena and {Ostriker}, Jeremiah P. and {Somerville}, Rachel S.},
        title = "{Synthetic nebular emission from massive galaxies - I: origin of the cosmic evolution of optical emission-line ratios}",
      journal = {\mnras},
         year = 2017,
        month = dec,
       volume = {472},
       number = {2},
        pages = {2468-2495},
          doi = {10.1093/mnras/stx2180},
archivePrefix = {arXiv},
       eprint = {1706.00010},
 primaryClass = {astro-ph.GA},
       adsurl = {https://ui.adsabs.harvard.edu/abs/2017MNRAS.472.2468H}
}

@ARTICLE{Hirschmann_2019,
       author = {{Hirschmann}, Michaela and {Charlot}, Stephane and {Feltre}, Anna and {Naab}, Thorsten and {Somerville}, Rachel S. and {Choi}, Ena},
        title = "{Synthetic nebular emission from massive galaxies - II. Ultraviolet-line diagnostics of dominant ionizing sources}",
      journal = {\mnras},
         year = 2019,
        month = jul,
       volume = {487},
       number = {1},
        pages = {333-353},
          doi = {10.1093/mnras/stz1256},
archivePrefix = {arXiv},
       eprint = {1811.07909},
 primaryClass = {astro-ph.GA},
       adsurl = {https://ui.adsabs.harvard.edu/abs/2019MNRAS.487..333H}
}

@ARTICLE{Hirschmann_2023,
       author = {{Hirschmann}, Michaela and {Charlot}, Stephane and {Feltre}, Anna and {Curtis-Lake}, Emma and {Somerville}, Rachel S. and {Chevallard}, Jacopo and {Choi}, Ena and {Nelson}, Dylan and {Morisset}, Christophe and {Plat}, Adele and {Vidal-Garcia}, Alba},
        title = "{Emission-line properties of IllustrisTNG galaxies: from local diagnostic diagrams to high-redshift predictions for JWST}",
      journal = {\mnras},
         year = 2023,
        month = dec,
       volume = {526},
       number = {3},
        pages = {3610-3636},
          doi = {10.1093/mnras/stad2955},
archivePrefix = {arXiv},
       eprint = {2212.02522},
 primaryClass = {astro-ph.GA},
       adsurl = {https://ui.adsabs.harvard.edu/abs/2023MNRAS.526.3610H}
}

@ARTICLE{Holwerda_2022,
       author = {{Holwerda}, Benne W. and {Smith}, Dominic and {Porter}, Lori and {Henry}, Chris and {Porter-Temple}, Ren and {Cook}, Kyle and {Pimbblet}, Kevin A. and {Hopkins}, Andrew M. and {Bilicki}, Maciej and {Turner}, Sebastian and {Acquaviva}, Viviana and {Wang}, Lingyu and {Wright}, Angus H. and {Kelvin}, Lee S. and {Grootes}, Meiert W.},
        title = "{Galaxy and mass assembly (GAMA): Self-Organizing Map application on nearby galaxies}",
      journal = {\mnras},
         year = 2022,
        month = jun,
       volume = {513},
       number = {2},
        pages = {1972-1984},
          doi = {10.1093/mnras/stac889},
archivePrefix = {arXiv},
       eprint = {2203.15611},
 primaryClass = {astro-ph.GA},
       adsurl = {https://ui.adsabs.harvard.edu/abs/2022MNRAS.513.1972H}
}

@Article{Hunter_2007,
  Author    = {Hunter, J. D.},
  Title     = {Matplotlib: A 2D graphics environment},
  Journal   = {Computing in Science \& Engineering},
  Volume    = {9},
  Number    = {3},
  Pages     = {90--95},
  publisher = {IEEE COMPUTER SOC},
  doi       = {10.1109/MCSE.2007.55},
  year      = 2007
}

@ARTICLE{Iyer_2025,
       author = {{Iyer}, Kartheik G. and {Pacifici}, Camilla and {Calistro-Rivera}, Gabriela and {Lovell}, Christopher C.},
        title = "{The Spectral Energy Distributions of Galaxies}",
      journal = {arXiv e-prints},
         year = 2025,
        month = feb,
          eid = {arXiv:2502.17680},
        pages = {arXiv:2502.17680},
          doi = {10.48550/arXiv.2502.17680},
archivePrefix = {arXiv},
       eprint = {2502.17680},
 primaryClass = {astro-ph.GA},
       adsurl = {https://ui.adsabs.harvard.edu/abs/2025arXiv250217680I}
}

@article{Jia_2022,
	author = {Jia, Weikuan and Sun, Meili and Lian, Jian and Hou, Sujuan},
	date = {2022/06/01},
	doi = {10.1007/s40747-021-00637-x},
	id = {Jia2022},
	isbn = {2198-6053},
	journal = {Complex \& Intelligent Systems},
	number = {3},
	pages = {2663--2693},
	title = {Feature dimensionality reduction: a review},
	url = {https://doi.org/10.1007/s40747-021-00637-x},
	volume = {8},
	year = {2022}}

@ARTICLE{Juodzbalis_2023,
       author = {{Juod{\v{z}}balis}, Ignas and {Conselice}, Christopher J. and {Singh}, Maitrayee and {Adams}, Nathan and {Ormerod}, Katherine and {Harvey}, Thomas and {Austin}, Duncan and {Volonteri}, Marta and {Cohen}, Seth H. and {Jansen}, Rolf A. and {Summers}, Jake and {Windhorst}, Rogier A. and {D'Silva}, Jordan C.~J. and {Koekemoer}, Anton M. and {Coe}, Dan and {Driver}, Simon P. and {Frye}, Brenda and {Grogin}, Norman A. and {Marshall}, Madeline A. and {Nonino}, Mario and {Pirzkal}, Nor and {Robotham}, Aaron and {}, Russell E., Jr., Ryan and {Ortiz}, III, Rafael and {Tompkins}, Scott and {Willmer}, Christopher N.~A. and {Yan}, Haojing},
        title = "{EPOCHS VII: discovery of high-redshift (6.5 < z < 12) AGN candidates in JWST ERO and PEARLS data}",
      journal = {\mnras},
         year = 2023,
        month = oct,
       volume = {525},
       number = {1},
        pages = {1353-1364},
          doi = {10.1093/mnras/stad2396},
archivePrefix = {arXiv},
       eprint = {2307.07535},
 primaryClass = {astro-ph.GA},
       adsurl = {https://ui.adsabs.harvard.edu/abs/2023MNRAS.525.1353J}
}

@ARTICLE{Kamai_2025,
       author = {{Kamai}, Ilay and {Bronstein}, Alex M. and {Perets}, Hagai B.},
        title = "{Machine Learning Inference of Stellar Properties Using Integrated Photometric and Spectroscopic Data}",
      journal = {\apj},
         year = 2025,
        month = nov,
       volume = {994},
       number = {1},
          eid = {110},
        pages = {110},
          doi = {10.3847/1538-4357/ae0cbc},
archivePrefix = {arXiv},
       eprint = {2507.10666},
 primaryClass = {astro-ph.SR},
       adsurl = {https://ui.adsabs.harvard.edu/abs/2025ApJ...994..110K}
}

@ARTICLE{Katz_2025,
       author = {{Katz}, Harley and {Rey}, Martin P. and {Cadiou}, Corentin and {Agertz}, Oscar and {Blaizot}, Jeremy and {Cameron}, Alex J. and {Choustikov}, Nicholas and {Devriendt}, Julien and {Hauk}, Uliana and {Jones}, Gareth C. and {Kimm}, Taysun and {Laseter}, Isaac and {Martin-Alvarez}, Sergio and {Matsumoto}, Kosei and {Pearce}, Autumn and {Rodr{\'\i}guez Montero}, Francisco and {Rosdahl}, Joki and {Sanati}, Mahsa and {Saxena}, Aayush and {Slyz}, Adrianne and {Stiskalek}, Richard and {Storck}, Anatole and {Veenema}, Oscar and {Yee}, Wonjae},
        title = "{MEGATRON: Reproducing the Diversity of High-Redshift Galaxy Spectra with Cosmological Radiation Hydrodynamics Simulations}",
      journal = {arXiv e-prints},
         year = 2025,
        month = oct,
          eid = {arXiv:2510.05201},
        pages = {arXiv:2510.05201},
          doi = {10.48550/arXiv.2510.05201},
archivePrefix = {arXiv},
       eprint = {2510.05201},
 primaryClass = {astro-ph.GA},
       adsurl = {https://ui.adsabs.harvard.edu/abs/2025arXiv251005201K}
}

@ARTICLE{Katz_2026,
       author = {{Katz}, Harley and {Rey}, Martin P. and {Cadiou}, Corentin and {Kimm}, Taysun and {Agertz}, Oscar},
        title = "{The Impact of Star Formation and Feedback Recipes on the Stellar Mass and Interstellar Medium of High-Redshift Galaxies}",
      journal = {\oja},
         year = 2026,
        month = feb,
       volume = {9},
        pages = {56097},
          doi = {10.33232/001c.156097},
archivePrefix = {arXiv},
       eprint = {2411.07282},
 primaryClass = {astro-ph.GA},
       adsurl = {https://ui.adsabs.harvard.edu/abs/2026OJAp....956097K}
}

@ARTICLE{Kewley_2019,
       author = {{Kewley}, Lisa J. and {Nicholls}, David C. and {Sutherland}, Ralph S.},
        title = "{Understanding Galaxy Evolution Through Emission Lines}",
      journal = {\araa},
         year = 2019,
        month = aug,
       volume = {57},
        pages = {511-570},
          doi = {10.1146/annurev-astro-081817-051832},
archivePrefix = {arXiv},
       eprint = {1910.09730},
 primaryClass = {astro-ph.GA},
       adsurl = {https://ui.adsabs.harvard.edu/abs/2019ARA&A..57..511K}
}

@ARTICLE{Klypin_2016,
       author = {{Klypin}, Anatoly and {Yepes}, Gustavo and {Gottl{\"o}ber}, Stefan and {Prada}, Francisco and {He{\ss}}, Steffen},
        title = "{MultiDark simulations: the story of dark matter halo concentrations and density profiles}",
      journal = {\mnras},
         year = 2016,
        month = apr,
       volume = {457},
       number = {4},
        pages = {4340-4359},
          doi = {10.1093/mnras/stw248},
archivePrefix = {arXiv},
       eprint = {1411.4001},
 primaryClass = {astro-ph.CO},
       adsurl = {https://ui.adsabs.harvard.edu/abs/2016MNRAS.457.4340K}
}

@ARTICLE{Kovacevic_2022,
       author = {{Kova{\v{c}}evi{\'c}}, M. and {Pasquato}, M. and {Marelli}, M. and {De Luca}, A. and {Salvaterra}, R. and {Belfiore}, A.},
        title = "{Exploring X-ray variability with unsupervised machine learning. I. Self-organizing maps applied to XMM-Newton data}",
      journal = {\aap},
         year = 2022,
        month = mar,
       volume = {659},
          eid = {A66},
        pages = {A66},
          doi = {10.1051/0004-6361/202142444},
archivePrefix = {arXiv},
       eprint = {2202.08868},
 primaryClass = {astro-ph.IM},
       adsurl = {https://ui.adsabs.harvard.edu/abs/2022A&A...659A..66K}
}

@ARTICLE{Kovlakas_2021,
       author = {{Kovlakas}, K. and {Zezas}, A. and {Andrews}, J.~J. and {Basu-Zych}, A. and {Fragos}, T. and {Hornschemeier}, A. and {Kouroumpatzakis}, K. and {Lehmer}, B. and {Ptak}, A.},
        title = "{The Heraklion Extragalactic Catalogue (HECATE): a value-added galaxy catalogue for multimessenger astrophysics}",
      journal = {\mnras},
         year = 2021,
        month = sep,
       volume = {506},
       number = {2},
        pages = {1896-1915},
          doi = {10.1093/mnras/stab1799},
archivePrefix = {arXiv},
       eprint = {2106.12101},
 primaryClass = {astro-ph.GA},
       adsurl = {https://ui.adsabs.harvard.edu/abs/2021MNRAS.506.1896K}
}

@ARTICLE{Kron_1980,
       author = {{Kron}, R.~G.},
        title = "{Photometry of a complete sample of faint galaxies.}",
      journal = {\apjs},
         year = 1980,
        month = jun,
       volume = {43},
        pages = {305-325},
          doi = {10.1086/190669},
       adsurl = {https://ui.adsabs.harvard.edu/abs/1980ApJS...43..305K}
}

@ARTICLE{Leethochawalit_2022,
       author = {{Leethochawalit}, Nicha and {Trenti}, Michele and {Morishita}, Takahiro and {Roberts-Borsani}, Guido and {Treu}, Tommaso},
        title = "{A quantitative assessment of completeness correction methods and public release of a versatile simulation code}",
      journal = {\mnras},
         year = 2022,
        month = feb,
       volume = {509},
       number = {4},
        pages = {5836-5857},
          doi = {10.1093/mnras/stab3265},
archivePrefix = {arXiv},
       eprint = {2111.08528},
 primaryClass = {astro-ph.GA},
       adsurl = {https://ui.adsabs.harvard.edu/abs/2022MNRAS.509.5836L}
}

@ARTICLE{Lovell_2025,
       author = {{Lovell}, Christopher C. and {Starkenburg}, Tjitske and {Ho}, Matthew and {Angl{\'e}s-Alc{\'a}zar}, Daniel and {Dav{\'e}}, Romeel and {Gabrielpillai}, Austen and {Iyer}, Kartheik G. and {Matthews}, Alice E. and {Roper}, William J. and {Somerville}, Rachel S. and {Sommovigo}, Laura and {Villaescusa-Navarro}, Francisco},
        title = "{Learning the Universe: Cosmological and Astrophysical Parameter Inference with Galaxy Luminosity Functions and Colours}",
      journal = {\mnras},
         year = 2025,
        month = oct,
          doi = {10.1093/mnras/staf1888},
archivePrefix = {arXiv},
       eprint = {2411.13960},
 primaryClass = {astro-ph.GA},
       adsurl = {https://ui.adsabs.harvard.edu/abs/2025MNRAS.tmp.1777L}
}

@ARTICLE{Lovell_2025b,
       author = {{Lovell}, Christopher C. and {Roper}, William J. and {Vijayan}, Aswin P. and {Wilkins}, Stephen M. and {Newman}, Sophie and {Seeyave}, Louise},
        title = "{Synthesizer: a Software Package for Synthetic Astronomical Observables}",
      journal = {The Open Journal of Astrophysics},
         year = 2025,
        month = oct,
       volume = {8},
          eid = {152},
        pages = {152},
          doi = {10.33232/001c.145766},
archivePrefix = {arXiv},
       eprint = {2508.03888},
 primaryClass = {astro-ph.IM},
       adsurl = {https://ui.adsabs.harvard.edu/abs/2025OJAp....8E.152L}
}

@ARTICLE{Ma_2016_bin,
       author = {{Ma}, Xiangcheng and {Hopkins}, Philip F. and {Kasen}, Daniel and {Quataert}, Eliot and {Faucher-Gigu{\`e}re}, Claude-Andr{\'e} and {Kere{\v{s}}}, Du{\v{s}}an and {Murray}, Norman and {Strom}, Allison},
        title = "{Binary stars can provide the `missing photons' needed for reionization}",
      journal = {\mnras},
         year = 2016,
        month = jul,
       volume = {459},
       number = {4},
        pages = {3614-3619},
          doi = {10.1093/mnras/stw941},
archivePrefix = {arXiv},
       eprint = {1601.07559},
 primaryClass = {astro-ph.GA},
       adsurl = {https://ui.adsabs.harvard.edu/abs/2016MNRAS.459.3614M}
}

@ARTICLE{Madau_1996,
       author = {{Madau}, Piero and {Ferguson}, Henry C. and {Dickinson}, Mark E. and {Giavalisco}, Mauro and {Steidel}, Charles C. and {Fruchter}, Andrew},
        title = "{High-redshift galaxies in the Hubble Deep Field: colour selection and star formation history to z\raisebox{-0.5ex}\textasciitilde4}",
      journal = {\mnras},
         year = 1996,
        month = dec,
       volume = {283},
       number = {4},
        pages = {1388-1404},
          doi = {10.1093/mnras/283.4.1388},
archivePrefix = {arXiv},
       eprint = {astro-ph/9607172},
 primaryClass = {astro-ph},
       adsurl = {https://ui.adsabs.harvard.edu/abs/1996MNRAS.283.1388M}
}

@ARTICLE{Maiolino_2024,
       author = {{Maiolino}, Roberto and {{\"U}bler}, Hannah and {Perna}, Michele and {Scholtz}, Jan and {D'Eugenio}, Francesco and {Witten}, Callum and {Laporte}, Nicolas and {Witstok}, Joris and {Carniani}, Stefano and {Tacchella}, Sandro and {Baker}, William M. and {Arribas}, Santiago and {Nakajima}, Kimihiko and {Eisenstein}, Daniel J. and {Bunker}, Andrew J. and {Charlot}, St{\'e}phane and {Cresci}, Giovanni and {Curti}, Mirko and {Curtis-Lake}, Emma and {de Graaff}, Anna and {Egami}, Eiichi and {Ji}, Zhiyuan and {Johnson}, Benjamin D. and {Kumari}, Nimisha and {Looser}, Tobias J. and {Maseda}, Michael and {Nelson}, Erica and {Robertson}, Brant and {Rodr{\'\i}guez Del Pino}, Bruno and {Sandles}, Lester and {Simmonds}, Charlotte and {Smit}, Renske and {Sun}, Fengwu and {Venturi}, Giacomo and {Williams}, Christina C. and {Willmer}, Christopher N.~A.},
        title = "{JADES. Possible Population III signatures at z = 10.6 in the halo of GN-z11}",
      journal = {\aap},
         year = 2024,
        month = jul,
       volume = {687},
          eid = {A67},
        pages = {A67},
          doi = {10.1051/0004-6361/202347087},
archivePrefix = {arXiv},
       eprint = {2306.00953},
 primaryClass = {astro-ph.GA},
       adsurl = {https://ui.adsabs.harvard.edu/abs/2024A&A...687A..67M}
}

@ARTICLE{Mathis_1983,
       author = {{Mathis}, J.~S. and {Mezger}, P.~G. and {Panagia}, N.},
        title = "{Interstellar radiation field and dust temperatures in the diffuse interstellar medium and in giant molecular clouds}",
      journal = {\aap},
         year = 1983,
        month = nov,
       volume = {128},
        pages = {212-229},
       adsurl = {https://ui.adsabs.harvard.edu/abs/1983A&A...128..212M}
}

@ARTICLE{McInnes_2018,
       author = {{McInnes}, Leland and {Healy}, John and {Melville}, James},
        title = "{UMAP: Uniform Manifold Approximation and Projection for Dimension Reduction}",
      journal = {arXiv e-prints},
         year = 2018,
        month = feb,
          eid = {arXiv:1802.03426},
        pages = {arXiv:1802.03426},
          doi = {10.48550/arXiv.1802.03426},
archivePrefix = {arXiv},
       eprint = {1802.03426},
 primaryClass = {stat.ML},
       adsurl = {https://ui.adsabs.harvard.edu/abs/2018arXiv180203426M}
}

@ARTICLE{Meurer_1999,
       author = {{Meurer}, Gerhardt R. and {Heckman}, Timothy M. and {Calzetti}, Daniela},
        title = "{Dust Absorption and the Ultraviolet Luminosity Density at z \raisebox{-0.5ex}\textasciitilde 3 as Calibrated by Local Starburst Galaxies}",
      journal = {\apj},
         year = 1999,
        month = aug,
       volume = {521},
       number = {1},
        pages = {64-80},
          doi = {10.1086/307523},
archivePrefix = {arXiv},
       eprint = {astro-ph/9903054},
 primaryClass = {astro-ph},
       adsurl = {https://ui.adsabs.harvard.edu/abs/1999ApJ...521...64M}
}

@ARTICLE{Naidu_2025,
       author = {{Naidu}, Rohan P. and {Oesch}, Pascal A. and {Brammer}, Gabriel and {Weibel}, Andrea and {Li}, Yijia and {Matthee}, Jorryt and {Chisolm}, John and {Pollock}, Clara L. and {Heintz}, Kasper E. and {Johnson}, Benjamin D. and {Shen}, Xuejian and {Hviding}, Raphael E. and {Leja}, Joel and {Tacchella}, Sandro and {Ganguly}, Arpita and {Witten}, Callum and {Atek}, Hakim and {Belli}, Siro and {Bose}, Sownak and {Bouwens}, Rychard and {Dayal}, Pratika and {Decarli}, Roberto and {de Graaff}, Anna and {Fudamoto}, Yoshinobu and {Giovinazzo}, Emma and {Greene}, Jenny E. and {Illingworth}, Garth and {Inoue}, Akio K. and {Kane}, Sarah G. and {Labbe}, Ivo and {Leonova}, Ecaterina and {Marques-Chaves}, Rui and {Meyer}, Roman A. and {Nelson}, Erica J. and {Roberts-Borsani}, Guido and {Schaerer}, Daniel and {Simcoe}, Robert A. and {Stefanon}, Mauro and {Sugahara}, Yuma and {Toft}, Sune and {van der Wel}, Arjen and {van Dokkum}, Pieter and {Walter}, Fabian and {Watson}, Darrach and {Weaver}, John R. and {Whitaker}, Katherine E.},
        title = "{A Cosmic Miracle: A Remarkably Luminous Galaxy at zspec = 14.44 Confirmed with JWST}",
      journal = {The Open Journal of Astrophysics},
         year = 2026,
        month = jan,
       volume = {9},
        pages = {56033},
          doi = {10.33232/001c.156033},
archivePrefix = {arXiv},
       eprint = {2505.11263},
 primaryClass = {astro-ph.GA},
       adsurl = {https://ui.adsabs.harvard.edu/abs/2026OJAp....956033N}
}

@article{Narayanan_2024,
	author = {Narayanan, Desika and Lower, Sidney and Torrey, Paul and Brammer, Gabriel and Cui, Weiguang and Dav{\'e}, Romeel and Iyer, Kartheik G. and Li, Qi and Lovell, Christopher C. and Sales, Laura V. and Stark, Daniel P. and Marinacci, Federico and Vogelsberger, Mark},
	doi = {10.3847/1538-4357/ad0966},
	journal = {\apj},
	month = {jan},
	number = {1},
	pages = {73},
	publisher = {The American Astronomical Society},
	title = {Outshining by Recent Star Formation Prevents the Accurate Measurement of High-z Galaxy Stellar Masses},
	url = {https://dx.doi.org/10.3847/1538-4357/ad0966},
	volume = {961},
	year = {2024}}

@article{Pedregosa_2011,
  author  = {Fabian Pedregosa and Ga{{\"e}}l Varoquaux and Alexandre Gramfort and Vincent Michel and Bertrand Thirion and Olivier Grisel and Mathieu Blondel and Peter Prettenhofer and Ron Weiss and Vincent Dubourg and Jake Vanderplas and Alexandre Passos and David Cournapeau and Matthieu Brucher and Matthieu Perrot and {{\'E}}douard Duchesnay},
  title   = {Scikit-learn: Machine Learning in Python},
  journal = {Journal of Machine Learning Research},
  year    = {2011},
  volume  = {12},
  number  = {85},
  pages   = {2825-2830},
  url     = {http://jmlr.org/papers/v12/pedregosa11a.html}
}
